\documentclass[runningheads, envcountsame, a4paper]{llncs}

\usepackage[dvipsnames]{xcolor}
\usepackage[backgroundcolor=magenta!10!white]{todonotes}

\usepackage{microtype}

\usepackage{amsmath,amssymb,amsfonts}
\usepackage{thmtools,thm-restate}

\usepackage{enumitem}

\usepackage{graphicx}
\usepackage{grffile}

\usepackage{placeins}

\usepackage{subcaption}

\usepackage{ifthen}

\usepackage{numprint}
\npdecimalsign{.}
\npthousandsep{,}

\usepackage{hyperref}
\usepackage{cleveref}
\usepackage{tikz}
\usetikzlibrary{arrows,
automata,
calc,
decorations,
decorations.pathreplacing,
positioning,
shapes}

\usepackage{standalone}
\usepackage{pgfplots, pgfplotstable}
\pgfplotstableset{col sep=semicolon,columns/formula/.style={string type}}

\usepackage{calc}

\usepackage{caption}
\usepackage{lscape}
\usepackage{multirow}

\newcommand{\tudparagraph}[2]{\vspace*{#1}

\noindent
{\bf #2}
}

\def\<{\langle}
\def\>{\rangle}
\newcommand{\lbr}{\left\lbrace}
\newcommand{\rbr}{\right\rbrace}

\newcommand{\Nat}{\mathbb{N}}

\newcommand{\movetrans}[1]{\stackrel{#1}{\longrightarrow}\!\!^*}

\newcommand{\Dag}{\textsc{dag}}

\newcommand{\tgba}{t-GBA}
\newcommand{\vwaa}{VWAA}

\newcommand{\Inf}[1]{\mathrm{Inf}(#1)}
\newcommand{\Fin}[1]{\mathrm{Fin}(#1)}
\newcommand{\aut}{\mathcal{A}}

\newcommand{\AP}{\mathit{AP}}
\newcommand{\Acc}{\Phi}

\newcommand{\langinf}{\lang}
\newcommand{\autpath}{\pi}
\newcommand{\trim}{\mathsf{trim}}

\newcommand{\true}{\mathsf{true}}
\newcommand{\false}{\mathsf{false}}

\newcommand{\until}{\mathcal{U}}
\newcommand{\release}{\mathcal{R}}
\newcommand{\finally}{\Diamond}
\newcommand{\globally}{\Box}
\newcommand{\neXt}{\bigcirc}

\newcommand{\lang}{\mathcal{L}}

\newcommand{\pnodes}[1]{\nprounddigits{0}\numprint{#1}}

\newcommand{\prism}{\texttt{PRISM}}

\newcommand{\spot}{\texttt{SPOT}}
\newcommand{\Spot}{\spot}
\newcommand{\ltlthreeba}{\texttt{LTL3BA}}

\newcommand{\ltltotgba}{\texttt{ltl2tgba}}

\newcommand{\duggi}{\texttt{Duggi}}

\begin{document}

\pagestyle{headings}  

\title{From LTL to Unambiguous B\"uchi Automata via Disambiguation of Alternating Automata}
\titlerunning{From LTL to Unambiguous B\"uchi Automata}  

\author{Simon Jantsch\inst{1} \and
        David M\"uller\inst{1} \and
        Christel Baier\inst{1} \and
        Joachim Klein\inst{1} 
}

\authorrunning{Simon Jantsch et al.}

\institute{
Technische Universit\"at Dresden, Germany
\thanks{
        The authors are supported by the DFG through
        the Collaborative Research Centers CRC 912 (HAEC),
        the DFG grant 389792660 as part of \href{https://perspicuous-computing.science}{TRR~248},
        the DFG-project BA-1679/12-1,
        the Cluster of Excellence EXC 2050/1 (CeTI, project ID 390696704, as part of Germany's Excellence Strategy), and
        the Research Training Group QuantLA (GRK 1763).
}
}
\maketitle

\begin{abstract}

This paper proposes a new algorithm for the generation of unambiguous B\"uchi automata (UBA) from LTL formulas. Unlike existing tableau-based LTL-to-UBA translations,
our algorithm deals with very weak alternating automata
 (VWAA) as an intermediate representation. It relies on a new notion of
 unambiguity for VWAA and a disambiguation procedure for VWAA. We introduce
 optimizations on the VWAA level and new LTL simplifications targeted at
 generating small UBA.
We 
report on an implementation of the construction in our tool \duggi{}
and discuss experimental results that compare the automata sizes and 
computation times of \duggi{} with the 
tableau-based LTL-to-UBA translation of the \spot{} tool set. Our experiments
also cover the analysis of Markov chains under LTL specifications, which is an important application of UBA.
\end{abstract}
 
\section{Introduction}
Translations from linear temporal logic (LTL) to non-deterministic B\"uchi
automata (NBA) have been studied intensively as they are a core ingredient in
the classical algorithmic approach to LTL model checking (see, e.g. \cite{VardiWolper86,BK08,CGP01}).
In the worst case, such translations produce automata that are exponentially larger than the input formula.
However, a lot of effort has been put into optimizing the general case, which has
turned LTL-to-NBA translations feasible in practice. Two classes of algorithms have emerged
as being especially well suited: tableau-based decomposition of
the LTL formula into an automaton (see, e.g. \cite{GPVW95,Couvreur99}), as
represented by the \Spot{} family of tools~\cite{Duret16}, and translations via
very weak alternating automata (VWAA) \cite{GO01}, 
where \ltlthreeba{}~\cite{BKRS12} is the leading tool currently.

A property that has been studied in many areas of automata theory is \emph{unambiguity}~\cite{Colcombet15}.
It allows non-deterministic branching but requires that each input word has at most one accepting run.
Prominent cases in which unambiguity can be utilized include the universality check for automata (``Is every word accepted?'') on finite words, which is PSPACE-complete for arbitrary non-deterministic finite automata (NFA), but in P for unambiguous finite automata (UFA)~\cite{StearnsHunt}.
Another example is model checking of Markov chains, which is in P if the specification is given as an unambiguous B\"uchi automaton (UBA)~\cite{BKKKMW16}, and PSPACE-hard for arbitrary NBA \cite{Vardi85}.
Thus, using UBA leads to a single-exponential algorithm for LTL model checking of Markov chains, whereas using deterministic automata always involves a double-exponential lower bound in time complexity.

Every $\omega$-regular language is expressible by UBA~\cite{Arnold85}, but NBA may be exponentially more succinct than UBA~\cite{KJC13} and UBA may be exponentially more succinct than any deterministic automaton~\cite{BousLoed10}.
Universality and language inclusion are in P for subclasses of UBA~\cite{BousLoed10,IsaLoed12}, but the complexity is open for general UBA.

Although producing UBA was not the goal of the early translation
from LTL to NBA by Vardi and Wolper~\cite{VardiWolper86}, their construction is asymptotically optimal and produces \emph{separated} automata, a subclass of UBA where the languages of the states are pairwise disjoint.
Separated automata can express all $\omega$-regular languages~\cite{CartonM03}, but UBA may be exponentially more succinct~\cite{BousLoed10}.
LTL-to-NBA translations have been studied intensively~\cite{GO01,GPVW95,EH00,Duret13}, but the generation of UBA from LTL formulas has not received much attention so far.
We are only aware of three approaches targeted explicitly at generating UBA or subclasses.
The first approach by Couvreur et al. \cite{CouSahSut03} adapts the algorithm of \cite{VardiWolper86}, but still generates separated automata. LTL-to-UBA translations that attempt to exploit the advantages of 
UBA over separated automata have been presented by 
Benedikt et al. \cite{BLW13} and Duret-Lutz \cite{Duret17}.
These adapt tableau-based 
LTL-to-NBA algorithms
(\cite{GPVW95} in the case of \cite{BLW13} and
 \cite{Couvreur99} in the case of \cite{Duret17})
and rely on transformations of the form 
$\varphi \lor \psi \leadsto \varphi \lor (\neg \varphi \land \psi)$
to enforce that splitting disjunctive formulas generates states with disjoint languages, thus ensuring unambiguity.

To the best of our knowledge, 
the only available tool that supports the translation of LTL formulas 
to UBA is \ltltotgba{}, which is part of the \Spot{} tool set 
and implements the
LTL-to-UBA algorithm of \cite{Duret17}.

Proofs of all theorems and lemmas can be found in the appendix.

\tudparagraph{0.3 em}{Contribution.}
We describe a novel LTL-to-UBA construction.
It relies on an intermediate representation of LTL formulas using VWAA and adapts the known translation from VWAA to NBA by Gastin and Oddoux \cite{GO01}.
We introduce a notion of unambiguity for VWAA, show that the subsequent translation steps preserve it and that checking whether a VWAA is unambiguous is PSPACE-complete (\Cref{sec:unambigvwaa}).
To the best of our knowledge, unambiguity for alternating automata has not been considered before.

\begin{figure}[tbp]
    \centering
    \begin{tikzpicture}[->,>=stealth',shorten >=1pt,auto,node distance=1cm, semithick]
      \node[draw,rectangle] (VWAA) {VWAA $\mathcal{A}$};
      \node[draw,rectangle] (TGBA) [right =1.8cm of VWAA] {\tgba{} $\mathcal{G}_{\mathcal{A}}$};
      \node[draw,rectangle] (PROD) [right =1.8cm of TGBA] {\(\trim(\mathcal{G}_\aut \otimes \mathcal{G}_\aut)\)};
      \node[draw,rectangle] (UBA) [below =1.8cm of PROD] {UBA $\mathcal{U}$};

      \node[inner sep=2pt,outer sep=0pt,fill=black] (TMP) [below=0.3cm of PROD] {};
      \node[scale=0.8,align=center] (ITE) [right =0.1cm of TMP] {$\exists$ ambiguous \\ state $s$?};

      \draw[<-] (VWAA) -- ++(-1.8,0) node[scale=1,above=5pt,midway] {LTL $\varphi$};
      \draw[->] (VWAA) -- (TGBA) node[scale=0.8,midway,above] {as in~\cite{GO01}};
      \draw[->] (TGBA) -- (PROD) node[scale=0.8,midway,above=3pt,align=center] {product\\construction};
      \draw[-] (PROD) -- ($(TMP) + (0,-0.1cm)$);
      \draw[->] (TMP) edge[bend left] node[scale=0.8,midway,below,align=center] {``yes'' \\ disambiguate $s$} (VWAA.270);
      \draw[->] (TMP) -- (UBA) node[pos=0.6,right=6pt,scale=0.8,align=left] {``no'' \\ degeneralize($\mathcal{G}_{\mathcal{A}}$)};
    \end{tikzpicture}
    \caption{The LTL-to-UBA step. A sequence of unambiguity checks and disambiguation transformations are applied and ultimately a UBA is returned. We use \(\trim(\mathcal{G}_\aut \otimes \mathcal{G}_\aut)\) to check whether unambiguity is achieved or more iterations are necessary.}
  \label{fig:ltltouba}
\end{figure}
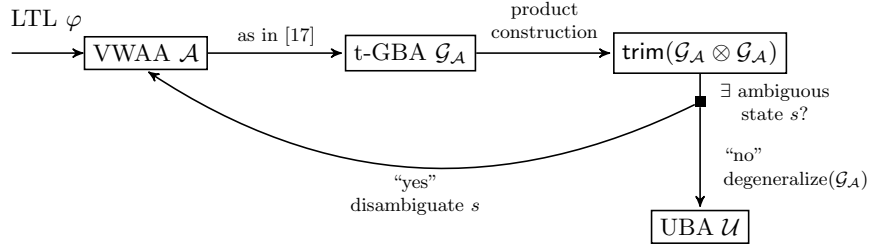

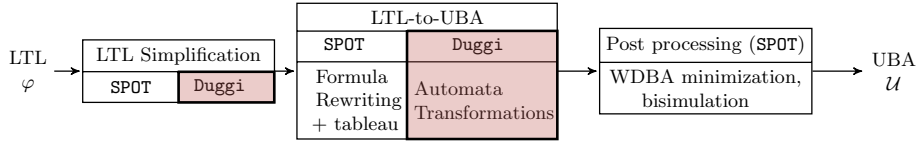
\begin{figure}[tbp]
    \centering
    \resizebox{\textwidth}{!}{\begin{tikzpicture}[->,>=stealth',shorten >=1pt, semithick]
        \node[align=center]                           (phi) at(-0.1 \textwidth,0)     {LTL\\\(\varphi\)};
\node[rectangle, draw, rectangle split, rectangle split parts=2, align=center] (preprocessing)  at(0.1 \textwidth,0)   {LTL Simplification
                \nodepart{second}\parbox[c]{\widthof{LTL Simplification}/2}{\centering \spot{}}
                    \parbox[c]{\widthof{LTL Simplification}/2}{ \centering \duggi{}}
                };
        \draw[-]   (preprocessing.center)  --  (preprocessing.south);
        \draw[-,very thick,fill=Maroon!80,fill opacity = 0.3] (preprocessing.center)  --  ($(preprocessing.south) + (0,0.01cm)$) -- ++(1.55cm,0) -- ++(0,0.5cm) -- ++(-1.6cm,0);
        \node[rectangle, draw, rectangle split, rectangle split parts=3] (LTLtoUBA)   at(0.43 \textwidth, 0)   {\nodepart{one}\parbox{4cm}{\centering LTL-to-UBA}\nodepart{two}\parbox{4cm}{\quad\spot{} \hfill \duggi{} \quad\quad\quad}\nodepart{three}
          \parbox{1.5cm}{Formula\\Rewriting\\+ tableau}
          \parbox{2cm}{Automata\\Transformations}
                };
        \draw[-]   ($(LTLtoUBA.center) + (-1 em, 2 em)$)  --  ($(LTLtoUBA.south) + (-1 em, 0)$);

        \draw[-,very thick,fill=Maroon!80,fill opacity = 0.3] ($(LTLtoUBA.center) + (-1 em, 2.064 em)$)  --  ($(LTLtoUBA.south) + (-1 em, 0.01cm)$) -- ++(2.435cm,0) -- ++(0,1.765cm) -- ++(-2.47cm,0);
        \node[rectangle, draw, rectangle split, rectangle split parts=2, align=center] (postprocessing) at(0.8 \textwidth, 0)   {Post processing (\spot{})
                \nodepart{second}WDBA minimization,\\
                \hspace*{-1 em}bisimulation
                };
        \node[align=center]                           (aut)               at(1.05 \textwidth,0) {UBA\\{\(\mathcal{U}\)}};

        \draw[->]   (phi)   to  (preprocessing.west);
        \draw[->]   (preprocessing.east)   to  (LTLtoUBA.west);
        \draw[->]   (LTLtoUBA.east)   to  (postprocessing.west);
        \draw[->]   (postprocessing.east)   to  (aut);
    \end{tikzpicture}
    }
    \caption{Overview of the general LTL-to-UBA generation algorithm. The LTL
    simplification step, the actual LTL-to-UBA translation step, and the automaton
post processing step can be combined freely. We propose novel rewriting rules for LTL and a LTL-to-UBA translation, both implemented in our tool \duggi{}.}
    \label{fig:overview}
\end{figure}

We present a disambiguation procedure for VWAA that relies on intermediate unambiguity checks to identify ambiguous states and local disambiguation transformations for the VWAA (\Cref{sec:disambiguate_vwaa}).
It has the property that an already unambiguous VWAA is not changed.
\Cref{fig:ltltouba} gives an overview of our LTL-to-UBA algorithm.
Apart from the main construction, we introduce novel LTL rewrite rules and a heuristic, both of which are aimed at producing small UBA and may also benefit existing tools (see \Cref{fig:overview}).
The heuristic is targeted at states with a certain structure, defined using the concepts of \emph{purely-universal} and \emph{alternating} formulas (\Cref{sec:liftnondet}).
Finally, we report on an implementation of our construction in our tool \duggi{} and compare it to the existing LTL-to-UBA translator \ltltotgba{}. We also compare \duggi{} with \ltltotgba{} in the context of Markov chain analysis under LTL specifications (\Cref{sec:experiments}).

 \section{Preliminaries}
\label{sec:prelim}

This section introduces our notation and standard definitions.
The set of infinite words over a finite alphabet $\Sigma$ is denoted by $\Sigma^{\omega}$ and we write $w[i]$ to denote the $i$-th position of an infinite word $w \in \Sigma^{\omega}$, and $w[i..]$ to denote the suffix $w[i]w[i{+}1]\ldots$.
We write $\mathcal{B}^+(X)$ to denote the set of positive Boolean formulas over a finite set of variables $X$.
A \emph{minimal model} of a formula $f \in \mathcal{B}^+(X)$ is a set $M \subseteq X$ such that $M \models f$, but no $M' \subset M$ satisfies $M' \models f$.
LTL is defined using \(\until\) (``Until'')and \(\neXt\) (``Next'').
Additionally we use syntactical derivations \(\finally\)
(``Finally''), \(\globally\) (``Globally''), and \(\release\) (``Release'') (see~\cite{BK08,GraedelThomasWilke02} for details).

\tudparagraph{0.3 em}{Alternating automata on infinite words.} An alternating \(\omega\)-automaton
\(\aut\) is a tuple \((Q, \Sigma, \Delta, \iota, \Acc)\) where \(Q\) is a
non-empty, finite set of states, \(\Sigma\) is a finite alphabet, \(\Delta : Q
\times \Sigma \to \mathcal{B}^+(Q)\) is the transition function, \(\iota \in \mathcal{B}^+(Q)\) is the initial condition and \(\Acc\) is the acceptance condition.
Additionally, we define the function $\delta : Q \times \Sigma \to 2^{2^Q}$ which assigns to a pair $(q,a) \in Q \times \Sigma$ the set of minimal models of $\Delta(q,a)$ and the set $I \subseteq 2^Q$ as the set of minimal models of $\iota$.
We denote by $\mathcal{A}(\iota')$ the automaton $(Q,\Sigma,\delta,\iota',\Phi)$
and we write $\mathcal{A}(Q_0)$ for $\mathcal{A}(\bigwedge_{q \in Q_0} q)$, if
$Q_0 \subseteq Q$. We call the number of the reachable states of an automaton
\(\aut\) its size. 

A run of \(\aut\) for \(w \in \Sigma^{\omega}\) is a directed acyclic graph (\Dag{}) \((V, E)\)~\cite{LodingT00},  where
\begin{enumerate}
    \item \(V \subseteq Q \times \Nat\), and $E \subseteq \bigcup_{0 \leq l}(Q \times \lbr l \rbr) \times (Q \times \lbr l {+} 1 \rbr)$,
    \item \(\lbr q \, : \, (q,0) \in V\rbr \in I\),
    \item for all \((q,l) \in V\) : \(\lbr q' \, : \, ((q, l), (q', l{+}1)) \in
        E\rbr \in \delta(q, w[l])\),
    \item for all \((q,l) \in V \setminus (Q \times \lbr 0 \rbr)\) there is a
        \(q'\) such that \(((q', l{-}1), (q, l)) \in E\).
\end{enumerate}
We define $V(i) = \{ s \, : \, (s,i) \in V\}$, called the $i$-th \emph{layer} of $V$.
A run is called accepting if every infinite path in it meets the acceptance condition.

A word is accepted by \(\aut\) if there exists an accepting run
for it. We denote the set of accepted words of \(\aut\) by \(\langinf(\aut)\).
We distinguish between B\"uchi, generalized B\"uchi and co-B\"uchi acceptance conditions.
A B\"uchi condition is denoted by \(\Inf{Q_f}\) for a set \(Q_f \subseteq Q\).
An infinite path \(\autpath = q_0\, q_1 \, \ldots\) meets \(\Inf{Q_f}\) if \(Q_f \cap \mathrm{inf}(\autpath) \neq \varnothing\), where $\mathrm{inf}(\autpath)$ denotes the set of infinitely occurring states in $\autpath$.
A co-B\"uchi condition is denoted by \(\Fin{Q_f}\) and $\autpath$ meets \(\Fin{Q_f}\) if $Q_f \cap \mathrm{inf}(\autpath) = \varnothing$.
An infinite path $\pi$ meets a generalized B\"uchi condition $\bigwedge_{i \in F}\Inf{Q_i}$ if it meets $\Inf{Q_i}$ for all $i \in F$.
A \emph{transition-based} acceptance condition uses sets of transitions $T \subseteq Q \times \Sigma \times Q$ instead of sets of states to define acceptance of paths.

We call a subset $C \subseteq Q$ a configuration and say that $C$ is reachable if it is a layer of some run.
A configuration $C$ is reachable from a state \(q\), also written as \(q \movetrans{} C\), if $C$ is a reachable configuration of $\mathcal{A}(q)$.
Analogously, \(C' \subseteq Q\) is reachable from \(C \subseteq Q\), or \(C \movetrans{} C'\),
if \(C'\) is a reachable configuration of $\mathcal{A}(C)$.
A configuration $C$ is \emph{reachable via u} if there is a run $(V,E)$ for a word $uw$, with $u \in \Sigma^*, w \in \Sigma^{\omega}$, such that $C = V(|u|)$.
We extend this notion to reachability from states and configurations via finite words in the expected way and write $q \movetrans{u} C'$ and $C \movetrans{u} C'$.
We define $\langinf(C) = \langinf(\aut(C))$.

The \emph{underlying graph} of $\mathcal{A}$ has vertices $Q$ and edges $\{ (q,q') \, : \, \exists a \in \Sigma. \exists S \in \delta(q,a). \, q' \in S\}$.
We say that $\mathcal{A}$ is \emph{very weak} if every strongly connected component of its underlying graph consists of a single state and \(\aut\) has a co-B\"uchi acceptance.
If \(\vert C_0 \vert = 1\) for every \(C_0 \in
I\), and \(\vert C_\delta \vert = 1\) for every \(C_\delta \in \delta(q, a)\) with $(q,a) \in Q \times \Sigma$, we call $\mathcal{A}$ non-deterministic.
As a non-deterministic automaton has only singleton successor sets, its runs are infinite sequences of states.
Finally, an automaton \(\aut\) is \emph{trimmed} if \(\langinf(q) \neq \varnothing\) holds for every state \(q\) in \(\aut\), and we write $\trim(\aut)$ for the automaton that we get by removing all states with empty language in $\aut$.
For the non-alternating automata types that we consider, $\trim(\aut)$ can be computed in linear time using standard graph algorithms.

\tudparagraph{0.3 em}{From LTL to NBA.}
We use the standard translation from LTL to \vwaa{} where the states of the
\vwaa{} correspond to subformulas of $\varphi$ and the transition relation follows the Boolean structure of the state and the LTL expansion laws~\cite{Vardi94,MullerSS88}.
It has been used as a first step in an LTL-to-NBA translation in~\cite{GO01},
whose construction we follow. 
We recall this construction in \Cref{app:ltl-to-vwaa}.
Additionally, we use the optimizations proposed in~\cite{BKRS12}.
We also maintain the following invariant, as proposed in~\cite{GO01}: for all $(q,a) \in Q \times \Sigma$ and successor sets $S_1,S_2 \in \delta(q,a)$, such that $S_1 \neq S_2$, it holds that $S_1 \not\subseteq S_2$.

A VWAA $\aut$ can be transformed into a transition-based generalized B\"uchi automaton (\tgba{}) by a powerset-like construction, where the
non-deterministic choices of $\aut$ are captured by non-deterministic choices of the
\tgba{}, and the universal choices are captured by the powerset.

\begin{definition}
    \label{vwaa2tgba}
    Let \(\aut = (Q, \Sigma, \Delta, \iota, \Fin{Q_f})\) be a VWAA. The \tgba{}
    \(\mathcal{G}_\aut\) is the tuple \((2^Q, \Sigma, \delta',I, \bigwedge_{f \in
    Q_f} \Inf{\mathcal{T}_f})\),
    where
    \begin{itemize}[itemsep=3pt]
        \item \(\delta'(C,a) = \bigotimes_{q \in C} \delta(q,a)\), where $T_1 \otimes T_2 = \{ C_1 \cup C_2 \, : \, C_1 \in T_1, C_2 \in T_2\}$
\item \(\mathcal{T}_f = \lbr (C,a,C') \, : \, f \not\in C' \text{ or there exists } Y \in \delta(f, a) \text{ and } f\not\in Y \subseteq C'\rbr\)
    \end{itemize}
\end{definition}

\begin{theorem}[Theorem 2 of \cite{GO01}]
    Let \(\aut\) be a VWAA and \(\mathcal{G}_\aut\) be as in \Cref{vwaa2tgba}. Then,
    \(\langinf(\aut) = \langinf(\mathcal{G}_\aut)\).
\end{theorem}

The size of $\mathcal{G}_{\mathcal{A}}$ may be exponential in $|Q|$ and the number of B\"uchi conditions of $\mathcal{G}_{\mathcal{A}}$ is $|Q_f|$.
Often a B\"uchi automaton with a (non-generalized) B\"uchi acceptance is desired.
For this step we follow the construction of~\cite{GO01},
which translates $\mathcal{G}_{\aut}$ into an NBA \(\mathcal{N}_{\mathcal{G}_\aut}\) of at most $|Q_f| \cdot
2^{|Q|}$ reachable states.
  \section{Unambiguous VWAA}
\label{sec:unambigvwaa}

In this section we introduce a notion of unambiguity for VWAA and show that
unambiguous VWAA are translated to UBA by the translation presented in \Cref{sec:prelim}.
We define unambiguity in terms of configurations of the VWAA, which are strongly related to the states of the resulting NBA.
Let $\aut = (Q,\Sigma,\Delta,\iota,\Fin{Q_f})$ be a fixed VWAA for the rest of this section.

\begin{definition}
  $\mathcal{A}$ is unambiguous if it has no distinct configurations $C_1, C_2$ that are reachable via the same word $u \in \Sigma^*$ and such that $\lang(C_1) \cap \lang(C_2) \neq \varnothing$.
\end{definition}
The standard definition of unambiguity is that an automaton is unambiguous if it has at most one accepting run for any word.
In our setting runs are \Dag{}'s and we do allow multiple accepting runs for a word, as long as they agree on the configurations that they reach for each prefix.
In this sense it is a weaker notion.
However, the notions coincide on non-deterministic automata as the edge relation of the run is then induced by the sequence of visited states.
\begin{theorem}
    \label{thm:vwaatouba}
    Let \(\mathcal{N}_{\mathcal{G}_{\mathcal{A}}}\) be the NBA for $\aut$, obtained by the translation from \Cref{sec:prelim}. If $\mathcal{A}$ is unambiguous, then $\mathcal{N}_{\mathcal{G}_{\mathcal{A}}}$ is unambiguous.
\end{theorem}
We show that every step in the translation from VWAA to NBA preserves unambiguity.
First, we establish the following correspondance:
\begin{restatable}{lemma}{gbatovwaarun}
  \label{lemm:vwaa_gba_run_corr}
  If $\mathcal{A}$ is unambiguous, then for every accepting run $r = Q_0Q_1\ldots$ of $\mathcal{G}_{\mathcal{A}}$ for $w \in \Sigma^\omega$ there exists an accepting run $\rho = (V,E)$ of $\mathcal{A}$ for $w$ such that $Q_i = V(i)$ for all $i \geq 0$.
\end{restatable}

Intuitively, the lemma states that if $\aut$ is unambiguous, then every accepting run $r$ of $\mathcal{G}_{\aut}$ can be matched by an accepting run $\rho$ of $\aut$ such that the states of $r$ are the layers of $\rho$.
The proof is not immediate and requires $\mathcal{A}$ to be unambiguous.

A direct consequence of \Cref{lemm:vwaa_gba_run_corr} is that if $\aut$ is unambiguous, then so is $\mathcal{G}_{\aut}$.
The degeneralization construction in~\cite{GO01} makes $|Q_f| + 1$ copies of $\mathcal{G}_{\mathcal{A}}$.
As the next copy is uniquely determined by the current state and word label, it preserves unambiguity.
In combination with \Cref{lemm:vwaatotgbaunamb} we obtain \Cref{thm:vwaatouba}.

We now show that deciding whether a VWAA is unambiguous is PSPACE-complete.
The idea for proving hardness is to reduce LTL satisfiability, which is known to be PSPACE-hard, to VWAA emptiness (this follows directly by the LTL $\to$ VWAA translation) and VWAA emptiness to VWAA unambiguity.
The second step uses the following trick:
a VWAA $\aut$ accepts the empty language if and only if the disjoint union of $\aut$ with itself is unambiguous.

To check wether \vwaa{} is unambiguous we first show that for every accepting run of $\aut$, we find a matching accepting run of $\mathcal{G}_\aut$, which follows directly from the definition of $\mathcal{G}_\aut$:
\begin{restatable}{lemma}{vwaatogbarun}
  \label{lemm:gba_vwaa_run_corr}
For every accepting run $\rho = (V,E)$ of $\aut$ for $w \in \Sigma^{\omega}$ there exists an accepting run $r=Q_0Q_1\ldots$ of $\mathcal{G}_{\aut}$ for $w$, such that $Q_i = V(i)$ for all $i \geq 0$.
\end{restatable}
\Cref{lemm:vwaa_gba_run_corr} and \Cref{lemm:gba_vwaa_run_corr} give us the following:
\begin{restatable}{lemma}{unambispreserved}
  \label{lemm:vwaatotgbaunamb}
  $\mathcal{A}$ is unambiguous if and only if $\mathcal{G}_{\mathcal{A}}$ is unambiguous.
\end{restatable}
However, checking whether $\mathcal{G}_{\mathcal{A}}$ is unambiguous can be done in space polynomial in the size of $\aut$, and we conclude:
\begin{restatable}{theorem}{checkunambpspace}
Deciding whether a \vwaa{} is unambiguous is PSPACE-complete.
\end{restatable}
 \section{Disambiguating VWAA}
\label{sec:disambiguate_vwaa}

Our disambiguation procedure is inspired by the idea of ``separating'' the language of successors for every non-deterministic branching.
A disjunction $\varphi \lor \psi$ is transformed into $\varphi \lor (\neg \varphi \land \psi)$ by this principle.
The rules for $\until$ and $\release$ are derived by applying the disjunction rule to the expansion law of the corresponding operator (see \Cref{tab:ltltrans}).
These rules are applied by \ltltotgba{} in its tableau-based algorithm to guarantee that the resulting automaton is unambiguous, and have also been proposed in~\cite{BLW13}.
\begin{table}[tbp]
  \caption{The adapted expansion laws for $\until$ and $\release$ are the result of applying the disjunction rule to the classic expansion laws.}
  \centering
  \begin{tabular}{l | c | c}
    & expansion law & adapted expansion law \\
    \hline
    $\varphi \, \until \, \psi$ & $\Gamma \equiv \psi \lor (\varphi \land \neXt \Gamma)$ & $\Gamma \equiv \psi \lor (\varphi \land \neg \psi \land \neXt \Gamma)$ \\
    $\varphi \, \release \, \psi$ & $\Gamma \equiv \psi \land (\varphi \lor \neXt \Gamma)$ & $\Gamma \equiv \psi \land (\varphi \lor (\neg \varphi \land \neXt \Gamma))$ \\
  \end{tabular}
  \label{tab:ltltrans}
\end{table}

In our approach we define corresponding transformations for non-deterministic branching in the \vwaa{}.
Furthermore, we propose to do this in an ``on-demand'' manner: instead of applying these transformation rules to every non-deterministic split, we identify ambiguous states during the translation and only apply the transformations to them.
This guarantees that we return the automaton produced by the core translation, without disambiguation, in case it is already unambiguous.

The main steps of our disambiguation procedure are the following:
\begin{enumerate}
\item A preprocessing step that computes a complement state $\tilde{s}$ for every state $s$.
\item \label{enum:identify} A procedure that identifies ambiguous states.
\item Local transformations that remove the ambiguity.
\end{enumerate}
If no ambiguity is found in step \ref{enum:identify}, the \vwaa{} is unambiguous.
The high-level overview is also depicted in \Cref{fig:ltltouba}.
In what follows we fix a VWAA $\aut = (Q,\Sigma,\Delta,\iota,\Fin{Q_f})$ and assume that it has a single initial state.

\tudparagraph{0.3 em}{Complement states.}
The transformations we apply for disambiguation rely on the following precondition: for every state $s$ of $\mathcal{A}$ there should be another state $\tilde{s}$ such that $\lang(\tilde{s}) = \overline{\lang(s)}$.
We compute these complement states in a preprocessing step and add them to $\mathcal{A}$.
Complementing alternating automata can be done without any blow up by dualizing both the acceptance condition and transition structure, as shown by Muller and Schupp~\cite{MS87}.
As dualizing the acceptance condition and complementing the set of final states yields an equivalent \vwaa{}, we can keep the co-B\"uchi acceptance when complementing.

The complement automaton has the same underlying graph and is therefore also very weak.
Furthermore, no state $s$ is reachable from its own complement state $\tilde{s}$, which is an invariant that we maintain and which ensures that very weakness is preserved in the construction.

\tudparagraph{0.3 em}{Source configurations and source states.}
To characterize ambiguous situations we define \emph{source configurations} and \emph{source states}.
A source configuration of $\mathcal{A}$ is a reachable configuration $C$ such that there exist two different configurations $C_1, C_2$ that are reachable from $C$ via some $a \in \Sigma$ and $\lang(C_1) \cap \lang(C_2) \neq \varnothing$.
By definition, $\mathcal{A}$ is not unambiguous if a source configuration exists.

Let $C$ be a source configuration of $\mathcal{A}$ and let $C_1, C_2$ be the successor configurations as described above.
A source state of $C$ is a state $s \in C$ with two transitions $S_1,S_2 \in \delta(s,a)$ such that $S_i \subseteq C_i$, for $i \in \{1,2\}$, $S_1 \neq S_2$
and $(S_1 \cup S_2) \setminus (C_1 \cap C_2) \neq \varnothing$.
The last condition ensures that either $S_1$ or $S_2$ contains a state that is not common to $C_1$ and $C_2$.
By \Cref{vwaa2tgba}, $C_i = \bigcup_{q \in C} S_q$ with $S_q \in \delta(a,q)$ for all $q \in C$, and thus $C$ must contain a source state.

\tudparagraph{0.3 em}{Ambiguity check and finding source states.}
For the analysis of source configurations and source states we use the standard
product construction \(\mathcal{G}_1 \otimes \mathcal{G}_2\), which returns a \tgba{} such that \(\langinf(\mathcal{G}_1 \otimes \mathcal{G}_2) = \langinf(\mathcal{G}_1) \cap \langinf(\mathcal{G}_2)\) for two given \tgba{}
\(\mathcal{G}_1\) and \(\mathcal{G}_2\).
Specifically, we consider the self product $\mathcal{G}_{\aut} \otimes \mathcal{G}_{\aut}$ of $\mathcal{G}_{\aut}$.
It helps to identify ambiguity: \(\mathcal{G}_{\aut}\) is not unambiguous if and only if there exists a reachable state \((C_1, C_2)\) in \(\trim(\mathcal{G}_\aut \otimes \mathcal{G}_{\aut})\) with \(C_1 \neq C_2\).

The pair of configurations $(C_1,C_2)$ is a witness to ambiguity of $\aut$.
We look for a symbol $a \in \Sigma$ and a configuration $C$ such that $(C,C) \xrightarrow{a} (C_1',C_2') \rightarrow^\ast (C_1, C_2)$ is a path in $\trim(\mathcal{G}_\aut \otimes \mathcal{G}_{\aut})$ and $C_1' \neq C_2'$.
Such a configuration must exist as we have assumed that $\aut$ has a single initial state $q_i$, which implies that \(\trim(\mathcal{G}_\aut \otimes \mathcal{G}_\aut)\) has a single initial state $(\{q_i\},\{q_i\})$.
\(C\) is a source configuration and therefore must contain a source state which we can find by inspecting all pairs of transitions of states in $C$.

\begin{figure}[tbp]
    \centering
    \begin{tikzpicture}[->,>=stealth',shorten >=1pt,auto,node distance=1.5cm, semithick]
        \node[state] (s)     at (0.75,1.5)      {\(s\)}; 
        \node[state] (s1)    at (0,0)      {\(s_1\)}; 
        \node[state] (s2)    at (1.5,0)      {\(s_2\)}; 

        \draw[->]   (s)  to node[left]  {\(a\)} (s1);
        \draw[->]   (s)  to node[right]  {\(a\)} (s2);
    \end{tikzpicture}
    \qquad\parbox{1cm}{\vspace{-2cm}\hspace{0.25cm}\large$\boldsymbol{\mapsto}$}\qquad
    \begin{tikzpicture}[->,>=stealth',shorten >=1pt,auto,node distance=1.5cm, semithick]
        \node[state] (s)        at (0.75,1.5)   {\(s\)}; 
        \node[state] (s1)       at (0,0)        {\(s_1\)}; 
        \node[state] (s2)       at (1.2,0)        {\(s_2\)}; 
        \node[state] (negs1)    at (2.2,0)      {\(\tilde{s_1}\)};   

        \draw[->]   (s)  to node[left]  {\(a\)} (s1);
        \draw[->]   plot[mark=*,mark size=2pt](s.286)  to  (s2);
        \draw[->,bend left]   (s)  to[in=130, out=332] node[above]  {\(a\)} (negs1);
    \end{tikzpicture}
    \caption{Disambiguation scheme for a source state \(s\) with successors \(s_1\) and \(s_2\) in the \vwaa{}.
      Transitions with successor set of size $\geq 1$ are conjoined by a $\bullet$.}
\label{fig:disamb_source_states}
\end{figure}
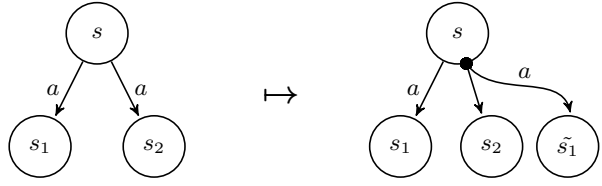

\tudparagraph{0.3 em}{Disambiguating a source state.}
\label{sub:disamb_source_state}
The general scheme for disambiguating source states is depicted in \Cref{fig:disamb_source_states}.
Assume that we have identified a source state \(s\) with successor sets \(S_1\) and
\(S_2\) as explained above.
The LTL-to-VWAA construction guarantees $S_1 \not\subseteq S_2$ and $S_2 \not\subseteq S_1$.
We need to distinguish the looping successor sets (i.e. those $S_i$ that contain $s$) from the non-looping.
Technically, we consider two cases: either $S_1$ or $S_2$ do not contain $s$ or both sets contain $s$.
In the first case we assume, w.l.o.g., that $s \notin S_1$.
The successor set $S_2$ is split into the \(\vert S_1\vert\) new successor sets $\{ (S_2 \cup \{ \tilde{s_1} \}) \, : \, s_1 \in S_1\}$.
The new sets of states are added to $\delta(s,a)$ and the successor set $S_2$ is removed.
If both $S_1$ and $S_2$ contain $s$, we proceed as in the first case but do not add the successor set $S_2 \cup \{\tilde{s}\}$ to $\delta(s,a)$.

This transformation does not guarantee that $s$ is not a source state anymore.
However, it removes the ambiguity that stems from the non-deterministic choice of transitions $S_1,S_2 \in \delta(a,s)$.
If $s$ is still a source state it will be identified again for another pair of transitions.
After a finite number of iterations all successor sets of $s$ for any symbol in $\Sigma$ will accept pairwise disjoint languages, in which case $s$ cannot be a source state anymore.
The transformation preserves very weakness as it only adds transitions from $s$ to complement states of successors of $s$ and by assumption there is no path between a state and its complement state.

 \tudparagraph{0.3 em}{Iterative algorithm.} 
Putting things together, our algorithm works as follows: it searches for source configurations of $\mathcal{A}$ (using $\mathcal{G}_{\mathcal{A}}$), applies the local disambiguation transformations to $\mathcal{A}$ as described and recurses (see \Cref{fig:ltltouba}).
As rebuilding the t-GBA may become costly, in our implementation we identify which part of the t-GBA has to be recomputed due to the changes in $\mathcal{A}$, and rebuild only this part.
If no source configuration is found, we know that both $\aut$ and $\mathcal{G}_{\mathcal{A}}$ are unambiguous and we can apply degeneralization to obtain a UBA.

\tudparagraph{0.3 em}{Complexity of the procedure.}
The \vwaa{}-to-\tgba{} translation that we adapt produces a \tgba{} $\mathcal{G}_{\aut}$ of size at most $2^n$ for a \vwaa{} $\mathcal{A}$ of size $n$.
In our disambiguation procedure we enlarge $\mathcal{A}$ by adding complement states for every state in the original automaton, yielding a \vwaa{} of size $2n$.
Thus, a first size estimate of $\mathcal{G}_{\aut}$ in our construction is $4^n$.
However, no state in $\trim(\mathcal{G}_{\aut})$ can contain both $s$ and $\tilde{s}$ for any state $s$ of $\mathcal{A}$.
The reason is that the language of a state in $\mathcal{G}_{\aut}$ is the intersection of the languages of the \vwaa{}-states it contains, and $\lang(s) \cap \lang(\tilde{s}) = \varnothing$.
Thus, $\trim(\mathcal{G}_{\aut})$ has at most $3^n$ states.

The amount of ambiguous situations that we identify is bounded by the number of non-deterministic splits in the VWAA, which may be exponential in the length of the input LTL formula.
In every iteration we check ambiguity of the new VWAA, which can be done in exponential time.
Thus, our procedure computes a UBA in time exponential in the length of the formula.

 \section{Heuristics for purely-universal formulas}
\label{sec:liftnondet}

In this section we introduce alternative disambiguation transformations for
special source states representing formulas $\varphi \until \nu$, where $\nu$ is
\emph{purely-universal}.
The class of purely-universal formulas is a syntactically defined subclass of LTL-formulas with suffix-closed languages.
These transformations reduce the size of the resulting UBA and often produce automata of a simpler structure.
The idea is to decide whether $\nu$ holds whenever moving to a state representing $\varphi \until \nu$ and, if not, finding the \emph{last} position where it does not hold.

\begin{figure}[tbp]
  \begin{subfigure}[t]{0.32\textwidth}
    \centering
    \begin{tikzpicture}[->,>=stealth',shorten >=1pt,auto,node distance=1.5cm, semithick]
      \node[scale=0.8, state,accepting] (FGa) {$\finally \globally a$};
      \node[scale=0.8, state] (Ga) [right = 0.9cm of FGa] {$\globally a$};

      \draw[<-] (FGa) -- ++(-0.55,0.55);
\draw[->] (FGa.east) -- (Ga) node [scale=0.8,midway,above=3pt] {$a$};

      \path (Ga) edge[loop below] node[scale=0.8, below] {$a$} (Ga)
            (FGa) edge[loop below] node[scale=0.8, below] {$\true$} (FGa);
    \end{tikzpicture}
    \caption{VWAA for $\finally \globally a$.}
    \label{subfig:originalaut}
  \end{subfigure}
  \hspace*{\fill}
  \begin{subfigure}[t]{0.32\textwidth}
    \centering
      \begin{tikzpicture}[->,>=stealth',shorten >=1pt,auto,node distance=1.5cm, semithick]
      \node[scale=0.8, state,accepting] (FGa) {$\finally \globally a$};
      \node[scale=0.8, state] (Ga) [right = 0.9cm of FGa] {$\globally a$};
      \node[scale=0.8, state,accepting] (Fna) [below = 0.6cm of FGa] {$\finally \neg a$};

      \draw[<-] (FGa) -- ++(-0.55,0.55);
      \draw[->] (Fna) -- ++(0.55,-0.6) node [scale=0.8,near start,right=2pt] {$\neg a$};
      \draw[->] (Fna) to[out=250,in=210,looseness=5]
      node [scale=0.8,midway, below = 1pt] {$a$}  (Fna);
      \draw[->] (FGa) to[out=240,in=200,looseness=5] node [scale=0.8,below=2pt] {$\neg a$} (FGa);
      \draw[->] (FGa) to[out=270,in=310,looseness=5] (FGa);
      \draw[->] plot[mark=*,mark size=2pt] (FGa.south) -- (Fna)
      node [scale=0.8,near end,right=2pt] {$a$};
      
      \draw[->] (FGa.east) -- (Ga) node [scale=0.8,midway,above=3pt] {$a$};

      \path (Ga) edge[loop below] node[scale=0.8, below] {$a$} (Ga);
    \end{tikzpicture}
    \caption{Standard disambiguation.}
    \label{subfig:standarddis}
  \end{subfigure}
  \hspace*{\fill}
  \begin{subfigure}[t]{0.32\textwidth}
    \centering
      \begin{tikzpicture}[->,>=stealth',shorten >=1pt,auto,node distance=1.5cm, semithick]
      \node[scale=0.8, state] (In) {$\finally \globally a$};
      \node[scale=0.8, state,accepting] (FGa) [below left = 0.6 and 0.25 of In] {$\varphi$};
      \node[scale=0.8, state] (Ga) [right = 0.9cm of FGa] {$\globally a$};

      \draw[<-] (In) -- ++(-0.55,0.55);
      \draw[->] (In) -- (FGa) node[pos=0.2,scale=0.8,left=3pt] {$\true$};
      \draw[->] (In) -- (Ga) node[pos=0.2,scale=0.8,right=3pt] {$\true$};
\draw[->] (FGa.east) -- (Ga) node [scale=0.8,pos=0.4,above=1pt] {$\neg a$};

      \path (Ga) edge[loop below] node[scale=0.8,below] {$a$} (Ga)
            (FGa) edge[loop below] node[scale=0.8,below] {$\true$} (FGa);
    \end{tikzpicture}
    \caption{Modified transformation. Here $\varphi = \finally (\neg a \land \neXt \globally
      a)$.}
    \label{subfig:optdis}
  \end{subfigure}
  \caption{Three \vwaa{} for $\finally \globally a$.
    The automaton in (\subref{subfig:standarddis}) is the result of standard disambiguation and (\subref{subfig:optdis}) is the result of the modified transformation applied to (\subref{subfig:originalaut}).
  The automaton in (\subref{subfig:optdis}) is non-deterministic and has two looping states, whereas (\subref{subfig:standarddis}) is not non-deterministic and has three looping states.}
  \label{fig:ugexample}
\end{figure}
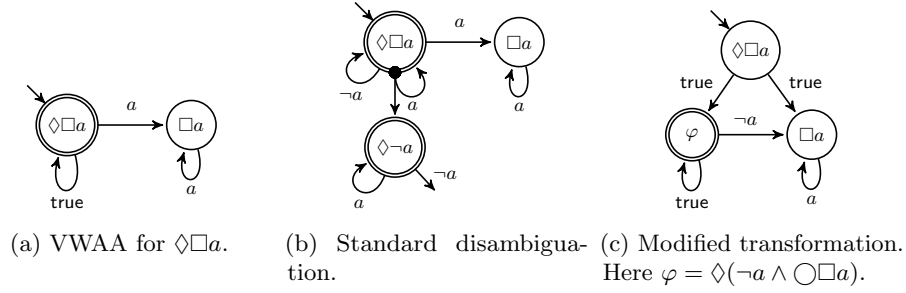

\begin{example}
  \label{ex:ug}
  Consider the formula $\finally \globally a$. A \vwaa{} for it is shown in~\Cref{subfig:originalaut}.
  It is ambiguous, as a word satisfying $\globally a$ may loop in the initial state for an arbitrary amount of steps before moving to the next state.

  In the standard disambiguation transformation the state $\finally \neg a$ is added to the self loop of the initial state (\Cref{subfig:standarddis}).
  The automaton in \Cref{subfig:optdis}, on the other hand, makes the following case distinction: either a word satisfies $\globally a$, in which case we move to that state directly, or there is a suffix that satisfies $\neg a$ and $\neXt \globally a$.
  The state $\varphi$ is used to find the \emph{last} occurrence of $\neg a$, which is unique.
  
\end{example}

To generalize this idea and identify the situations where it is applicable we use the syntactically defined subclasses of \emph{purely-universal} ($\nu$), \emph{purely-eventual} ($\mu$) and \emph{alternating} ($\xi$) formulas (\cite{EH00,BKRS12}).
In the following definition $\varphi$ ranges over arbitrary LTL formulas:
\begin{align*}
  \nu &::= \globally\varphi \mid \nu \lor \nu \mid \nu \land \nu \mid \neXt \nu \mid \nu \until \nu \mid \varphi \release \nu \mid \finally \nu \\
  \mu &::= \finally\varphi \mid \mu \lor \mu \mid \mu \land \mu \mid \neXt \mu \mid \varphi \until \mu \mid \mu \release \mu \mid \globally \mu \\
  \xi &::= \globally\mu \mid \finally \nu \mid \xi \lor \xi \mid \xi \land \xi \mid \neXt \xi \mid \varphi \until \xi \mid \varphi \release \xi \mid \finally \xi \mid \globally \xi
\end{align*}
Formulas that fall into these classes define suffix closed ($\nu$), prefix closed ($\mu$) and prefix invariant ($\xi$) languages respectively:

\begin{lemma}[\cite{EH00,BKRS12}]
  For all $u \in \Sigma^*$ and $w \in \Sigma^{\omega}$: \begin{itemize}
    \item If $\nu$ is purely-universal, then $uw \models \nu \implies w \models \nu$.
    \item If $\mu$ is purely-eventual, then $w \models \mu \implies uw \models \mu$.
    \item If $\xi$ is alternating, then $w \models \xi \iff uw \models \xi$.
  \end{itemize}
\end{lemma}

Let $\nu$ be purely-universal.
We want to find a formula $\mathfrak{g}(\nu)$, called the \emph{goal} of $\nu$, that is simpler than $\nu$ and satisfies $\mathfrak{g}(\nu) \land \neXt \nu \equiv \nu$.
If $\nu$ does not hold initially for some word $w$ we can identify the last suffix $w[i..]$ where it does not hold, given that such an $i$ exists, by checking if $w[i..]$ satisfies $\neg \mathfrak{g}(\nu) \land \neXt \nu$.

It is not clear how to define $\mathfrak{g}(\nu)$ for purely-universal formulas of the form $\nu_1 \lor \nu_2$ or $\nu_1 \until \nu_2$.
We therefore introduce the concept of \emph{disjunction-free} purely-universal formulas in which all occurrences of $\lor$ and $\until$ appear in the scope of some $\globally$.
As $\varphi \release \nu \equiv \nu$ if $\nu$ is purely-universal, we assume that all occurences of $\release$ are also in the scope of some $\globally$ for purely-universal formulas.

\begin{restatable}{lemma}{disjfreenutransf}
\label{lemm:disjfreenutransf}
  Every purely-universal formula $\nu$ can be rewritten into a formula $\nu_1 \lor \ldots \lor \nu_n$, where $\nu_i$ is disjunction-free for all $1 \leq i \leq n$.
\end{restatable}
Disjunction-free purely-universal formulas have a natural notion of ``goal''.

\begin{definition}
  Let $\nu$ be a disjunction-free and purely-universal formula.
  We define $\mathfrak{g}(\nu)$ inductively as follows:
  \[\begin{array}{clclclcl}
    \mathfrak{g}(\globally \varphi) &= \varphi & \hspace{7 em} & \mathfrak{g}(\neXt \nu) &= &\neXt \mathfrak{g}(\nu) \\
    \mathfrak{g}(\nu_1 \land \nu_2) &= \mathfrak{g}(\nu_1) \land \mathfrak{g}(\nu_2) & & \mathfrak{g}(\finally \nu) &= &\true
  \end{array}\]
\end{definition}

The reason for defining $\mathfrak{g}(\finally \nu)$ as $\true$ is that $\finally \nu$ is an alternating formula and checking its validity can thus be temporarily suspended.
Indeed, the definition satisfies the equivalence that we aimed for:

\begin{restatable}{lemma}{disjfreenu}
  \label{lemm:disjfreenu}
  Let $\nu$ be a disjunction-free and purely-universal formula.
  Then $\mathfrak{g}(\nu) \land \neXt \nu \equiv \nu$.
\end{restatable}

In \Cref{ex:ug} $\neg \mathfrak{g}(\nu) \land \neXt \nu$ corresponds to $\neg a \land \neXt \globally a$, which is realized by the transition from state $\varphi$ to state $\globally a$ in \Cref{subfig:optdis}.

\Cref{lemm:unutrans} shows the general transformation scheme (applied left to right). It introduces non-determinism, but we show that it is not a cause of ambiguity as the languages of the two disjuncts are disjoint.
An important difference to the known rule for $\until$ is that the left-hand side of the $\until$-formula stays unchanged.
This is favorable as it is the left-hand side that may introduce loops in the automaton.

\begin{restatable}{lemma}{unutrans}
  \label{lemm:unutrans}
  Let $\nu$ be a disjunction-free and purely-universal formula.
  Then
  \[\text{1.} \enspace \varphi \, \until (\nu \lor \psi) \equiv \nu \lor \gamma \text{\enspace and \enspace} \text{2.} \enspace \lang(\nu) \cap \lang(\gamma) = \varnothing\]
  where $\gamma = \varphi \, \until \, ((\varphi \land \neg \mathfrak{g}(\nu) \land \neXt \nu) \lor (\psi \land \neg \nu))$.
\end{restatable}

LTL formulas may become larger when applying this transformation.
However, they are comparable to the LTL formulas produced by the standard disambiguation transformations in terms of the number of subformulas.
If all occurrences of $\neXt$ in $\nu$ are in the scope of some $\globally$, then no subformulas are added.
Otherwise, $\mathfrak{g}(\nu)$ and $\neXt \nu$ may introduce new $\neXt$-subformulas.

 \section{Implementation and Experiments}
\label{sec:experiments}

The tool \duggi{} is an LTL-to-UBA translator based on the construction introduced in the foregoing sections.\footnote{\duggi{} and the \prism{} implementation, together with all experimental data, are available at \url{https://wwwtcs.inf.tu-dresden.de/ALGI/TR/FM19-UBA/}.}
It reads LTL formulas in a prefix syntax and produces (unambiguous) automata in the HOA format~\cite{Hanoi-CAV15}.
In the implementation we deviate from or extend the procedure described above in the following ways:
\begin{itemize}
\item{We make use of the knowledge given by the \vwaa{}-complement states in the translation steps to \tgba{} $\mathcal{G}_{\aut}$ and the product \(\mathcal{G}_\aut \otimes \mathcal{G}_\aut\). It allows an easy emptiness check: if $s$ and $\tilde{s}$ are present in some $\mathcal{G}_\aut$ or \(\mathcal{G}_\aut \otimes \mathcal{G}_\aut\) state, then it accepts the empty language and does not have to be further expanded.}
\item{We have included the following optimization of the LTL-to-VWAA procedure: when translating a formula $\globally \mu$, where $\mu$ is purely-eventual, we instead translate $\globally \neXt \mu$.
This results in an equivalent state with fewer transitions.
It is close to the idea of suspension as introduced in~\cite{BKRS12}, but is not covered by it.}
\item{Additionally, \duggi{} features an LTL rewriting procedure that uses many of the
LTL simplification rules in the
literature~\cite{SomBloem00,EH00,BKRS12,MS17}.
We have included the following rules that are not used by \spot:
\[\text{I} \,\, (\globally \finally \varphi) \wedge (\finally\globally \psi) \mapsto \globally \finally (\varphi \wedge \globally\psi) \hspace{1cm} \text{II} \,\, (\finally\globally \varphi) \vee (\globally\finally \psi) \mapsto \finally\globally  (\varphi \vee \finally \psi)\]
These rewrite rules are more likely to produce formulas of the form $\finally \globally \varphi$, to which the heuristic of \Cref{sec:liftnondet} can be applied.
 They stem from~\cite{MS17}, where the reversed rules have been used to achieve a normal form.}
\end{itemize}

\tudparagraph{0.3 em}{LTL benchmarks from the literature.}
\begin{figure}[tbp]
  \begin{subfigure}[t]{0.47\textwidth}
    \centering
    \scalebox{.37}{\includegraphics{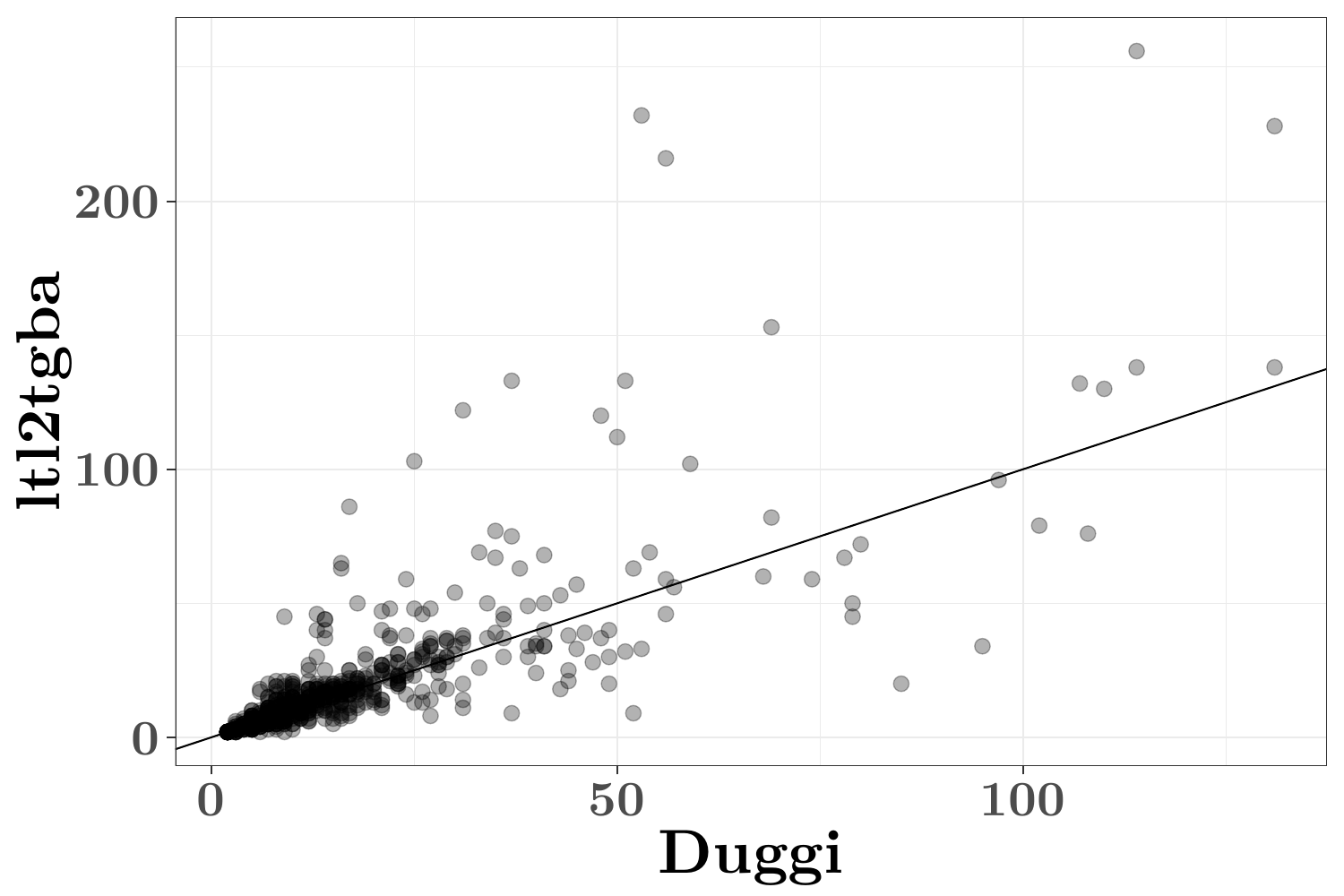}
    }
    \caption{Entire set}
  \end{subfigure}
  \hfill
  \begin{subfigure}[t]{0.47\textwidth}
    \centering
    \scalebox{0.37}{\includegraphics{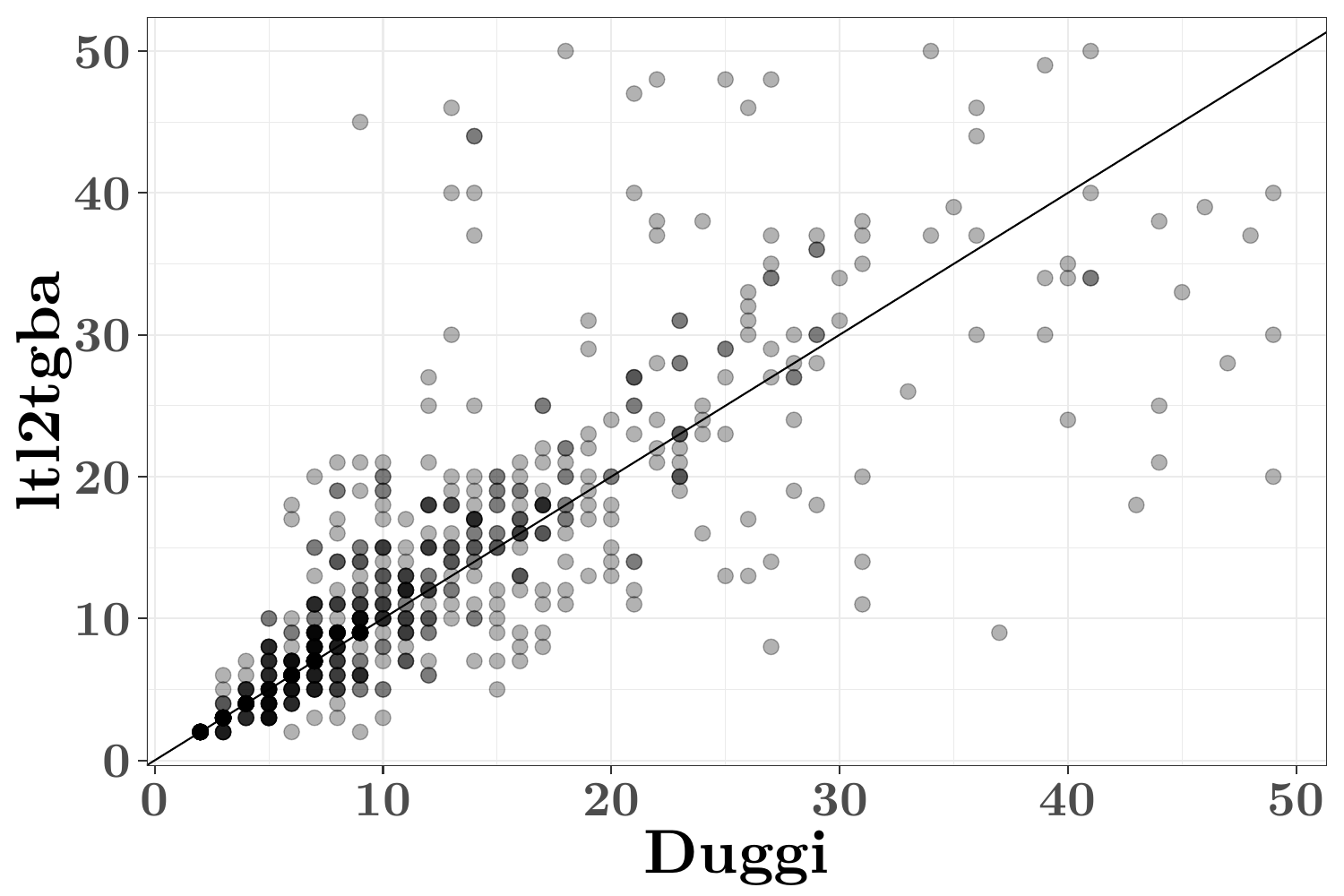}
    }
    \caption{Instances where both tools produced automata with at most 50 states}
  \end{subfigure}
  \caption{Non-WDBA-recognizable fragment of \textsc{ltlstore} (948 formulas). Every point stands for a formula where the according automaton size for \duggi{} is the abcissa, the automaton size of \ltltotgba{} the ordinate. Points above the line stand for formulas where \duggi{} performed better.}\label{fig:ltlstore}
\end{figure}
We now compare the UBA sizes for LTL formulas of the benchmark set \textsc{ltlstore}~\cite{KMS18}.
It collects formulas from various case studies and tool evaluation papers in different contexts.
We include the negations of all formulas and filter out duplicates, leaving 1419 formulas.

Languages that are recognizable by \emph{weak deterministic B\"uchi automata} (WDBA) can be efficiently minimized~\cite{Loeding01} and \ltltotgba{} applies this algorithm as follows: it computes the minimal deterministic B\"uchi automaton and the UBA and returns the one with fewer states.
Our formula set contains 472 formulas that are WDBA-recognizable and for which we could compute the minimal WDBA within the bounds  of 30 minutes and 10\,GB of memory using \ltltotgba{}.
Of these 472 formulas we found 11 for which the UBA generated by either \duggi{} or \ltltotgba{} was smaller than the minimal WDBA, and only two where the difference was bigger than 3 states.
On the other hand, the minimal WDBA were smaller than the UBA produced by \ltltotgba{} (\duggi{}) for 164 (203) formulas.
This supports the approach by \ltltotgba{} to apply WDBA minimization when possible and in what follows we focus on the fragment of the \textsc{ltlstore} that does not fall into this class.
In~\cite{Duret17} it was noted that WDBA minimization often leads to smaller automata than the LTL-to-NBA translation of \ltltotgba{}.

We consider the following configurations: \duggi{} is the standard configuration, \duggi{}$_{\setminus \mathrm{(R,H)}}$ is \duggi{} without the new rewrite rules I and II (R) and/or without the heuristic introduced in \Cref{sec:liftnondet} (H).
For \spot{}, \ltltotgba{} is the standard configuration that produces UBA without WDBA-minimization, which is switched on in \ltltotgba{}$_{\text{WDBA}}$.
We use simulation-based postprocessing as provided by \spot{} in all \duggi{}-configurations (they are enabled by default in \ltltotgba{}).
We use \spot{} with version 2.7.2.
All computations, including the PMC experiments, were performed on a computer with two Intel E5-2680 8 cores at \(2.70\)\,GHz running Linux, with a time bound of 30 minutes and a memory bound of 10\,GB.

Scatter plots comparing the number of states of UBA produced by \ltltotgba{} and \duggi{} are shown in \Cref{fig:ltlstore}.
\Cref{tab:summary} gives cumulative results of different configurations on these formulas.
All configurations of \duggi{} use more time than \ltltotgba{}, but produce smaller automata on average.
One reason why \duggi{} uses more time is the on-demand nature of algorithm, which rebuilds the intermediate t-GBA several times while disambiguating.
The average number of disambiguation iterations per formula of \duggi{} on the entire \textsc{ltlstore} was $9.5$.

\begin{table}[tbp]
  \centering
  \caption{Cumulative results on the \textsc{ltlstore} benchmark set.}
  \resizebox{\textwidth}{!}{\begin{tabular}{r|r|r|r|r||r|r|r|r}
    & \multicolumn{4}{c||}{non-WDBA-recognizable} & \multicolumn{4}{c}{WDBA-recognizable} \\
    & states & $\varnothing$\,states & time in \(\mathrm{s}\) & timeouts & states & $\varnothing$\,states  & time in \(\mathrm{s}\) & timeouts\\
    \hline
\duggi{}\, & \,\pnodes{16169}\, & \,\(20.702\)\, & \,\pnodes{38932}\, & \,\(167\)\, & \,\pnodes{6866}\, & \,\(16.308\)\, & \,\pnodes{5958}\, & \,\(51\) \\ \duggi{}$_{\setminus \mathrm{R}}$\, & \,\pnodes{15450}\, & \,\(20.196\)\, & \,\pnodes{37803}\, & \,\(183\)\, & \,\pnodes{6857}\, & \,\(16.287\)\, & \,\pnodes{5978}\, & \,\(51\) \\ \duggi{}$_{\setminus \mathrm{RH}}$\, & \,\pnodes{14415}\, & \,\(19.323\)\, & \,\pnodes{39772}\, & \,\(202\)\, & \,\pnodes{6882}\, & \,\(16.346\)\, & \,\pnodes{5758}\, & \,\(51\) \\ \ltltotgba{}\, & \,\pnodes{19547}\, & \,\(24.618\)\, & \,\pnodes{6089}\, & \,\(154\)\, & \,\pnodes{9250}\, & \,\(20.240\)\, & \,\pnodes{3965}\, & \,\(15\) \\ \ltltotgba{}$_{\text{WDBA}}$\, & \,\pnodes{19411}\, & \,\(24.539\)\, & \,\pnodes{7309}\, & \,\(157\)\, & \,\pnodes{7632}\, & \,\(16.700\)\, & \,\pnodes{3814}\, & \,\(15\) \\ \end{tabular}}
  \label{tab:summary}
\end{table}

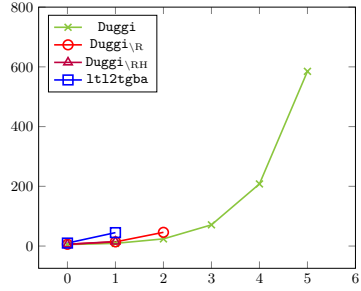
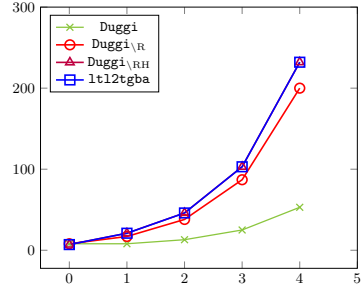
\begin{figure}[tbp]
  \begin{subfigure}{0.49 \textwidth}
    \centering
    \pgfplotstableread{./csv/fairness.csv}{\fairnesscsv}
\begin{tikzpicture}
  \begin{axis}[
    ymax=800,
    xmax=6,
    mark size=3.0pt,
    legend pos=north west
    ]
    \addplot +[line width=1.0pt,mark=x,color=LimeGreen] table  [x expr=\coordindex,y=duggi]{\fairnesscsv};
    \addplot +[line width=1.0pt,mark=o,color=red] table  [x expr=\coordindex,y=duggiNR]{\fairnesscsv};
    \addplot +[line width=1.0pt,mark=triangle,color=purple] table  [x expr=\coordindex,y=duggiNRH]{\fairnesscsv};
    \addplot +[line width=1.0pt,mark=square,color=blue] table  [x expr=\coordindex,y=spot]{\fairnesscsv};
    \legend{\texttt{Duggi},\texttt{Duggi}$_{\setminus \text{R}}$,\texttt{Duggi}$_{\setminus \text{RH}}$,\texttt{ltl2tgba}}
  \end{axis}
\end{tikzpicture}\caption{$\Phi_n = \bigwedge_{i \leq n} (\finally \globally p_{2i} \lor \globally \finally p_{2i+1})$}
    \label{sfig:fairness}
  \end{subfigure}
  \hfill
  \begin{subfigure}{0.49 \textwidth}
    \centering
    \pgfplotstableread{./csv/theta.csv}{\thetacsv}
\begin{tikzpicture}
  \begin{axis}[
    ymax=300,
    xmax=5,
    mark size=3.0pt,
    legend pos=north west
    ]
    \addplot +[line width=1.0pt,mark=x,color=LimeGreen,thick] table  [x expr =\coordindex,y=duggi]{\thetacsv};
    \addplot +[line width=1.0pt,mark=o,color=red] table  [x expr=\coordindex,y=duggiNR]{\thetacsv};
    \addplot +[line width=1.0pt,mark=triangle,color=purple] table  [x expr=\coordindex,y=duggiNRH]{\thetacsv};
    \addplot +[line width=1.0pt,mark=square,color=blue] table  [x expr=\coordindex,y=spot]{\thetacsv};
    \legend{\texttt{Duggi},\texttt{Duggi}$_{\setminus \text{R}}$,\texttt{Duggi}$_{\setminus \text{RH}}$,\texttt{ltl2tgba}}
  \end{axis}
\end{tikzpicture}\caption{$\theta_n = (\bigwedge_{i \leq n} \globally \finally p_i) \rightarrow \globally (req \rightarrow \finally res)$}
    \label{sfig:theta}
  \end{subfigure}
  \caption{UBA sizes for two sets of parametrized formulas.}
  \label{fig:parametrized}
\end{figure}

\tudparagraph{0.3 em}{LTL rewrites and the purely-universal heuristic.}
A formula that benefits from using the rewrite rules I and II is $\Phi_n = \bigwedge_{i \leq n} \finally \globally p_{2i} \lor \globally \finally p_{2i+1}$, which describes a \emph{strong fairness} condition.
Here \ltltotgba{} applies the rule $\finally \varphi \lor \globally \finally \psi \mapsto \finally (\varphi \lor \globally \finally \psi)$ which yields $\bigwedge_{i \leq n} \finally (\globally p_{2i} \lor \globally \finally p_{2i+1})$.
Applying rule II yields the formula $\Psi_n = \finally \globally (\bigwedge_{i \leq n} p_{2i} \lor \finally p_{2i+1})$.
\Cref{sfig:fairness} shows that \duggi{} produces smaller automata for $\Phi_n$. \Cref{sfig:theta} shows the corresponding results for the parametrized formula $\theta_n = (\bigwedge_{i \leq n} \globally \finally p_i) \rightarrow \globally (req \rightarrow \finally res)$ which is a request/response pattern under fairness conditions.

\begin{figure}[tbp]
    \begin{subfigure}[t]{0.49\textwidth}
        \centering
        \includegraphics[width=0.8\textwidth]{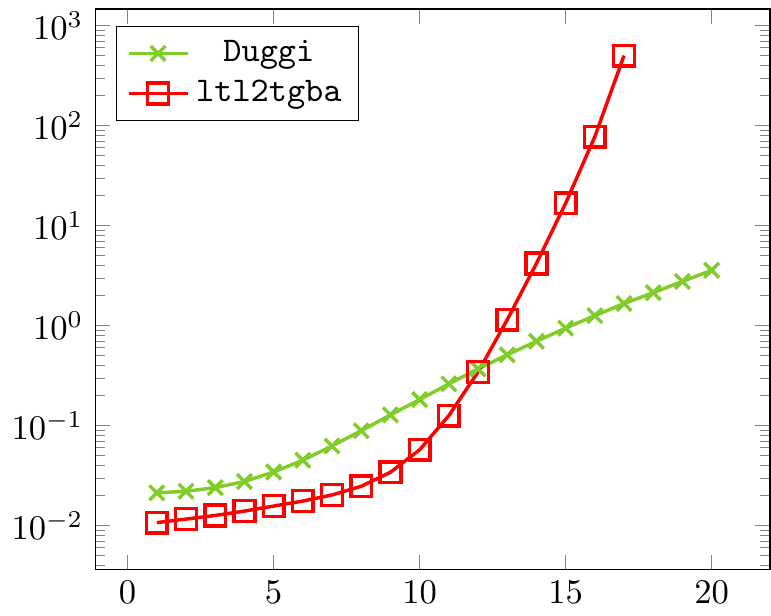}
        \caption{Time in seconds needed for the translation of \(\varphi^\mathrm{steps}_k\) into a UBA.}
        \label{fig:nsteps_ltl}
    \end{subfigure}\hfill
    \begin{subfigure}[t]{0.49\textwidth}
    \centering
\includegraphics[width=0.8\textwidth]{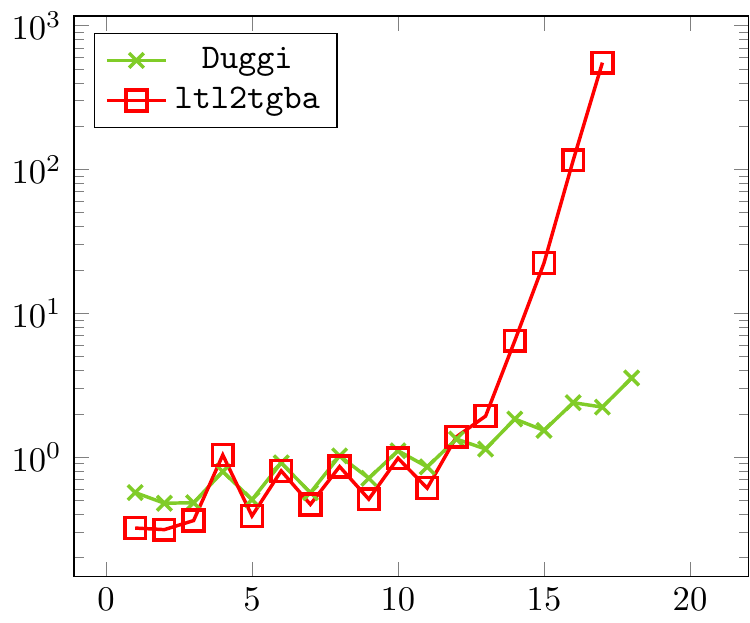}
        \caption{Time in seconds needed for model checking the BRP model with \(\varphi^\mathrm{steps}_k\).}
        \label{fig:nsteps_brp}
    \end{subfigure}
    \caption{Time consumption for translating and model checking
    \(\varphi^\mathrm{steps}_k\) (which includes building the automaton). }
\label{fig:n_steps}
\end{figure}

A property that profits from the ``on-demand'' disambiguation is: ``\(b\) occurs \(k\) steps before \(a\)''. We express it with the formula \(\varphi^\mathrm{steps}_k = \neg a \ \until \  \bigl( b \wedge \neg a \land \neXt \neg a \land \ldots \land \neXt^{k-1} \neg a \land \neXt^k a\bigr)\).
Both \duggi{} and \ltltotgba{} produce the minimal UBA, but \ltltotgba{} produces an exponential-sized automaton in an intermediate step, because it does not realize that the original structure is already unambiguous.
This leads to high run times for large $k$ (see \Cref{fig:nsteps_ltl}).

\tudparagraph{0.3 em}{Use case: probabilistic model checking.}\label{sec:pmc}
Now we look at an important application of UBA, the analysis of Markov chains.
We compare run times of an implementation of~\cite{BKKKMW16} for Markov chain model checking with UBA, using \prism{} (version 4.4) and either \duggi{} or \ltltotgba{} as automata generation backends.
We take two models of the \prism{} benchmark suite~\cite{prismBenchmark}, the
bounded retransmission protocol, and the cluster working protocol~\cite{HHK00}.

The bounded retransmission protocol (BRP) is a message transmission protocol,
where a sender sends a message and receives an acknowledgment if the transmission was successful.
We set the parameter \(\mathrm{N}\) (the
number of the message parts) to \(16\), and \(\mathrm{MAX}\) (the number of
maximal retries) to \(128\).
We reuse $\varphi^\mathrm{steps}_k$, which now means: ``\(k\) steps before an acknowledgment there was a
retransmit'', where we replace \(a\) by \texttt{ack\_received} and \(b\) by \texttt{retransmit}.
As expected, the faster automaton generation leads to lower model checking times when using \duggi{} (\Cref{fig:nsteps_brp}).
The reason for the spikes in
\Cref{fig:nsteps_brp} is that the probability of the property is zero in the BRP model for odd $k$.
This makes the model checking (which uses the numeric procedure of \cite{BKKKMW16}) easier.
For bigger \(k\) the automaton generation uses a bigger share of the time, making this effect less pronounced.

As second model we analyse the cluster working model with the LTL properties presented in \cite{HLSTZ15}.
It consists of a workstation cluster with two sub-clusters that are connected by a backbone and have $n = 16$ participants each.
\begin{figure}[tbp]
    \begin{subfigure}[t]{0.48 \textwidth}
    \centering
    \resizebox{0.8 \textwidth}{!}{\includegraphics{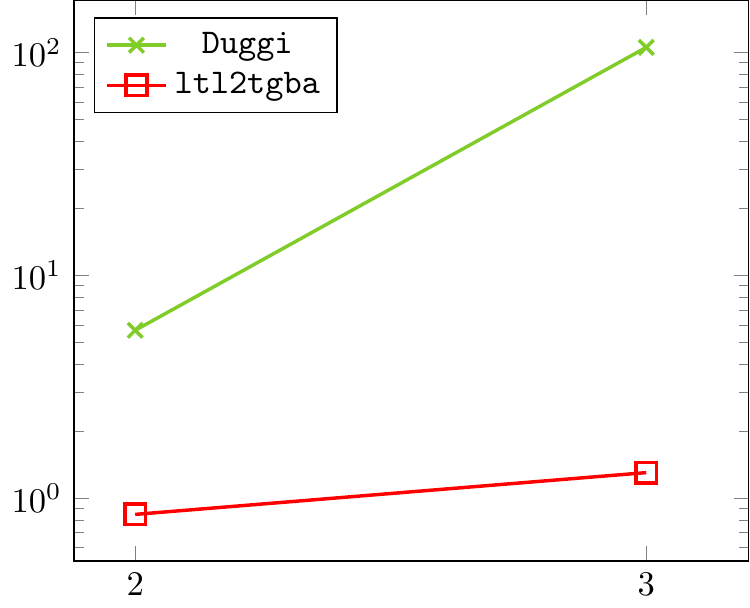}
    }
    \caption{Time consumption for \(\varphi_k\).}
    \label{fig:cluster_gf_and}
    \end{subfigure}\hfill \begin{subfigure}[t]{0.48 \textwidth}
    \centering
    \resizebox{0.8 \textwidth}{!}{\includegraphics{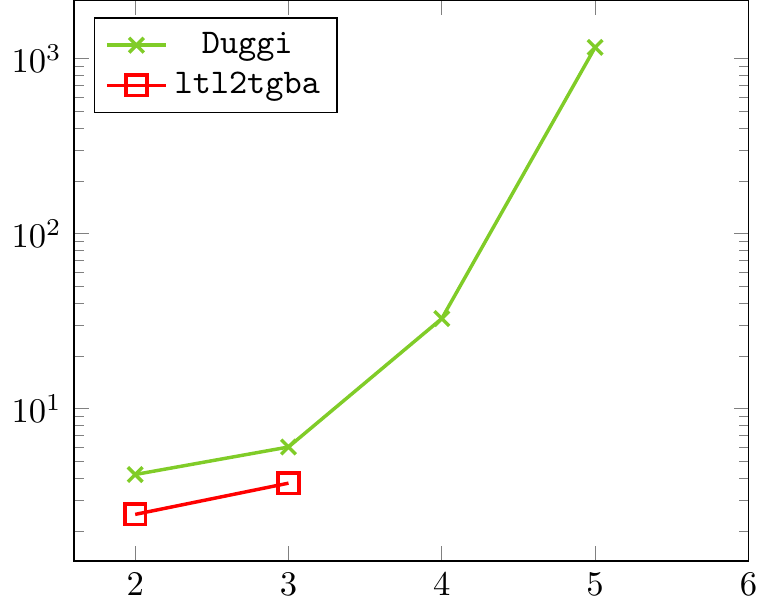}
    }
    \caption{Time consumption for \(\psi_k\).}
    \label{fig:cluster_gf_or}
    \end{subfigure}\caption{Model checking times for the cluster protocol with 
    \(\varphi_k\) and \(\psi_k\).}
\end{figure}
Let \(\texttt{fct}_i\) denote the number of functional working stations in sub-cluster \(i\).
We define \(\varphi_{\globally\finally} = \globally \finally (\texttt{fct}_1 = n)\), which expresses that the first cluster stays functional on the long run and \(\varphi_{\finally\globally} =\bigvee_{i \in \lbrace 0,\ldots,k\rbrace}\finally\globally (\texttt{fct}_2 = n - i)\), which expresses the property that from some point, the second cluster contains at least \(n-k\) functional working stations.
We check the two formula patterns \(\varphi_k = \varphi_{\globally\finally}
\wedge \varphi_{\finally\globally}\)  and  \(\psi_k = \varphi_{\globally\finally} \vee \varphi_{\finally\globally}\).
We leave out a third property described in~\cite{HLSTZ15}, which is WDBA-recognizable (see \Cref{app:experiments} for further details).

 The results for \(\varphi_k\) are depicted in
 \Cref{fig:cluster_gf_and}. Both tools have a time-out at \(k=4\), although, for smaller
 \(k\), the time consumption of \duggi{} was bigger than \ltltotgba{}. Comparing
 the automata size, \duggi{} produces smaller automata for both \(k=2\) and \(k=3\), e.g., \pnodes{32} (\duggi{}) vs. \pnodes{137} (\ltltotgba{}) states for \(k=3\).
The results for \(\psi_k\) can be seen in
\Cref{fig:cluster_gf_or}. \duggi{} performed better than \ltltotgba{}, as \duggi{} reached the time-out at \(k = 6\) (vs. \(k=4\) for \ltltotgba{}).
However, if no time-out was reached, \ltltotgba{} consumed
less time. Nevertheless, for \(k \leqslant 3\), model checking time of
both tools was below \(7\,\mathrm{s}\). Still, \duggi{} produced smaller automata, e.g., \pnodes{25} (\duggi{}) vs. \pnodes{59} (\ltltotgba{}) states for \(k=3\).
 \section{Conclusion}
\label{sec:conclusion}

In this paper we have presented a  novel LTL-to-UBA translation.
In contrast to other LTL-to-UBA translations \cite{CouSahSut03,BLW13,Duret17} we use alternating automata as an intermediate representation.
To adapt the VWAA-to-NBA construction of \cite{GO01} for the unambiguity setting, 
we introduced a notion of unambiguity for VWAA and a corresponding
disambiguation procedure.
This may be of independent interest when considering unambiguity for different types of alternating automata.
We devise heuristics that exploit structural properties of purely-universal and alternating formulas for disambiguation.
Furthermore, we identify LTL rewriting rules that benefit the construction of UBA.

Experimental analysis on a big LTL benchmark set shows that our tool \duggi{} produces smaller automata on average than the existing tools.
In particular, formulas containing nested $\finally$ and $\globally$ benefit from our heuristics and rewrite rules.
Such formulas occur often, for example when modelling fairness properties.
Experiments on Markov chain model checking indicate that the positive properties of our approach carry over to this domain.

Our approach opens up many possibilities for optimization, for example by processing multiple source states at once, or in a certain order.
This would let us decrease the number of disambiguation steps, and thus the run time.
It would be interesting to investigate intermediate strategies in our framework that allow for a trade-off between automata sizes and computation times.
Another promising direction is to identify more patterns on LTL or VWAA that allow special disambiguation transformations.
As many interesting properties stem from the safety-/cosafety-class, a combination of our approach with the ideas of the UFA generation described in \cite{Mohri13} seems to be beneficial.
The application of simulation-based automata reductions to UBA is also an open question.
Whereas bisimulation preserves unambiguity, it is unclear whether there exist simulation relations targeted specifically at shrinking unambiguous automata.

\bibliographystyle{splncs04}
\bibliography{main}

\newpage

\appendix

\section{From LTL to VWAA}
\label{app:ltl-to-vwaa}

We follow the translation given in~\cite{GO01} but use all subformulas as states, as proposed in~\cite{BKRS12}.
The main idea for the construction is to take advantage of the expansion
laws for \(\until\) and \(\release\):
\begin{align*}
    \varphi \until \psi &= \psi \vee \bigl(\varphi \wedge \neXt (\varphi \until \psi)\bigr)\\
    \varphi \release \psi &= \psi \wedge \bigl(\varphi \vee \neXt (\varphi \release \psi)\bigr)
\end{align*}
For a better presentation, we give here the transition function in an unusual
way, namely as a function: \(\delta : Q \rightarrow 2^{2^\Sigma \times 2^Q}\).
It is easy to transform a function \(\delta\) given in the above form into the
representation $\Delta : Q \times \Sigma \to \mathcal{B}^+$:
\[\Delta(q, a) = \bigvee_{\substack{(\alpha,C) \in \delta(q)\\ a \in \alpha}} ~\bigwedge_{s \in C} s \]

The construction takes an LTL formula in which all negations appear in front of atomic propositions (\emph{positive normal form}) as input.
If all dual operators are included in the syntax ($\release$ is the dual of $\until$), LTL formulas can be transformed into an equivalent formula in positive normal form of the same length.

\begin{definition}
    Let \(\varphi\) be an LTL formula in positive normal form over \(\AP\). We define the co-B\"uchi
    \(\omega\)-automaton \(\aut_\varphi\) as a tuple \((Q, \Sigma, \delta, \varphi,
    \Fin{Q_f})\) where $\Sigma = 2^{AP}$, \(Q\) is the set of subformulas of $\varphi$, \(Q_f = \lbr
    \psi_1 \until \psi_2 \, : \, \psi_1 \until \psi_2\in Q\rbr \) and \(\delta\) is
    defined as follows:
    \[\begin{array}{llcll}
            \delta(\true) &= \lbr(\Sigma,\varnothing)\rbr  & \hspace*{3 em} & \delta(\false) &=\varnothing\\
        \delta(a) &= \lbr(\Sigma_a,\varnothing)\rbr  & \qquad & \delta(\neg a) &=\lbr(\Sigma_{\neg a}, \varnothing)\rbr\\
        \delta(\varphi \wedge \psi) &= \delta(\varphi) \otimes \delta(\psi)  & \qquad & \delta(\varphi \vee \psi) &= \delta(\varphi) \cup \delta(\psi)\\
        \delta(\neXt \varphi) &= \lbr(\Sigma, \lbr \varphi\rbr)\rbr & & &\\
        \multicolumn{5}{c}{\delta(\varphi \, \until \, \psi) = \delta(\psi) \cup (\delta(\varphi) \otimes \lbr (\Sigma, \lbr \varphi \, \until \, \psi\rbr)\rbr)}\\
        \multicolumn{5}{c}{\delta(\varphi \, \release \, \psi) = \delta(\psi)\otimes (\delta(\varphi) \cup \lbr (\Sigma, \lbr \varphi \, \release \,\psi\rbr)\rbr)}
    \end{array}\]
    where
    \[\begin{array}{rl}
    M_1 \otimes M_2 &= \lbr (u \cap v, Q_1 \cup Q_2) \, : \, (u,Q_1) \in M_1, (v, Q_2) \in M_2\rbr \\
    \Sigma_a &= \{ m \in \Sigma \, : \, a \in m\} \\
    \Sigma_{\neg a} &= \Sigma \setminus \Sigma_a
    \end{array}\]
\end{definition}

The translation has been refined in \cite{BKRS12} to produce smaller automata
with less non-determinism. In our implementation, we use these optimizations,
but these optimizations do not alter or simplify our disambiguation algorithm.

\begin{lemma}[\cite{GO01,BKRS12}]
    Let \(\varphi\) and \(\aut_\varphi\) be as above. Then, \(\langinf(\varphi) = \langinf(\aut_\varphi)\).
\end{lemma}
 \section{Degeneralization}
\label{app_degen}

As some applications demand an automaton with a (non-generalized) B\"uchi acceptance on states, a
degeneralization procedure, which also converts transition-based acceptance to
state-based acceptance, is necessary.
We use the degeneralization construction as presented in~\cite{GO01}.

\begin{definition}[Degeneralization]
    Let \(\mathcal{G} = (Q, \Sigma, \delta, Q_0, \Inf{F_1} \wedge \ldots \wedge
\Inf{F_n})\) be a \tgba{}. Then we define \(\mathcal{N}_\mathcal{G}\) to be \((Q
\times \lbrace 0, \ldots, n\rbrace, \Sigma, \delta_\mathcal{N}, Q_0 \times
\lbrace 0\rbrace,
Q \times \lbrace n\rbrace )\) where
\begin{align*}
    \delta_\mathcal{N}(\langle q, i\rangle, a) &= \lbrace \langle q',
i'\rangle \, : \, q' \in \delta(q, a) \textrm{ and } i' = \mathrm{next}(i, q \xrightarrow{a} q')\rbrace \textrm{ and}\\
    \mathrm{next}(i, q \xrightarrow{a} q') &=
    \begin{cases}
        \max(\lbrace i \leqslant j \leqslant n \, : \, \forall k \in \lbrace i+1, \ldots , j \rbrace. q \xrightarrow{a} q' \in F_k\rbrace\rbrace) & \textrm{ if } i \neq n \\
        \max(\lbrace 0 \leqslant j \leqslant n \, : \, \forall k \in \lbrace 0, \ldots , j \rbrace. q \xrightarrow{a} q' \in F_k\rbrace\rbrace) & \textrm{ if } i = n \\
    \end{cases}
\end{align*}
\end{definition}

This construction creates copies of \(\mathcal{G}\) for every B\"uchi acceptance
set \(\Inf{F_i}\) and switches from copy \(i\) to copy \(i'\) if and only if
the current transition satisfies all acceptance sets \(\Inf{F_j}\) with \( i < j \leqslant i'\).
Therefore:

\begin{lemma}[Theorem 3 of \cite{GO01}]
    Let \(\mathcal{G}\) and \(\mathcal{N}_\mathcal{G}\) be as above. Then
    \(\langinf(\mathcal{G}) = \langinf(\mathcal{N}_\mathcal{G})\).
\end{lemma}

Obviously, the number of states in \(\mathcal{B}_\mathcal{G}\) is \(\vert Q\vert
\cdot n\).
 \section{Proof for \Cref{sec:unambigvwaa}}
\label{app:unamb_vwaa_lemm}

The correctness proof of the original construction from VWAA to t-GBA in~\cite{GO01} features a similar proof, but here we want to show that we can find corresponding runs such that the layers of the VWAA are equal to the states of the t-GBA.
To show this, we need the precondition that the VWAA is unambiguous. 

The first part shows that every accepting run of $\mathcal{G}_{\aut}$ is matched by an accepting run of $\aut$.
It is complicated by the way that accepting transitions are defined in $\mathcal{G}_{\aut}$: they require that a final state $q_f$ must \emph{have the option} to choose a non-looping transition infinitely often (see \Cref{vwaa2tgba}).
Thus, there might exist runs of $\mathcal{G}_{\aut}$ that represent runs of $\mathcal{A}$ where $q_f$ has this option infinitely often, but never takes the ``good'' transition.
However, this situation cannot occur if $\mathcal{A}$ is unambiguous as it would imply multiple accepting runs for the same word that differ on some layer.

In the following proofs we denote the successors of a node $q \in V$ in a graph $(V,E)$ by $E(q)$.
As before, we use $\delta(q,a)$ to denote the set of minimal models of $\Delta(q,a)$, and $I$ to denote the set of minimal models of $\iota$.

\gbatovwaarun*
\begin{proof}
  We show this lemma by induction over the number of elements in $F$.

  Base case: $F = \varnothing$.
  We define $V = \{ (q,i) \, : \, q \in Q_i\}$.
  $Q_0 \in I$ must hold as $r$ is a run of $\mathcal{G}_{\aut}$ and thus the initial condition is satisfied.
  For every transition $Q_i \xrightarrow{w[i]} Q_{i+1}$ we know that $Q_{i+1} \in \bigotimes_{q \in Q_i} \delta(q,w[i])$ and we define the edges between $V(i)$ and $V(i+1)$ to match the corresponding successor sets.
  The result is a run of $\aut$ for $w$ and it is accepting as $F = \varnothing$.

  Now consider $F = F' \cup \{q_f\}$. Let $\mathcal{A} =
  (Q,\Sigma,\Delta,\iota,\Fin{F})$ and $\mathcal{A}' = (Q,\Sigma,\Delta,\iota,\Fin{F'})$ be
  $\mathcal{A}$ where $q_f$ is not marked final.
  Let $w$ and $r = Q_0Q_1 \ldots$
  be an accepting run of $\mathcal{G}_{\mathcal{A}}$ for $w$.
  By \Cref{vwaa2tgba} the only
  difference between $\mathcal{G}_{\mathcal{A}'}$ and
  $\mathcal{G}_{\mathcal{A}}$ is that $\mathcal{G}_{\mathcal{A}}$ has the
  additional acceptance set $T_{q_f}$, which adds an obligation for runs to be
  accepting. This implies that any accepting run
  of $\mathcal{G}_{\mathcal{A}}$ is also an accepting run of
  $\mathcal{G}_{\mathcal{A}'}$. So $r$ is an accepting run of
  $\mathcal{G}_{\mathcal{A}'}$ and we make use of the induction hypothesis to
  get an accepting run $\rho = (V,E)$ of $\mathcal{A}'$ such that for all $i$:
  $Q_i = V(i)$. If we can show that $\rho$ is also an accepting run of
  $\mathcal{A}$ for $w$ we are done.

  So suppose that it is not accepting. Then there exists a rejecting path $\pi$
  through $\rho$ that ultimately stabilizes on a state $f \in F$. But $f$ can only be $q_f$, as any other rejecting path would contradict
  the fact that $\rho$ is an accepting run of $\mathcal{A}'$. Hence there is
  some $k$ such that for all $j > k$: $q_f \in V(j)$ and $((q_f,j),(q_f,j+1)) \in E$.

  As $q_f \in F$ we know that there are infinitely many $i$ such that
  $(Q_i,w[i],Q_{i+1}) \in T_{q_f}$.
$\mathcal{G}_{\mathcal{A}}$ has only a
  finite amount of states, and $\Sigma$ is also finite, so there must exist a
  triple $(S,a,S')$ such that for infinitely many $i > k$: $Q_i = S$,
  $Q_{i+1} = S'$, $w[i] = a$ and $(S,a,S') \in T_{q_f}$.
  Furthermore, for
  infinitely many of these $i$ the edges of $\rho$ between $S$ and $S'$ will be
  the same.
  We fix an edge relation between $S$ and $S'$ in $\rho$ that occurs infinitely
  often, name the succesors of $q \in S$ by $e(q)$ and set:
  \[J = \{j \mid S = V(j), S' = V(j+1), w[j] = a \text{ and } E(q,j) = e(q)
    \text{ for all } q \in S\}\]
  It follows that $S' = \bigcup_{q \in S}e(q)$ and $J$ is infinite.
  By the definition of $T_{q_f}$ there exists a way for
  $q_f$ to take a non-looping transition, i.e. there exists a successor
  configuration $Y$, such that $Y \in \delta(q_f,a)$,
  $q_f \notin Y$ and $Y \subseteq S'$.

  Our aim is to show that we can choose $Y$ instead of $e(q_f)$ as successor set
  of $q_f$ in $S$ for all these edges, without losing the property that for all
  $i$: $Q_i = V(i)$.
  Clearly, we would
  not make the following layer bigger by choosing $Y$, as $Y \subseteq S'$.
  So our aim is to show that by choosing $Y$ as successor set for $q_f$, the
  following layer is not strictly smaller.
  We show that the following holds:
  \[S' = \underbrace{Y \cup (\displaystyle\bigcup_{q \in S \setminus \{q_f\}}
      e(q))}_{S''}\]
  This would mean that replacing $e(q_f)$ by $Y$ does not change the following
  layer in the run. 

  Suppose that this equality does not hold.
  Then $S''$ is strictly contained in $S'$, as $Y \subseteq S'$.
  Both $S''$ and $S'$ are possible successor configurations of $S$ and
  $a$ as $e(q) \in \delta(q,a)$ for every $q \in S$
  and both $Y$ and $e(q_f)$ are elements of
  $\delta(q_f,a)$.

  But then for all $j \in J$ we can construct an accepting run $\rho_j$ of
  $\aut$ on $w$ such that all pairs of runs in $\{\rho_j \mid j \in J\}$ differ on some layer.
  We construct $\rho_j = (V',E')$ by mimicking $\rho$ up till position $j$ and in position $j$ we choose $Y$ as successor configuration of $q_f$.
  For all following positions $k > j$ such that $k \in J$ we also choose $Y$
  as successor configuration of $q_f$, given that $q_f \in V'(k)$.
  All infinite paths of $\rho_j$ that do not visit $q_f$ infinitely often can be
  mapped to infinite paths in $\rho$, and thus in particular no $f \in F'$ is
  visited infinitely often.
  Furthermore, $q_f \notin E'(q_f,k)$ for all $k \in J$ such that $k \geq j$, and
  thus no infinite path stabilizes on $q_f$.
  Therefore $\rho_j$ is an accepting run of $\aut$ on $w$.

  Two runs $\rho_j, \rho_h$ with $j,h \in J$ and $j < h$ differ on depth $j+1$
  as $\rho_j$ chooses $Y$ as successor set of $q_f$ on level $j$ and
  $\rho_h$ chooses $e(q_f)$.
  This implies that the $(j+1)$'th layer of $\rho_j$ is $S''$ and the $(j+1)$'th
  is $S'$ and hence $\rho_j$ and $\rho_h$ differ on layer $j+1$, which contradicts the fact that $\aut$ is unambiguous.

  Thus $S' = S''$ and we get $\rho' = (V',E')$ by mimicking $\rho$ for all
  positions $k \notin J$, and setting $E(q_f,j) = Y$ for all positions $j \in
  J$.
  Edges of the other states are not changed for any position.
  As $S' = S''$ we can do this without changing the vertex set, i.e. $V' = V$.
  The new run is accepting as $q_f \notin E'(q_f,j)$ for all positions $j \in
  J$ and hence no infinite branch stabilizes at the state $q_f$.
  Furthermore, the property $Q_i = V'(i)$ holds for all $i \in \mathbb{N}$ as it
  already holds for $\rho$ by induction hypothesis.
\end{proof}

To find an accepting run of $\mathcal{G}_{\mathcal{A}}$ that corresponds to an accepting run of $\aut$ we can directly use the definition of the transition relation of $\mathcal{G}_{\mathcal{A}}$:

\vwaatogbarun*
\begin{proof}
Let $\aut$ and $\mathcal{G}_{\aut}$ be as above and $\rho = (V,E)$ be an accepting run of
$\aut$ for $w$.
We show that $r = V(0)V(1)\ldots$ is an accepting run of $\mathcal{G}_{\aut}$.
As $\rho$ is a run of $\aut$ we get $V(0) \in I$ and hence $r$ satisfies
the initial condition of a t-GBA run.

We show that for all $i: V(i+1) \in \bigotimes_{q \in V(i)}\delta(q,w[i])$.
As $\rho$ is a run of $\aut$, every $q \in V(i)$ must have successors $E(q,i)
\subseteq V(i+1)$ such that $E(q,i) \in \delta(q,w[i])$.
Furthermore, each $q' \in V(i+1)$ must have a predecessor in $V(i)$ which implies
that $\bigcup_{q \in V(i)} E(q,i) = V(i+1)$.
By definition of $\otimes$ we get
\[V(i+1) \in \bigotimes_{q \in V(i)} \delta(q,w[i])\]
and thus $(V(i),w[i],V(i+1))$ is a transition of $\mathcal{G}_{\aut}$.

Finally, we have to show that $r$ is accepting. We show for all $f \in F$ that
for infinitely many positions $i$: $(V(i), w[i], V(i+1)) \in T_f$. Suppose that this
does not hold. Then there exists a $k$ such that for all $n \geq k:
(V(n),w[n],V(n+1)) \notin T_f$. By the definition of $T_f$ we know that for all
$n > k: f \in V(n)$ and there is no $S \in \delta(f,w[n])$ such that $f
\notin S$ and $S \subseteq V(n+1)$. This implies, however, that infinitely often
$f$ is in the set of successors of $f$.
But then $\rho$ is not accepting which contradicts the assumption.
\end{proof}

\unambispreserved*
\begin{proof}
  Suppose that $\mathcal{A}$ is unambiguous and $\mathcal{G}_{\mathcal{A}}$ is not.
  Then there exists a word $w \in \Sigma^{\omega}$ and two accepting runs $r_1 = Q_0Q_1\ldots,r_2=P_0P_1\ldots$ of $\mathcal{G}_{\mathcal{A}}$ for $w$ such that for some $i \in \Nat$: $Q_i \neq P_i$.
  Choose $u \in \Sigma^*$ and $v \in \Sigma^{\omega}$ such that $w = uv$ and $|u| = i$.
This contradicts the assumption that $\mathcal{A}$ is unambiguous as $V_1(i)$ and $V_2(i)$ are both reachable via $u$ and accept the word $v$.

  To show the other direction, assume that $\mathcal{A}$ is ambiguous.
  Then, there exists a word $w \in \Sigma^{\omega}$ and two accepting runs $\rho_1 = (V_1,E_1)$, $\rho_2 = (V_2,E_2)$ of $\mathcal{A}$ for $w$ such that for some $i \in \Nat$: $V_1(i) \neq V_2(i)$.
  By \Cref{lemm:gba_vwaa_run_corr}, there exist two accepting runs $r_1,r_2$ of $\mathcal{G}_{\mathcal{A}}$ on $w$ such that $r_1(i) \neq r_2(i)$, and hence $\mathcal{G}_{\mathcal{A}}$ is ambiguous.
\end{proof}

\section{Unambiguity check on VWAA}
\label{app:unamb_vwaa}

We now turn to the complexity of deciding whether a VWAA is unambiguous.
First, we define the union of two automata which we need to prove hardness.

\begin{definition}
Let \(\aut_1 = (Q_1, \Sigma, \Delta_1, \iota_1, \Fin{F_1})\) and \(\aut_2 = (Q_2, \Sigma, \Delta_2, \iota_2, \Fin{F_2})\) be two alternating co-B\"uchi automata over the same alphabet. The union automaton \(\aut_1 \cup \aut_2\) is defined as a tuple
\((Q_\cup, \Sigma, \Delta_\cup, \iota_{\cup}, \Fin{F_\cup})\) where
\begin{itemize}
    \item \(Q_\cup = (Q_1 \times \lbr 1\rbr) \cup (Q_2 \times \lbr 2\rbr)\),
    \item \(\Delta_\cup((q,i), a) = 
      \begin{cases}
        \Delta_1(q,a) \times 1 & \text{if } i = 1 \\
        \Delta_2(q,a) \times 2 & \text{if } i = 2 \\
      \end{cases}
      \)
    \item \(\iota_{\cup} =  (\iota_1 \times 1) \lor (\iota_2 \times 2)\), and
    \item \(F_\cup = (F_1 \times \lbr 1\rbr) \cup (F_2 \times \lbr 2 \rbr)\)
\end{itemize}

and $f \times i$, for $f \in \mathcal{B}^+(X)$ and $i \in \mathbb{N}$, is the positive Boolean formula over $X \times \{i\}$ in which every variable $x \in X$ that occurs in $f$ is replaced by $(x,i)$.
\end{definition}

As the definition of the union automaton merely creates an automaton consisting
of two disjoint copies, it holds that \(\langinf(\aut_1 \cup \aut_2) =
\langinf(\aut_1) \cup \langinf(\aut_2)\).

\begin{lemma}
  \label{lemm:pspacehard}
    The problem of deciding whether a given \vwaa{} $\mathcal{A}$ is unambiguous is PSPACE-hard.
\end{lemma}

\begin{proof}
    We reduce the PSPACE-hard problem of LTL satisfiability \cite{SC85}
    to deciding unambiguity of VWAA.
    
    Assume that we are given an LTL formula \(\varphi\).
    We can construct a VWAA \(\aut_\varphi\) that is equivalent to $\varphi$ in time linear in the size of $\varphi$~\cite{Vardi94,MullerSS88}.
An accepting run in $\aut_{\varphi}$ corresponds to two distinct accepting runs in $\aut_{\varphi} \cup \aut_{\varphi}$ and thus we have:
    \[\langinf(\aut_\varphi \cup \aut_\varphi) = \varnothing \text{ iff } \aut_\varphi \cup \aut_\varphi \text{ is unambiguous.}\]

    As \(\langinf(\aut_\varphi \cup \aut_\varphi) = \langinf(\aut_\varphi) =
    \langinf(\varphi)\) holds, \(\aut_\varphi \cup \aut_\varphi\) is
    unambiguous if and only if \(\varphi\) is satisfiable.
\end{proof}

\begin{lemma}
  \label{lemm:inpspace}
    The problem whether a given \vwaa{} \(\mathcal{A}\) is unambiguous can be
    decided in PSPACE. 
\end{lemma}

\begin{proof}
    By the theorem of Savitch, we know that NPSPACE=PSPACE. 
We give an NPSPACE (in the size of \vwaa{}) algorithm for checking whether \(\mathcal{G}_\aut\) is unambiguous, which is enough by \Cref{lemm:vwaatotgbaunamb}.
    
    The algorithm guesses a lasso in the self product of \(\mathcal{G}_\aut\):
    a path that reaches a state \(( C_1, C_2)\) such that $C_1 \neq C_2$ and such that there exists an accepting loop starting in \(( C_1, C_2)\).
    To guess a successor of a state $(Q_1,Q_2)$ in \(\mathcal{G}_\aut \otimes \mathcal{G}_\aut\) for a symbol $a \in \Sigma$ the algorithm first guesses a successor configuration of $Q_1$ and $a$ in $\aut$.
    This can be done by guessing a set $S_q \subseteq Q$ for each $q \in Q_1$ and verifying that $S_q$ is a minimal model of $\Delta(q,a)$.
    Then $\bigcup_{q \in Q_1} S_q$ is chosen as successor configuration of $Q_1$ and the procedure is repeated for $Q_2$.
    So non-deterministically guessing a successor of $(Q_1,Q_2)$ in \(\mathcal{G}_\aut \otimes \mathcal{G}_\aut\) can be done in polynomial time.

    The algorithm only needs to remember the current state in \(\mathcal{G}_\aut \otimes \mathcal{G}_\aut\) and, for the loop, which of the acceptance sets in $\mathcal{T}$ have already been satisfied.
    Therefore, it requires at most space polynomial in the size of $\varphi$.

\end{proof}

By combining \Cref{lemm:pspacehard} and \Cref{lemm:inpspace} we get:

\checkunambpspace*
 \section{Proofs for \Cref{sec:liftnondet}}
\label{app:liftingnondet}

\disjfreenutransf*
\begin{proof}
  Let $\nu$ be a purely-universal formula.
  As a first step we remove all occurrences of $\until$ and $\release$ in $\nu$ that are not in the scope of some $\globally$ by applying the rules $\nu_1 \, \until \, \nu_2 \mapsto \nu_2 \lor (\nu_1 \land \finally \nu_2)$ and $\varphi \release \nu' \mapsto \nu'$.
  These transformation rules preserve equivalence if $\nu_1,\nu_2$ and $\nu'$ are purely-universal.
  Then, we proceed by induction on the structure of $\nu$.
  In the case that $\nu = \globally \varphi$ it already has the desired structure.
  For the cases $\lor, \land, \neXt$ and $\finally$ we apply the induction hypothesis to the subformulas and then lift the disjunction over the corresponding operators.
\end{proof}

\disjfreenu*
\begin{proof}
Let $\nu$ be a disjunction-free and purely-universal formula.
We show the claim by induction on the structure of $\nu$ ($\nu_1$, $\nu_2$ and $\nu'$ are assumed to be purely-universal).
\begin{itemize}
\item $\nu = \globally \varphi$: By instantiation we get the $\varphi \land \neXt \globally \varphi \equiv \globally \varphi$.
\item $\nu = \neXt \nu'$: By induction hypothesis we have $\mathfrak{g}(\nu') \land \neXt \nu' \equiv \nu'$, which implies $\neXt \mathfrak{g}(\nu') \land \neXt \neXt \nu' \equiv \neXt \nu'$.
\item $\nu = \nu_1 \land \nu_2$: By induction hypothesis we have $\mathfrak{g}(\nu_1) \land \neXt \nu_1 \equiv \nu_1$ and $\mathfrak{g}(\nu_2) \land \neXt \nu_2 \equiv \nu_2$.
  This implies $\mathfrak{g}(\nu_1) \land \mathfrak{g}(\nu_2) \land \neXt \nu_1 \land \neXt \nu_2 \equiv \nu_1 \land \nu_2$.
\item $\nu = \finally \nu'$: We have defined $\mathfrak{g}(\finally \nu') = \true$.
  As $\finally \nu'$ is an alternating formula it satisfies $\neXt \finally \nu' \equiv \finally \nu'$, which proves the claim.
\end{itemize}
\end{proof}

\unutrans*
\begin{proof}
  We show \emph{1.} by showing $\varphi \, \until (\nu \lor \psi) \implies \nu \lor \gamma$ and $\nu \lor \gamma \implies \varphi \, \until (\nu \lor \psi)$.
  \begin{itemize}
  \item $\varphi \, \until (\nu \lor \psi) \implies \nu \lor \gamma$:
    Take a word $w$ that satisfies $\varphi \, \until (\nu \lor \psi)$.
    If $w \models \nu$ we are done.
We know that either $w \models \varphi \, \until \nu$ or $w \models \varphi \, \until \psi$.

    In the first case there is a \emph{least} $i$ such that $w[i..] \models \nu$ and for all $j < i$: $w[j..] \models \varphi$.
    We get $w[(i-1)..] \models \varphi \land \neg \mathfrak{g}(\nu) \land \neXt \nu$ by \Cref{lemm:disjfreenu} and thus $w \models \gamma$.

    In the second case there is an $i$ such that $w[i..] \models \psi$ and for all $j < i$: $w[j..] \models \varphi$.
    We can assume that $w[i..] \not\models \nu$, as the first case applies otherwise, which implies $w[i..] \models \psi \land \neg \nu$ and thus proves $w \models \gamma$.

  \item $\nu \lor \gamma \implies \varphi \, \until (\nu \lor \psi)$:
    We have: \\
    \[\gamma \implies \varphi \, \until ((\varphi \land \neXt \nu) \lor \psi) \implies \varphi \, \until (\nu \lor \psi)\]
    which proves this case.
  \end{itemize}

  To show \emph{2.} we show that $\lang(\nu) \cap \lang(\gamma) = \varnothing$.
  By \Cref{lemm:disjfreenu} we have $\neg \nu \equiv \finally \neg \mathfrak{g}(\nu)$ or $\neg \nu \equiv \true$.
  If $\neg \nu \equiv \true$ we get $\lang(\nu) = \varnothing$, which proves the claim.
  Otherwise, we have $\neg \nu \equiv \finally \neg \mathfrak{g}(\nu)$.
  Both $\varphi \land \neg \mathfrak{g}(\nu) \land \neXt \nu$ and $\psi \land \neg \nu$ imply $\finally \neg \mathfrak{g}(\nu)$.
  As $\finally \neg \mathfrak{g}(\nu)$ is prefix invariant we get $\gamma \implies \neg \nu$ and thus $\lang(\nu) \cap \lang(\gamma) = \varnothing$.
\end{proof}

 \section{Implementation and Experiments}
\label{app:experiments}

\subsection{LTL to UBA}

In this section we give some more details on the experiments with the
\(1542\) benchmark formulas of the \textsc{LTLStore}
\cite{KMS18}. Those \(1542\) formulas include every formula in \cite{KMS18} and
its negation, as well as duplicates occurring over several files.
If one removes the duplicated formulas, the benchmark formula set
amounts to \(1419\) formulas (including negations). The odd number comes from
the fact, that \(p_0\) and \(\neg p_0\) are counted as duplicates, as they differ
only on the structure of the literal.
We use here the version of June 29, 2018, commit
\textsc{ad8b5cd7c9c30d1e65dbda676fdf41821c3a8adb}.

The \textsc{ltlstore} is grouped in sets of formulas, that contain either a parametrized family of formulas or a set of formulas used for a case study or tool evaluation.
The following figures show scatter plots for the individual formula sets, whose
names are given in the caption. We removed the plots where \duggi{} and
\ltltotgba{} showed similar behavior concerning automata sizes and time-outs to not enlarge the appendix unnecessarily. You find a description how to generate the missing plots at \url{https://wwwtcs.inf.tu-dresden.de/ALGI/TR/FM19-UBA/}.

\begin{figure}[h!]
  \begin{subfigure}[t]{0.47\textwidth}
    \centering
    \scalebox{.37}{\includegraphics{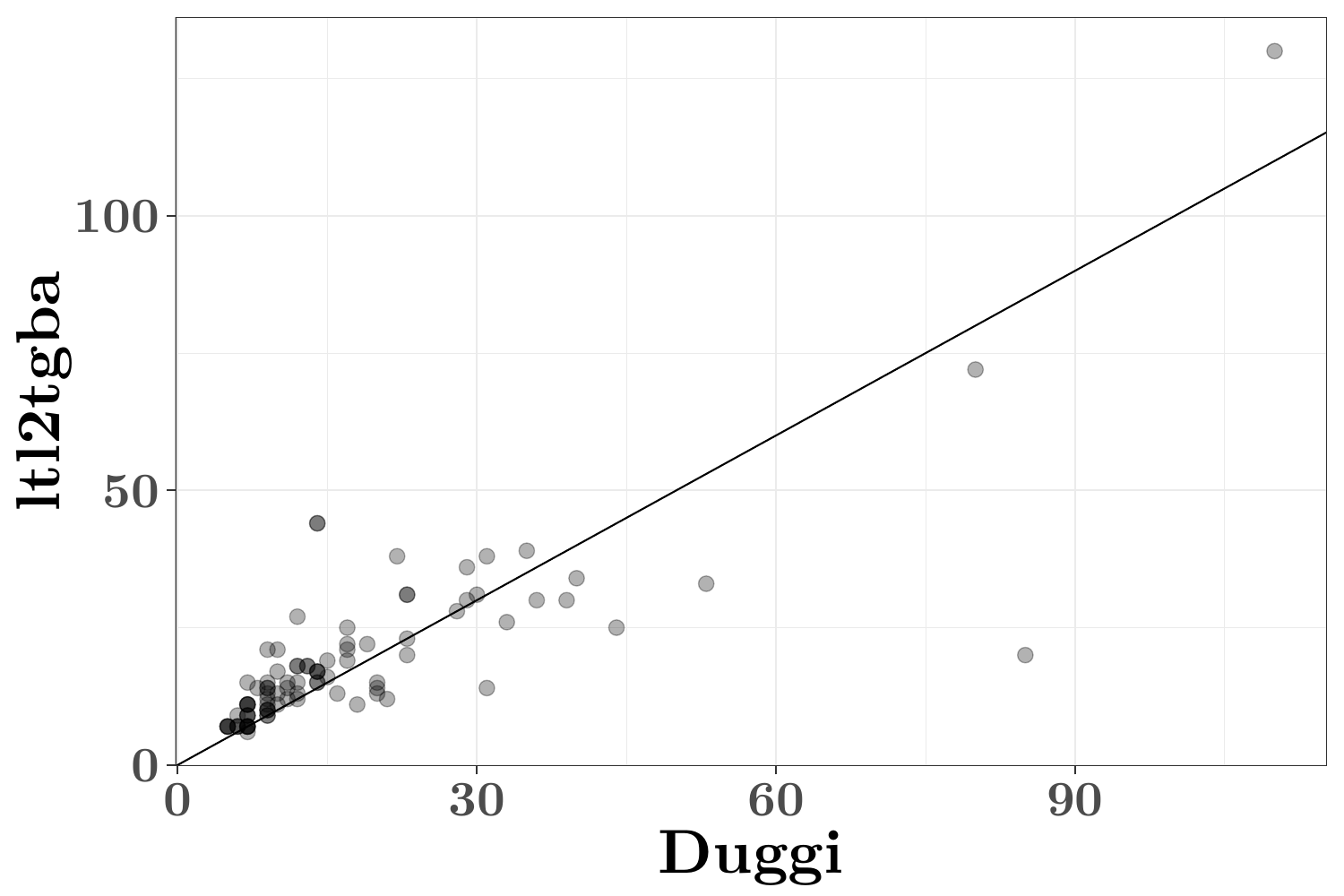}
    }
    \caption{acacia, timeouts: \duggi{}: 26, \ltltotgba{}: 32}
  \end{subfigure}
  \hfill
  \begin{subfigure}[t]{0.47\textwidth}
    \centering
    \scalebox{.37}{\includegraphics{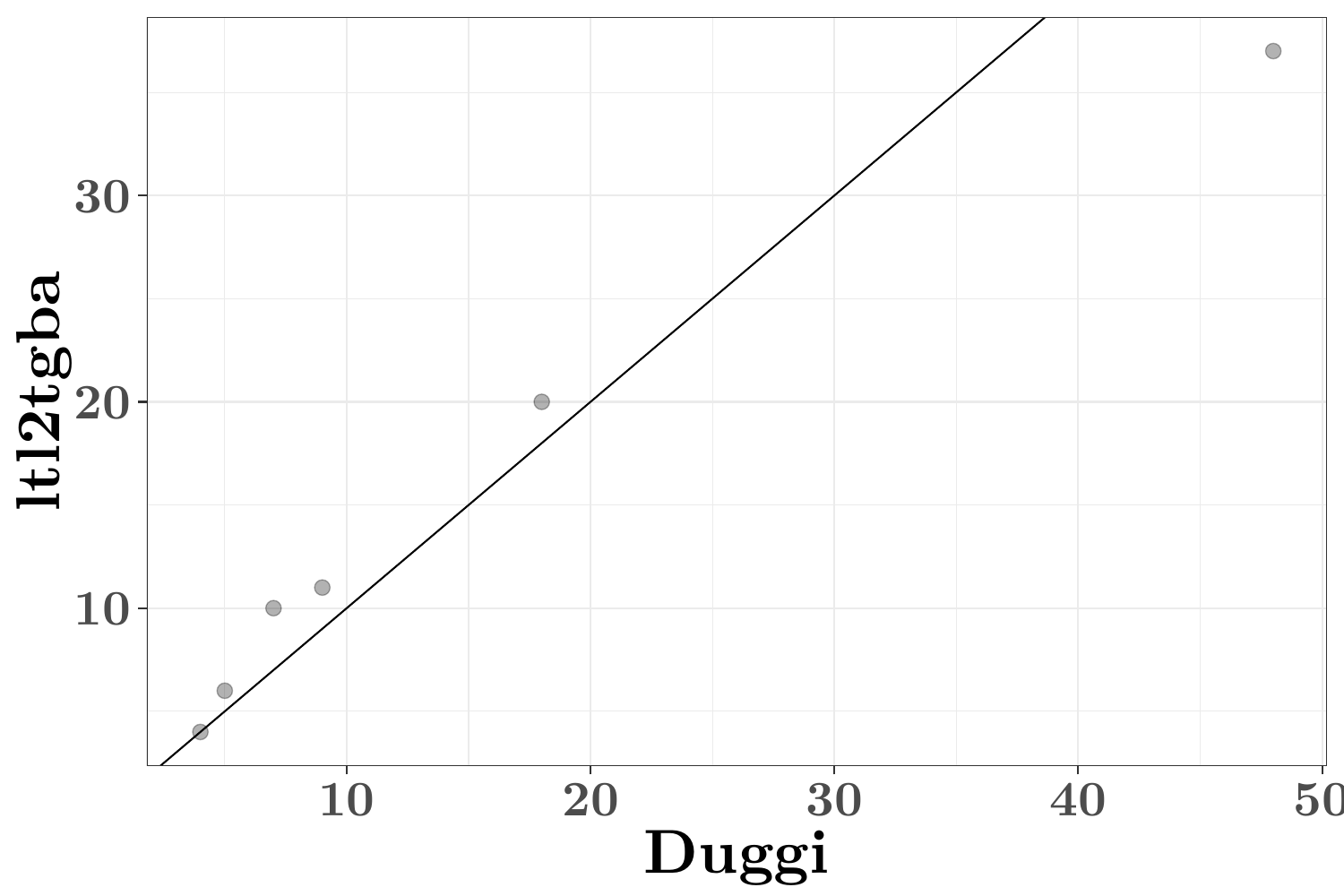}
    }
    \caption{chained, timeouts: \duggi{}: 2, \ltltotgba{}: 6}
  \end{subfigure}
\end{figure}

\begin{figure}[h!]
  \begin{subfigure}[t]{0.47\textwidth}
    \centering
    \scalebox{0.37}{\includegraphics{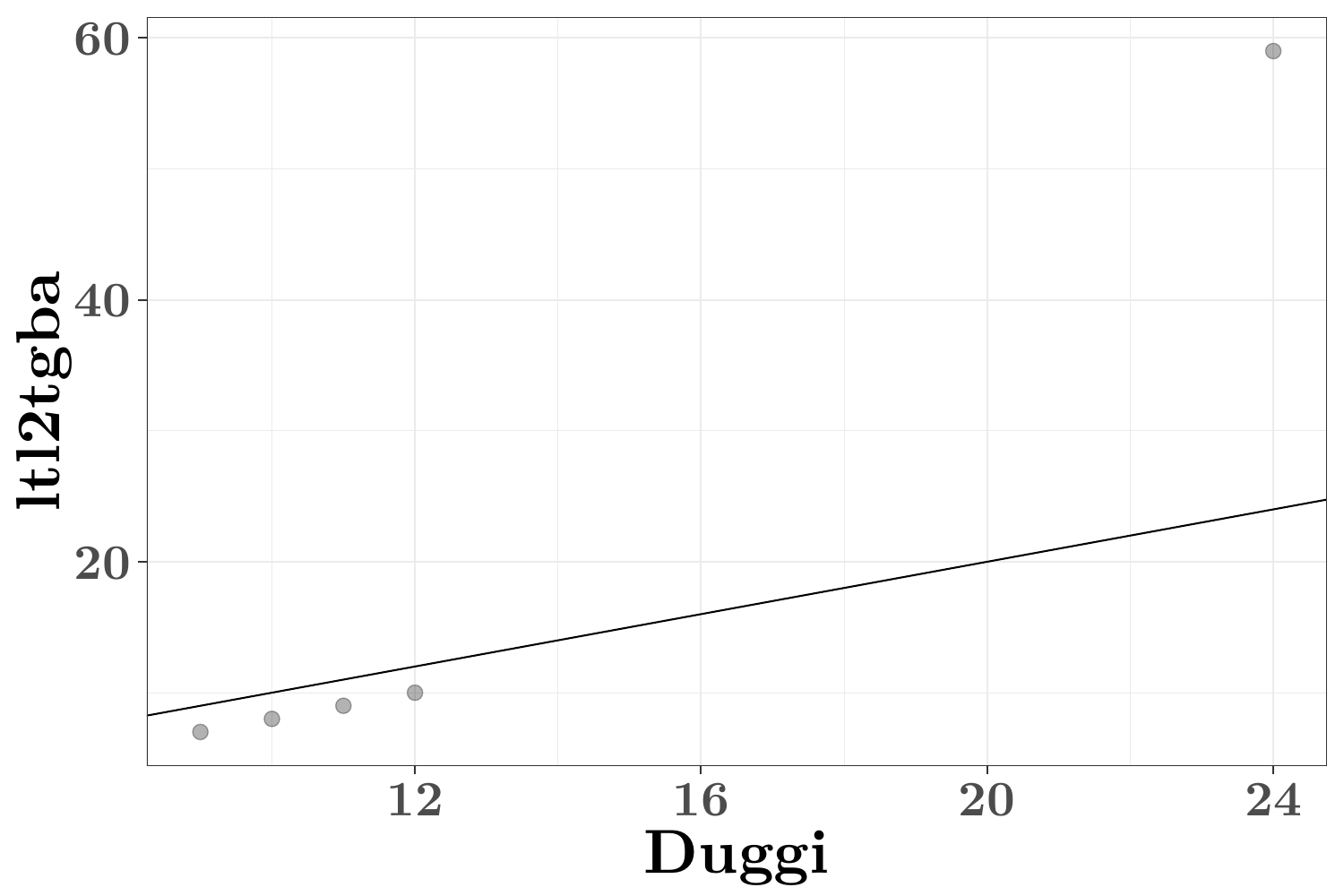}
    }
    \caption{cluster, timeouts: \duggi{}: 1, \ltltotgba{}: 3}
  \end{subfigure}
  \hfill
  \begin{subfigure}[t]{0.47\textwidth}
    \centering
    \scalebox{.37}{\includegraphics{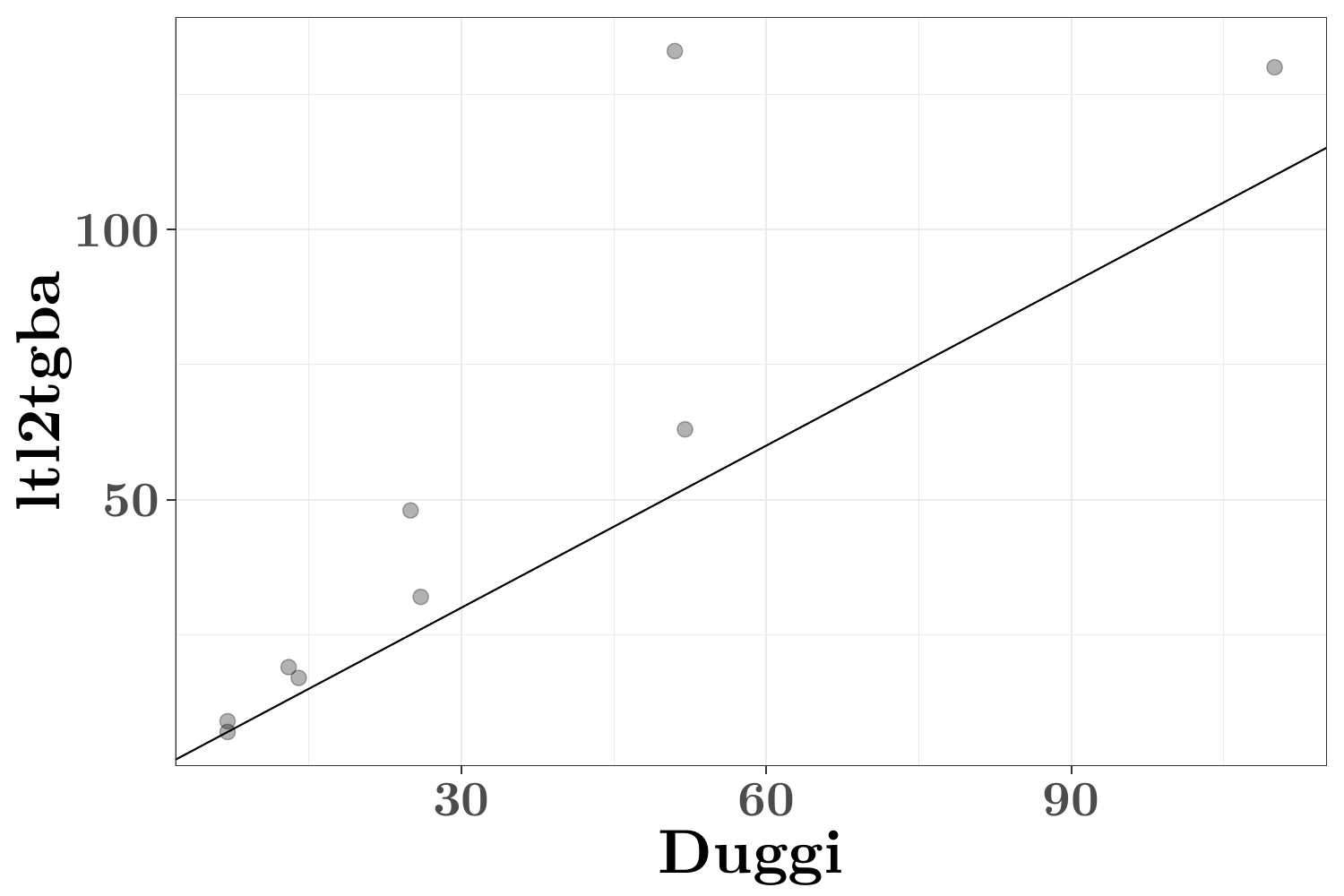}
    }
    \caption{detector\_1-5,10,20, timeouts: \duggi{}: 4, \ltltotgba{}: 5}
  \end{subfigure}
\end{figure}

\begin{figure}[h!]
  \begin{subfigure}[t]{0.47\textwidth}
    \centering
    \scalebox{.37}{\includegraphics{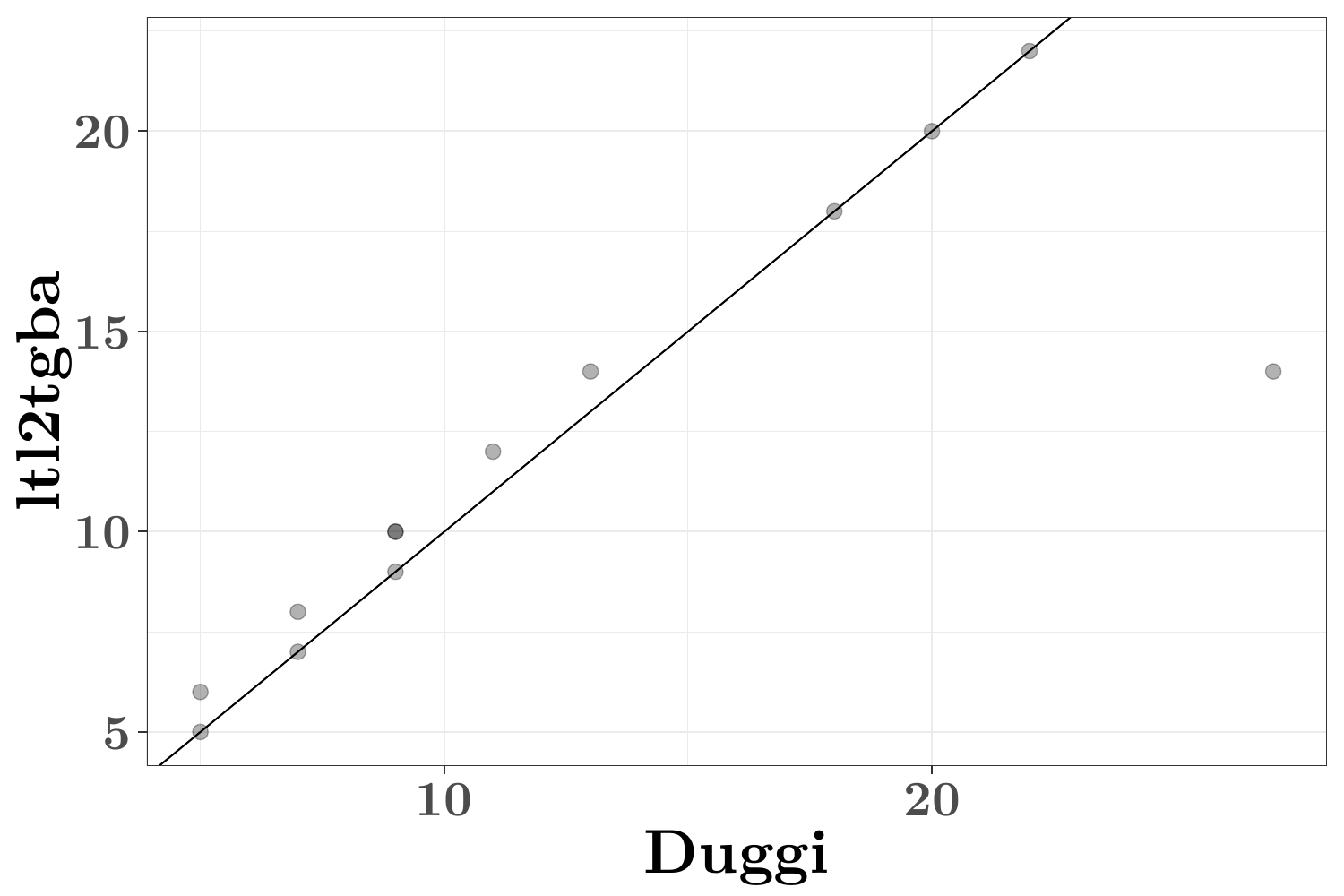}
    }
    \caption{family\_f, timeouts: \duggi{}: 5, \ltltotgba{}: 3}
  \end{subfigure}
  \hfill
  \begin{subfigure}[t]{0.47\textwidth}
    \centering
    \scalebox{0.37}{\includegraphics{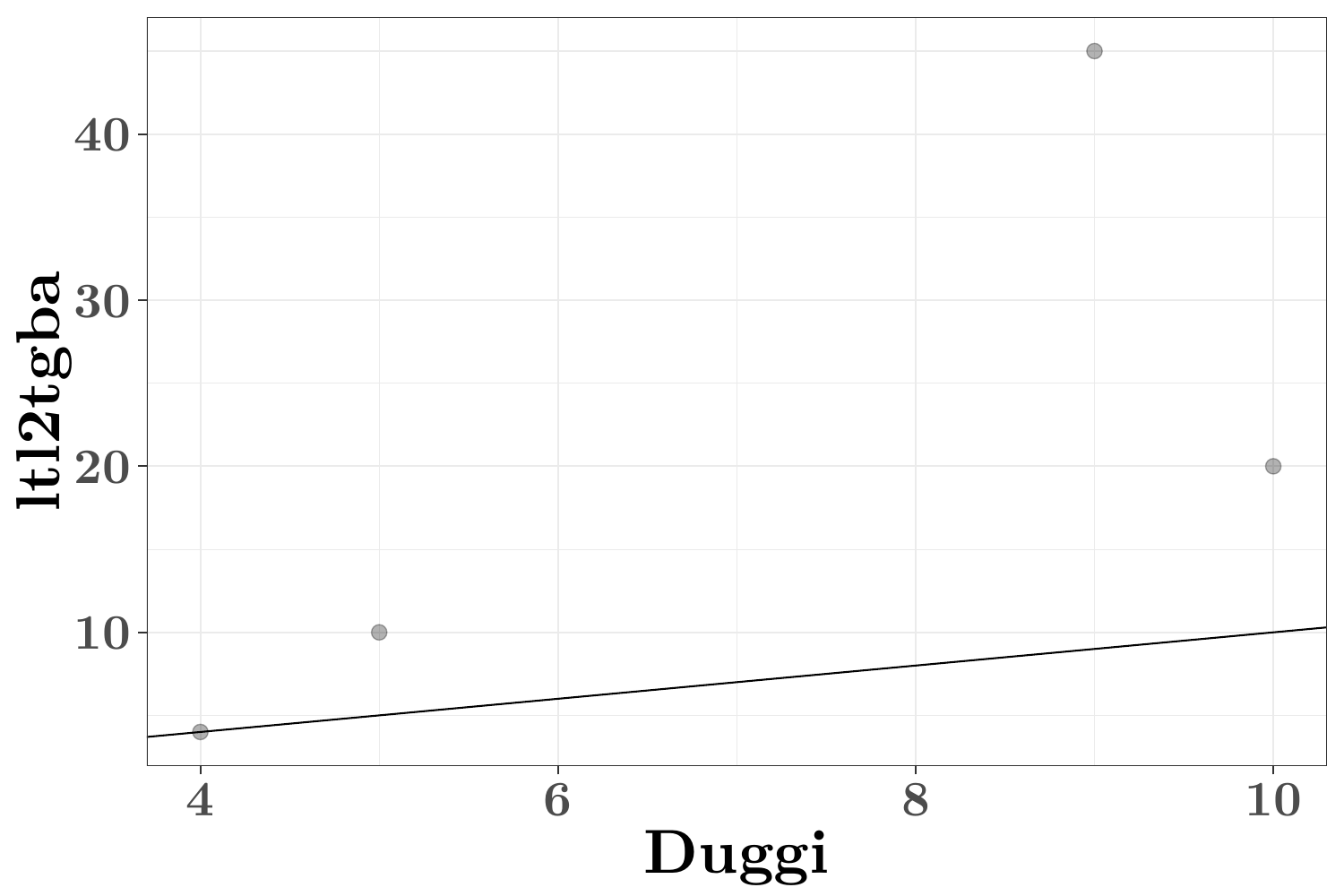}
    }
    \caption{fggf, timeouts: \duggi{}: 1, \ltltotgba{}: 6}
  \end{subfigure}
\end{figure}

\begin{figure}[h!]
  \begin{subfigure}[t]{0.47\textwidth}
    \centering
    \scalebox{.37}{\includegraphics{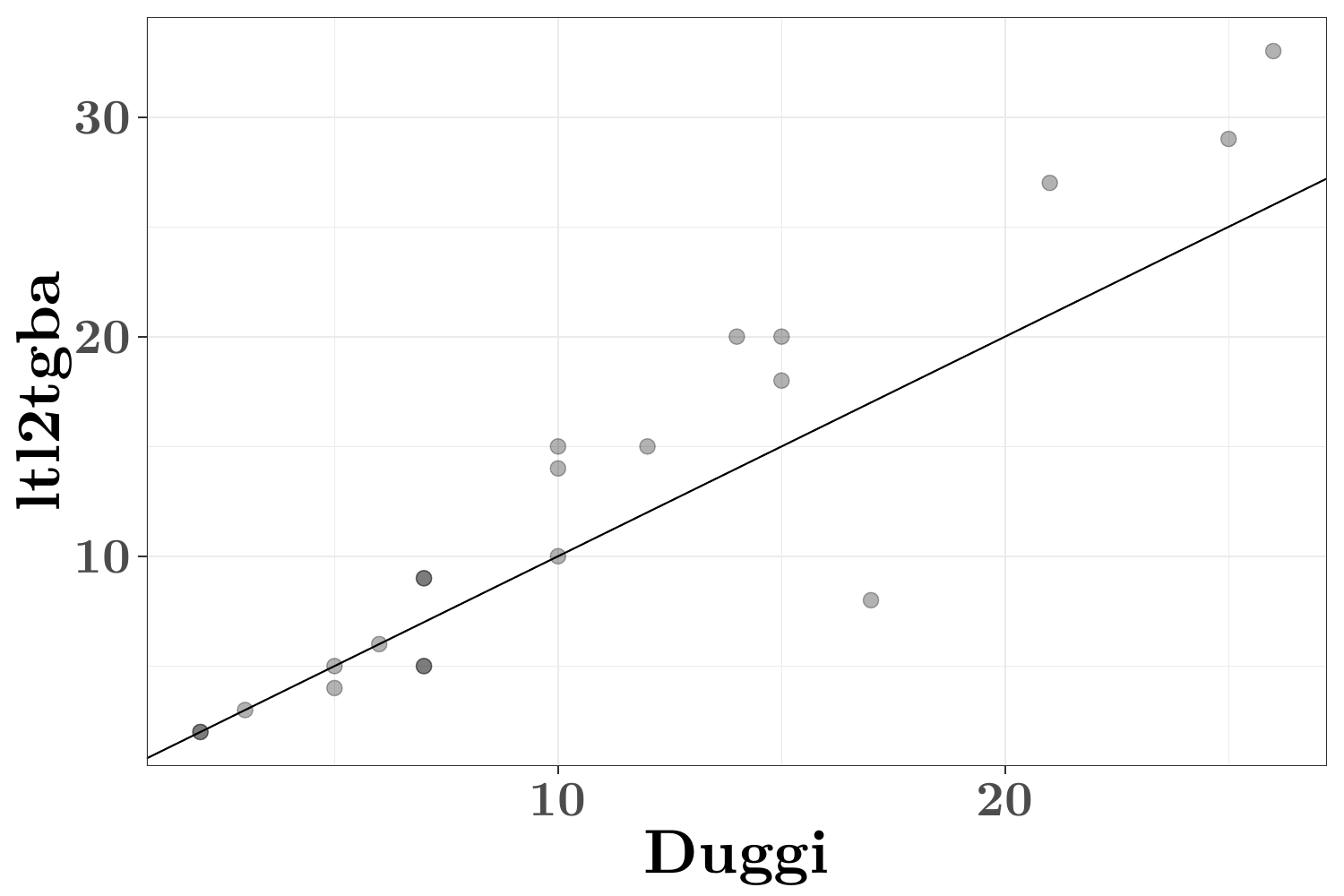}
    }
    \caption{further, timeouts: \duggi{}: 4, \ltltotgba{}: 7}
  \end{subfigure}
  \hfill
  \begin{subfigure}[t]{0.47\textwidth}
    \centering
    \scalebox{0.37}{\includegraphics{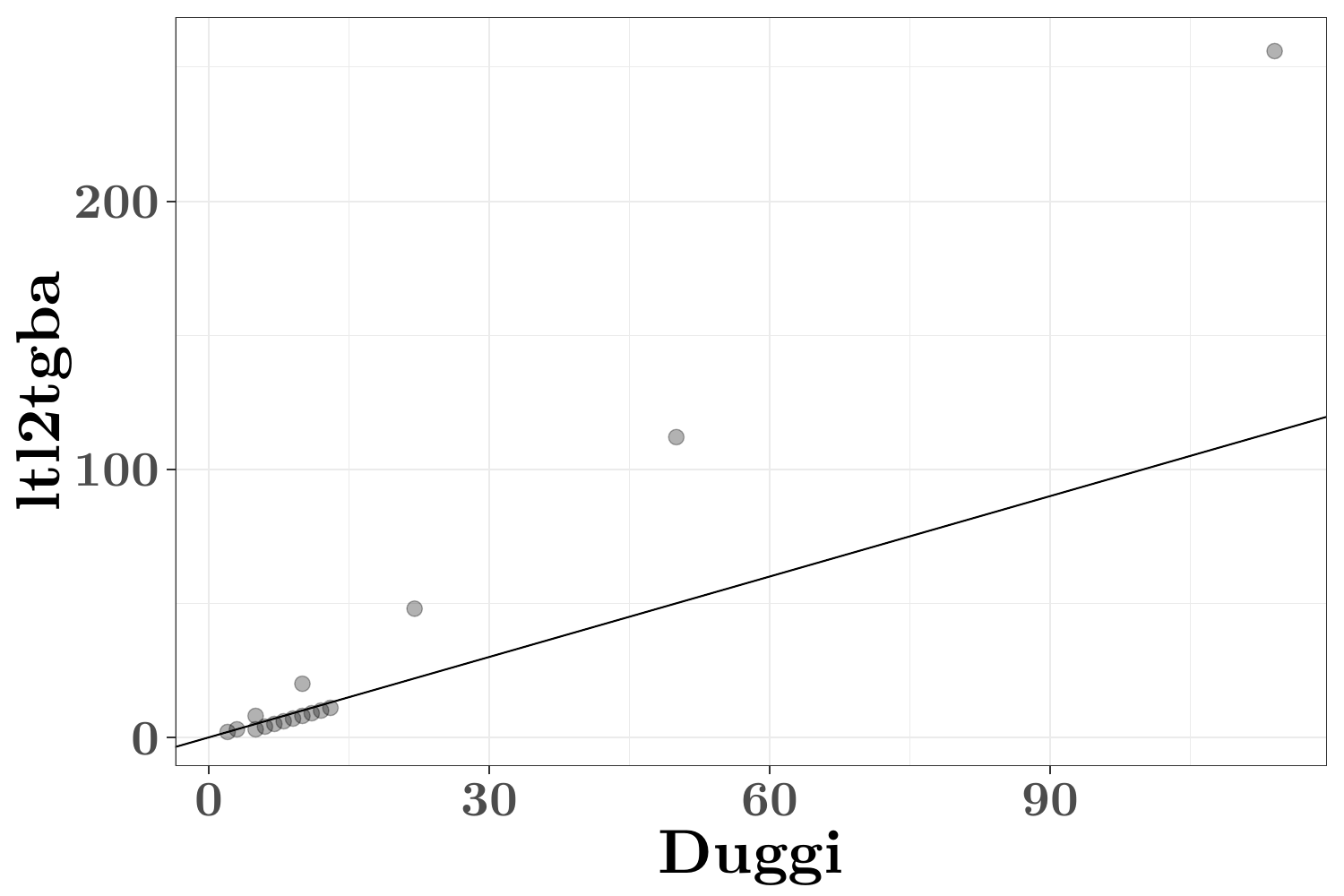}
    }
    \caption{gf\_and, timeouts: \duggi{}: 4, \ltltotgba{}: 4}
  \end{subfigure}
\end{figure}

\begin{figure}[h!]
  \begin{subfigure}[t]{0.47\textwidth}
    \centering
    \scalebox{0.37}{\includegraphics{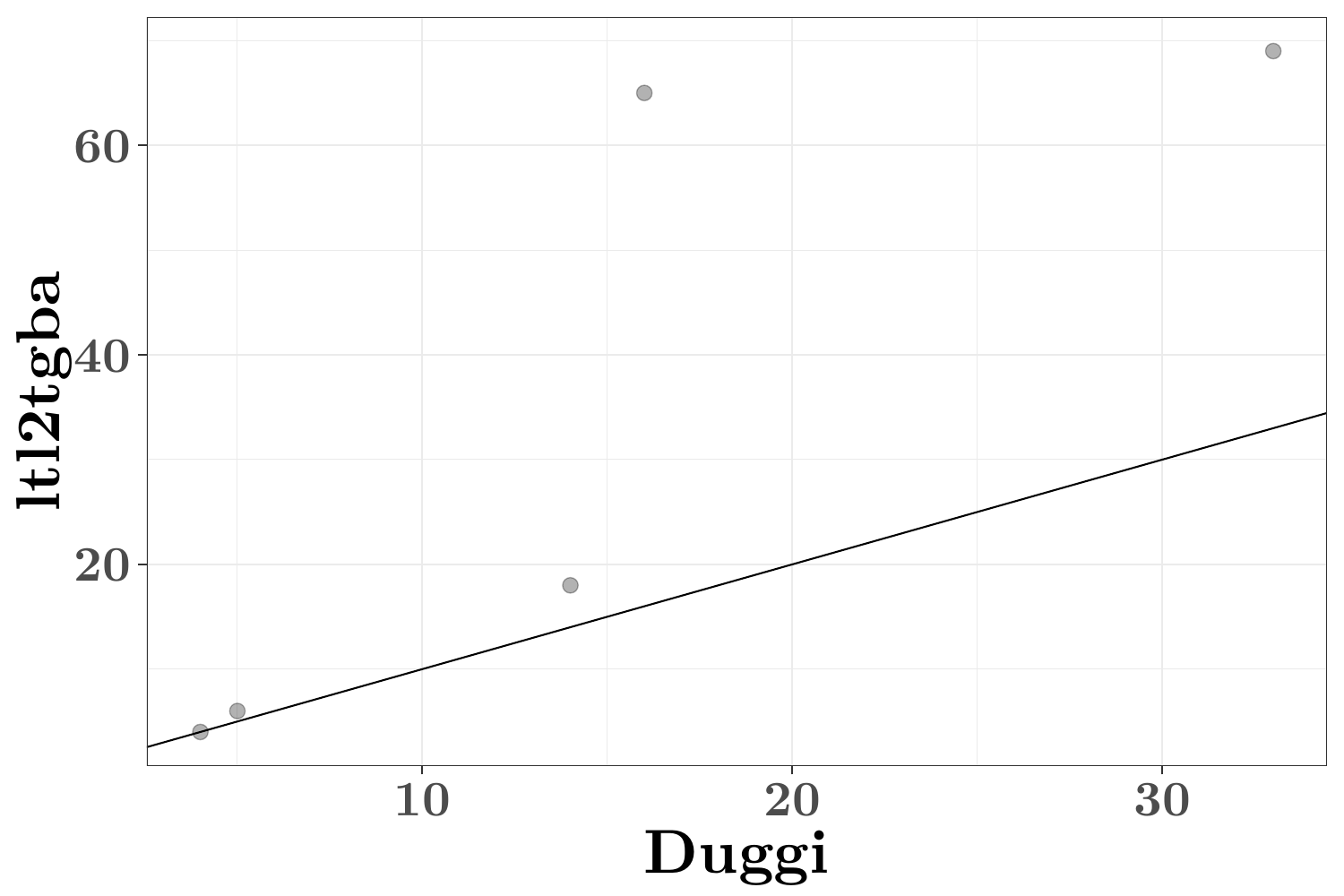}
    }
    \caption{gr1, timeouts: \duggi{}: 7, \ltltotgba{}: 9}
  \end{subfigure}
  \hfill
  \begin{subfigure}[t]{0.47\textwidth}
    \centering
    \scalebox{.37}{\includegraphics{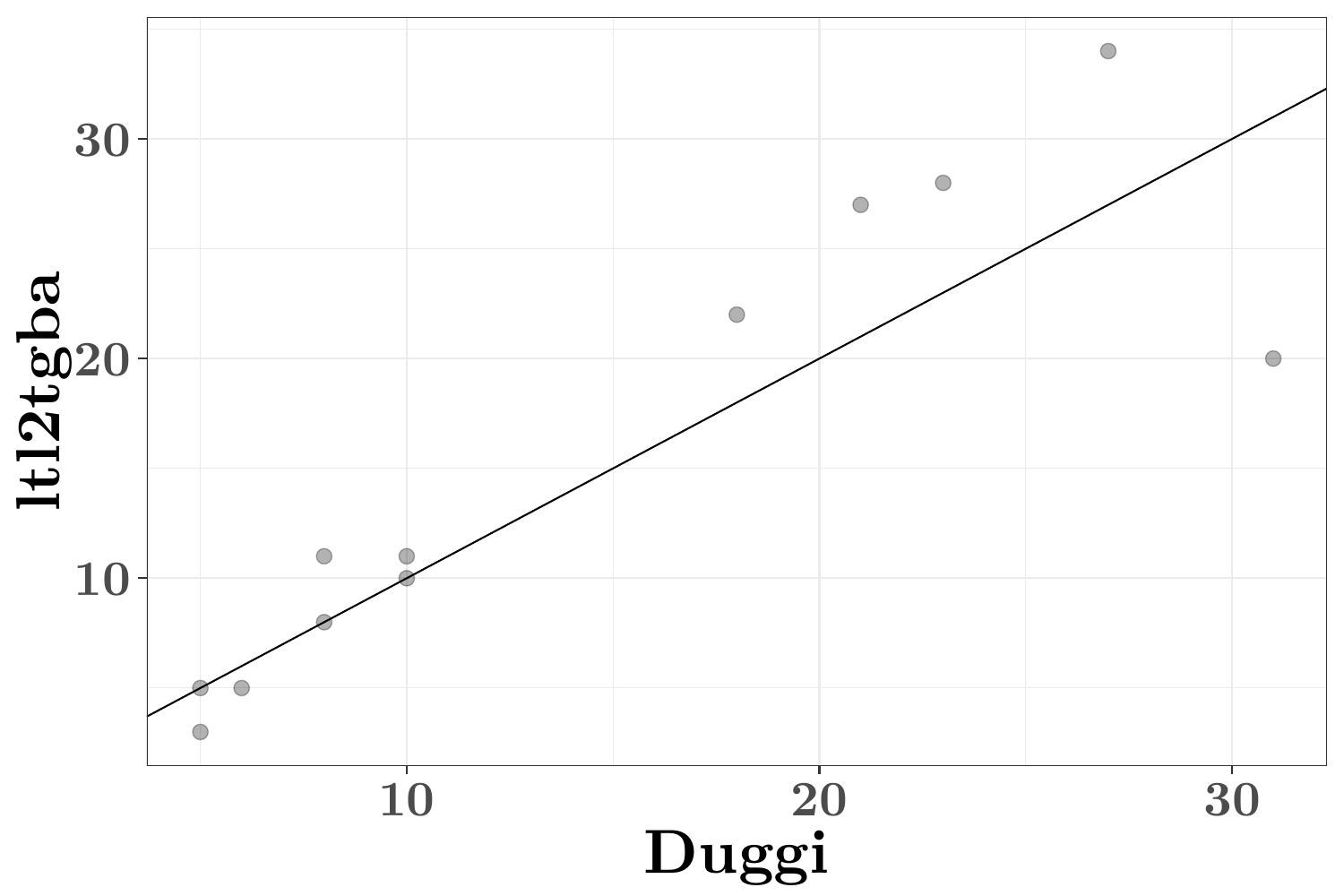}
    }
    \caption{libe\_router, timeouts: \duggi{}: 2, \ltltotgba{}: 0}
  \end{subfigure}
\end{figure}

\begin{figure}[h!]
  \begin{subfigure}[t]{0.47\textwidth}
    \centering
    \scalebox{.37}{\includegraphics{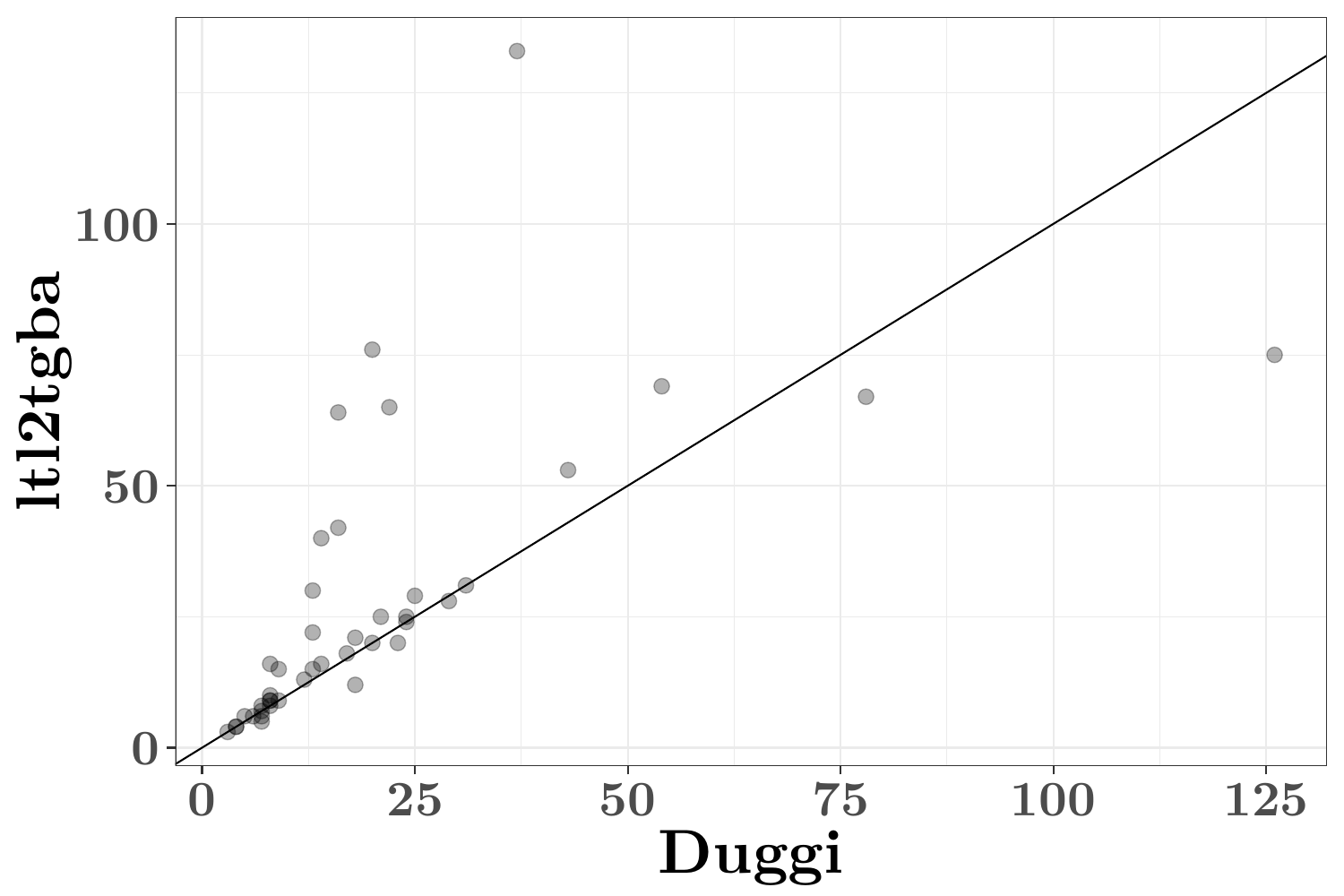}
    }
    \caption{lily2, timeouts: \duggi{}: 2, \ltltotgba{}: 4}
  \end{subfigure}
  \hfill
  \begin{subfigure}[t]{0.47\textwidth}
    \centering
    \scalebox{0.37}{\includegraphics{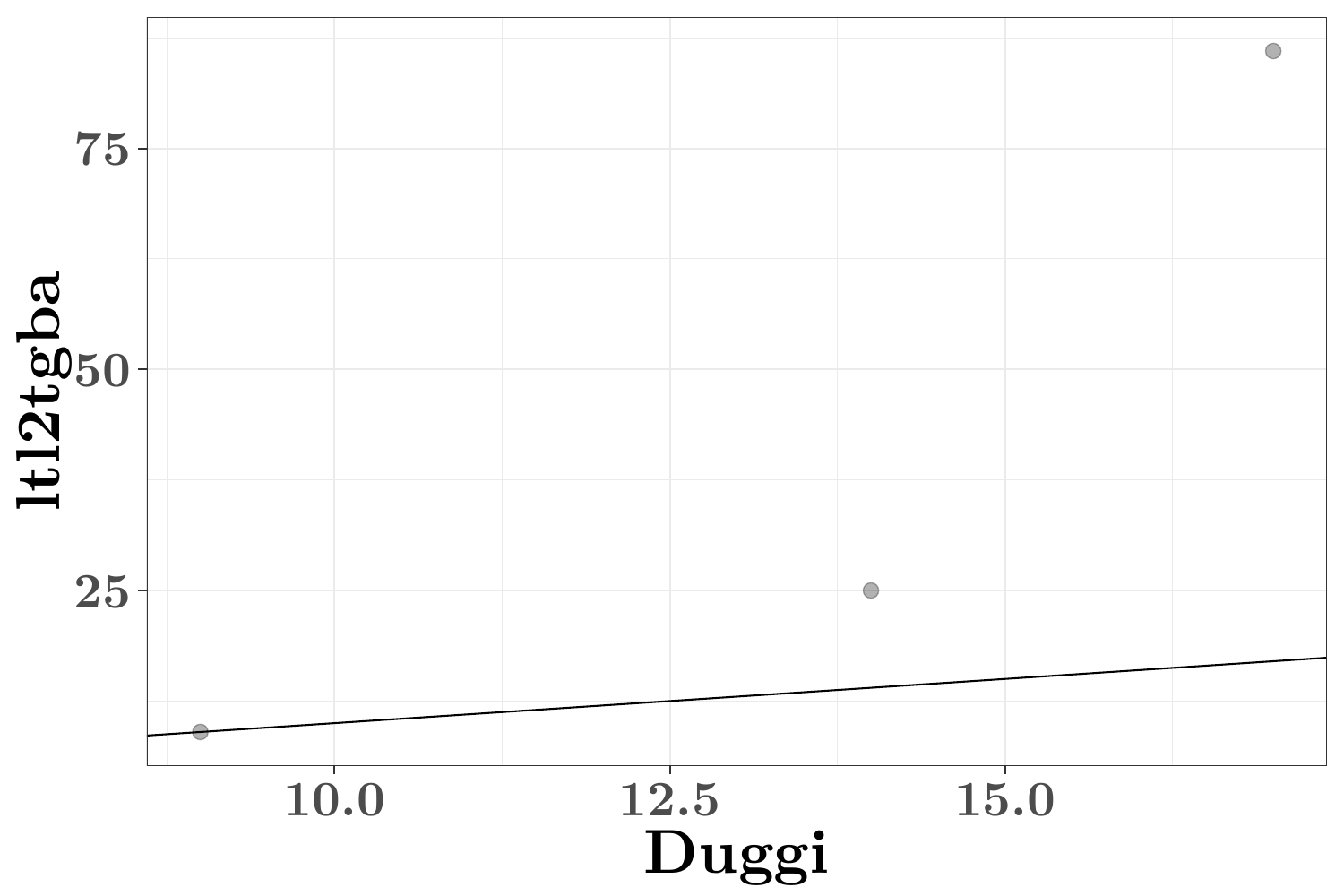}
    }
    \caption{load\_balancer\_1-5, timeouts: \duggi{}: 6, \ltltotgba{}: 5}
  \end{subfigure}
\end{figure}

\begin{figure}[h!]
  \begin{subfigure}[t]{0.47\textwidth}
    \centering
    \scalebox{.37}{\includegraphics{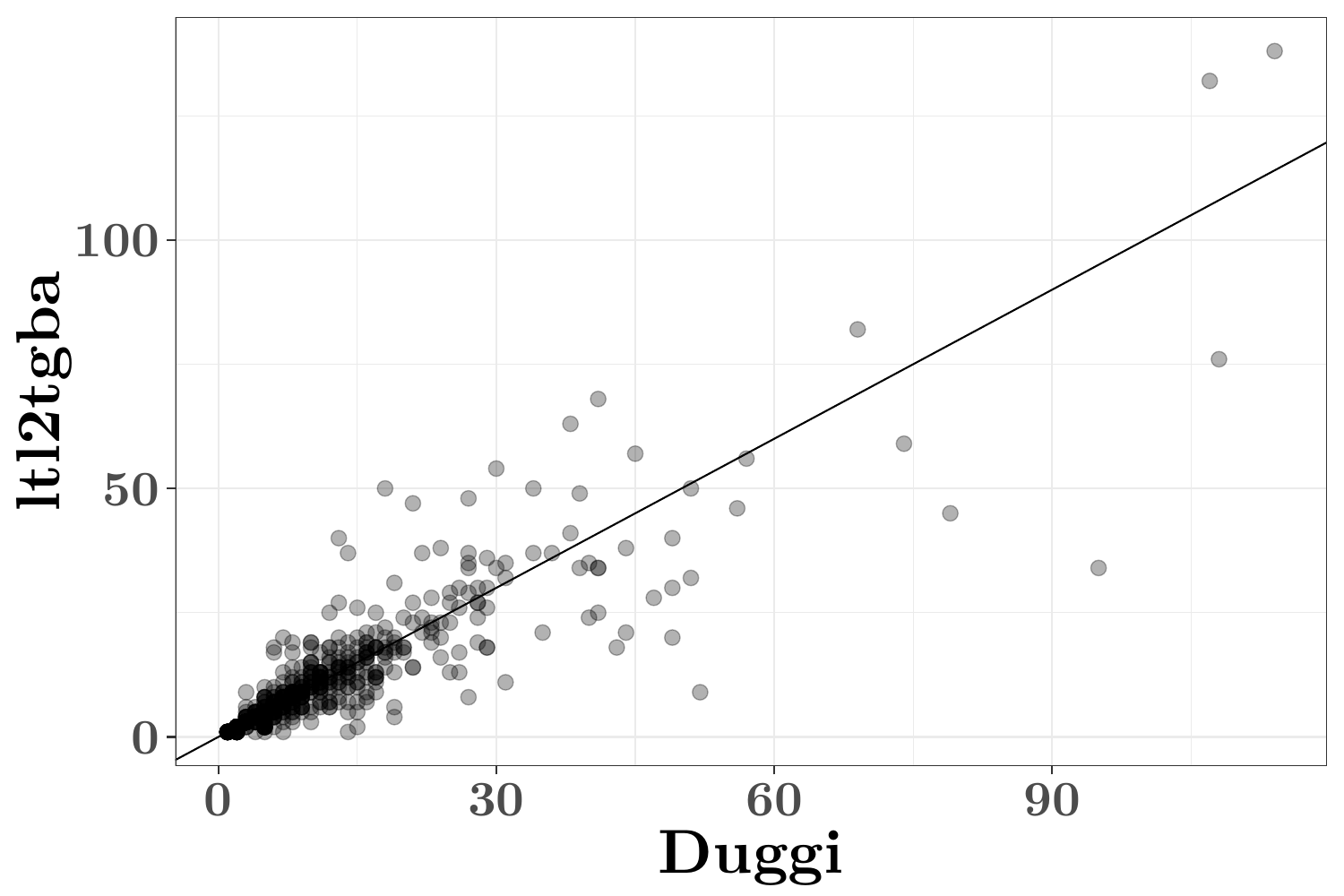}
    }
    \caption{lpar19\_all, timeouts: \duggi{}: 54, \ltltotgba{}: 2}
  \end{subfigure}
  \hfill
  \begin{subfigure}[t]{0.47\textwidth}
    \centering
    \scalebox{0.37}{\includegraphics{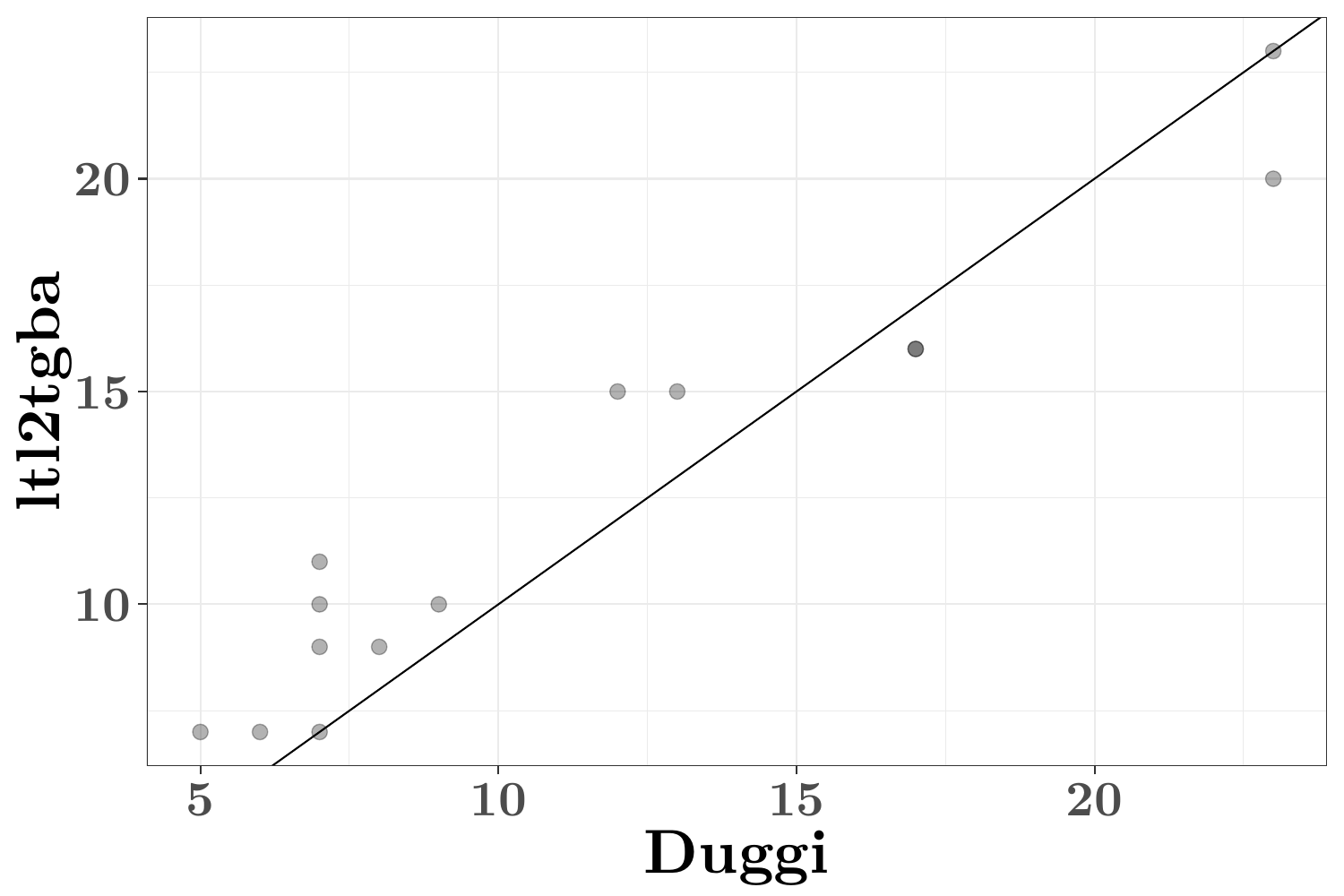}
    }
    \caption{LtlNfBa, timeouts: \duggi{}: 0, \ltltotgba{}: 0}
  \end{subfigure}
\end{figure}

\begin{figure}[h!]
  \begin{subfigure}[t]{0.47\textwidth}
    \centering
    \scalebox{0.37}{\includegraphics{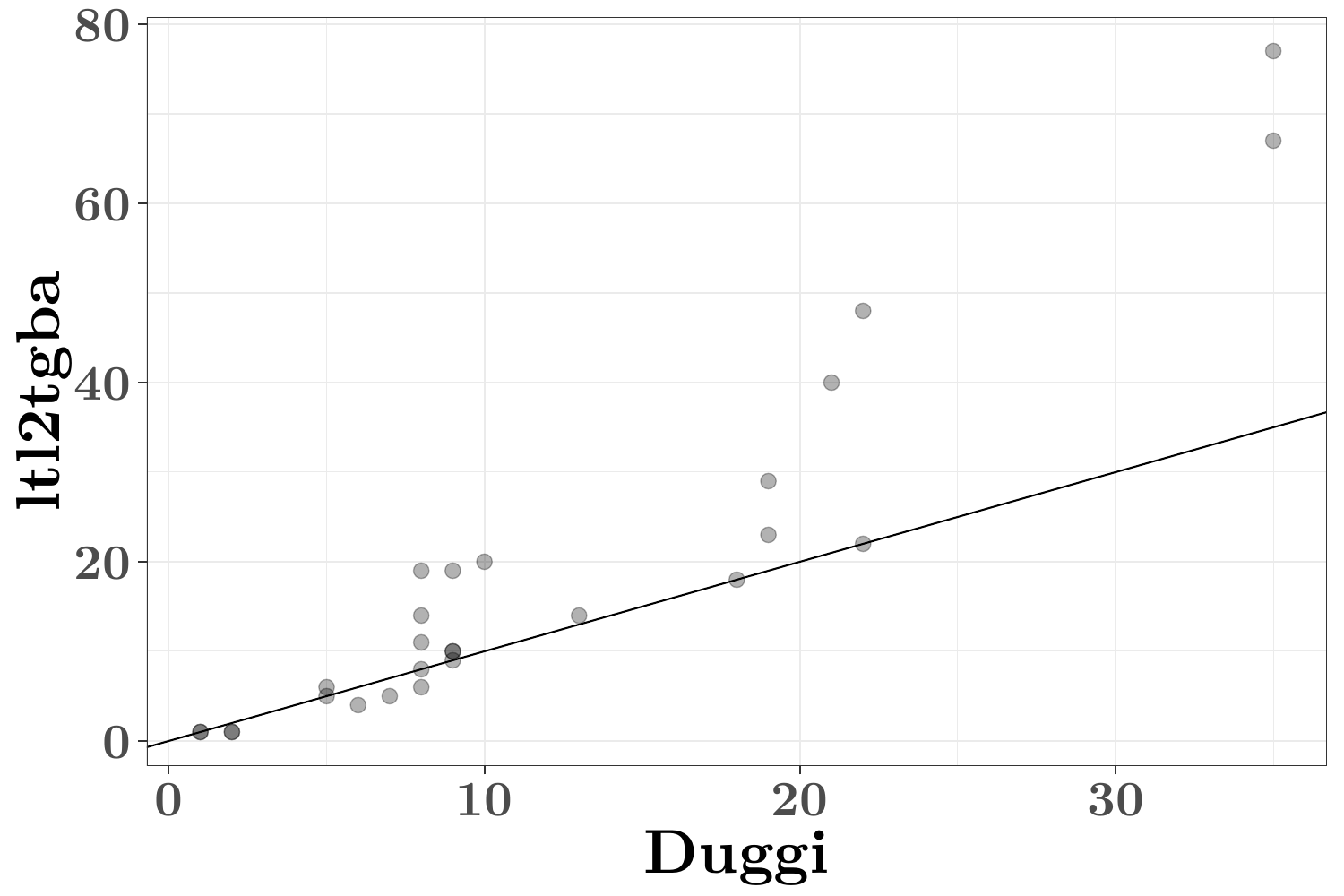}
    }
    \caption{mutual, timeouts: \duggi{}: 0, \ltltotgba{}: 1}
  \end{subfigure}
  \hfill
  \begin{subfigure}[t]{0.47\textwidth}
    \centering
    \scalebox{.37}{\includegraphics{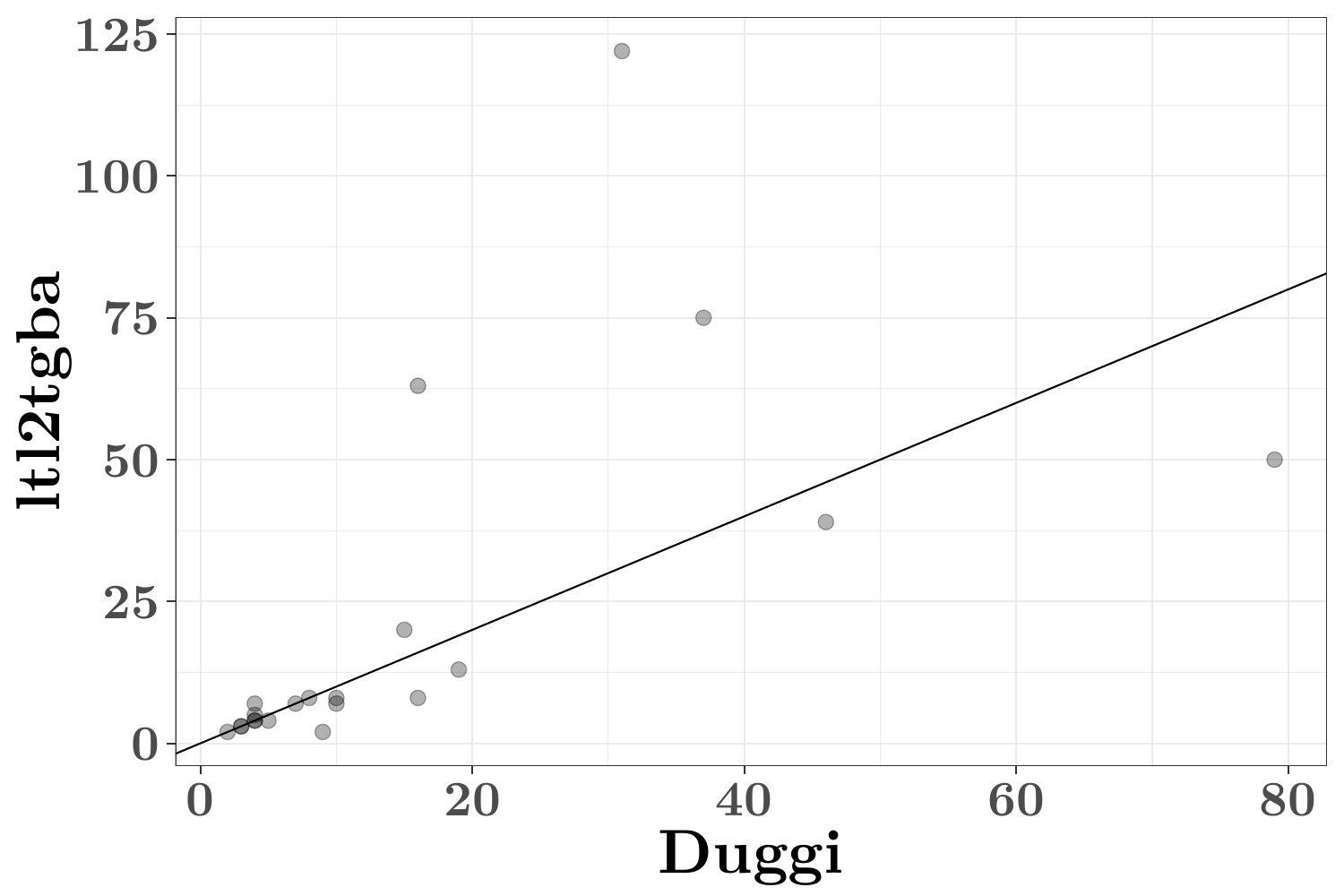}
    }
    \caption{rabinizer3, timeouts: \duggi{}: 0, \ltltotgba{}: 3}
  \end{subfigure}
\end{figure}

\begin{figure}[h!]
  \begin{subfigure}[t]{0.47\textwidth}
    \centering
    \scalebox{.37}{\includegraphics{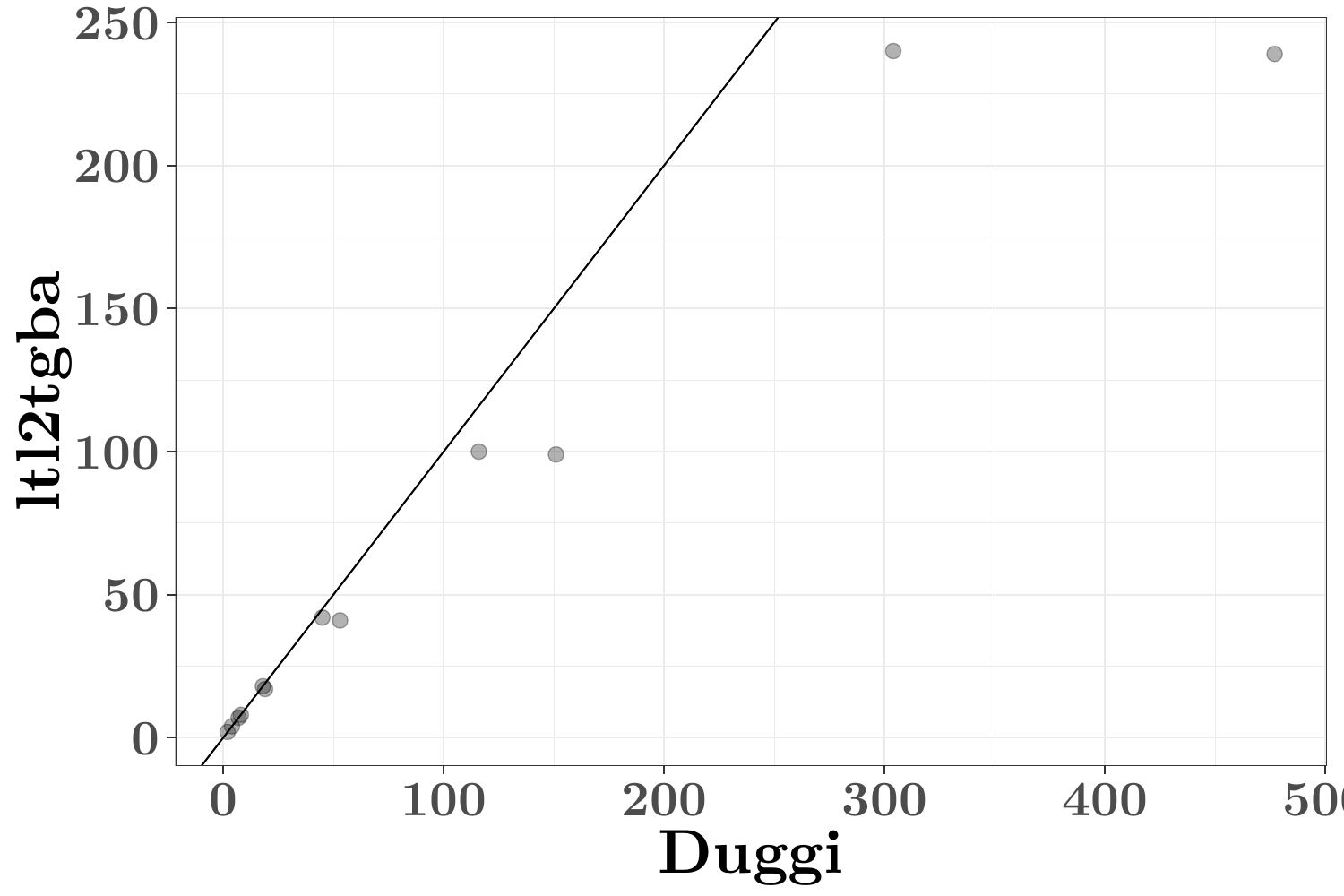}
    }
    \caption{Q, timeouts: \duggi{}: 8, \ltltotgba{}: 6}
  \end{subfigure}
  \hfill
  \begin{subfigure}[t]{0.47\textwidth}
    \centering
    \scalebox{0.37}{\includegraphics{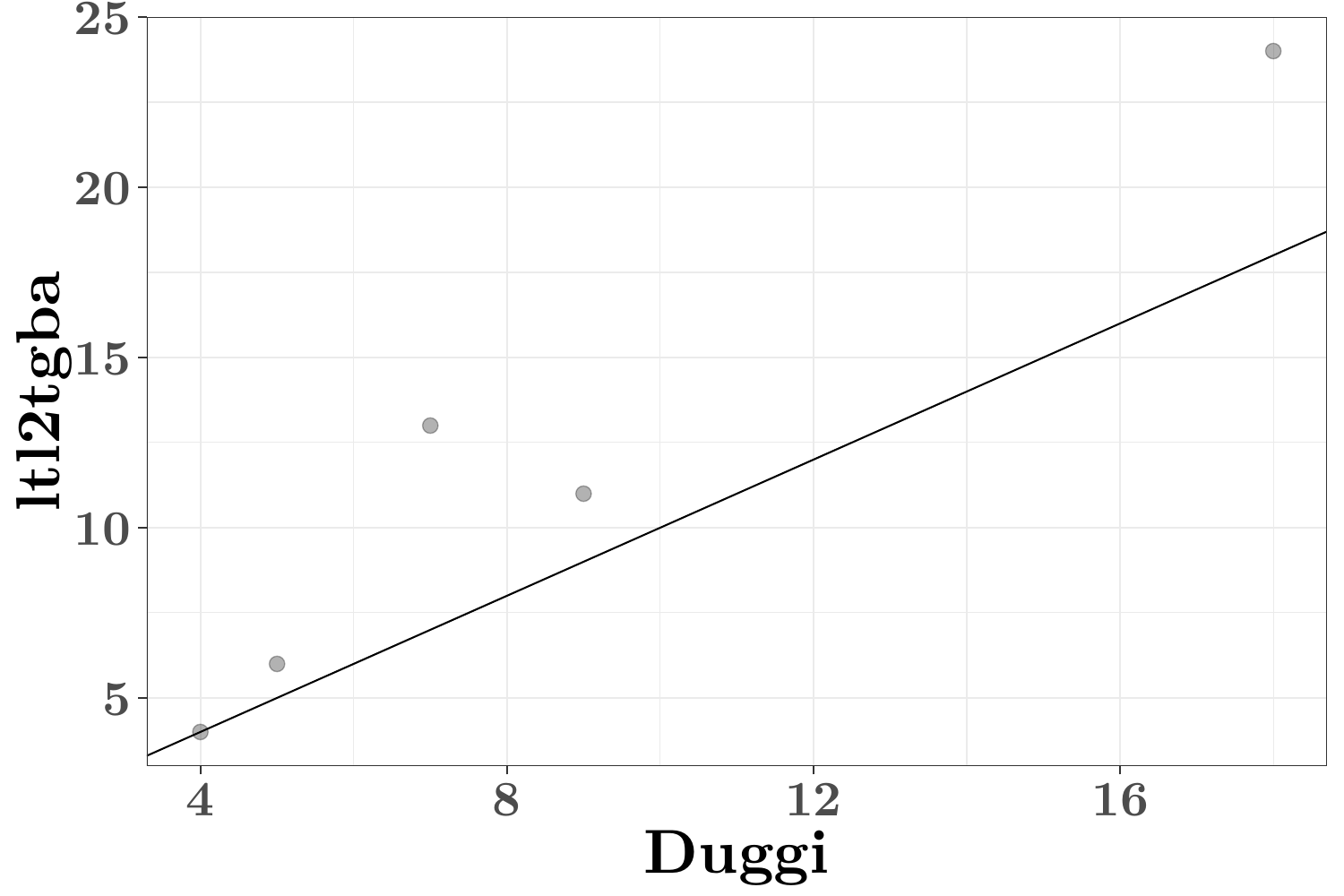}
    }
    \caption{R, timeouts: \duggi{}: 10, \ltltotgba{}: 15}
  \end{subfigure}
\end{figure}

\begin{figure}[h!]
  \begin{subfigure}[t]{0.47\textwidth}
    \centering
    \scalebox{.37}{\includegraphics{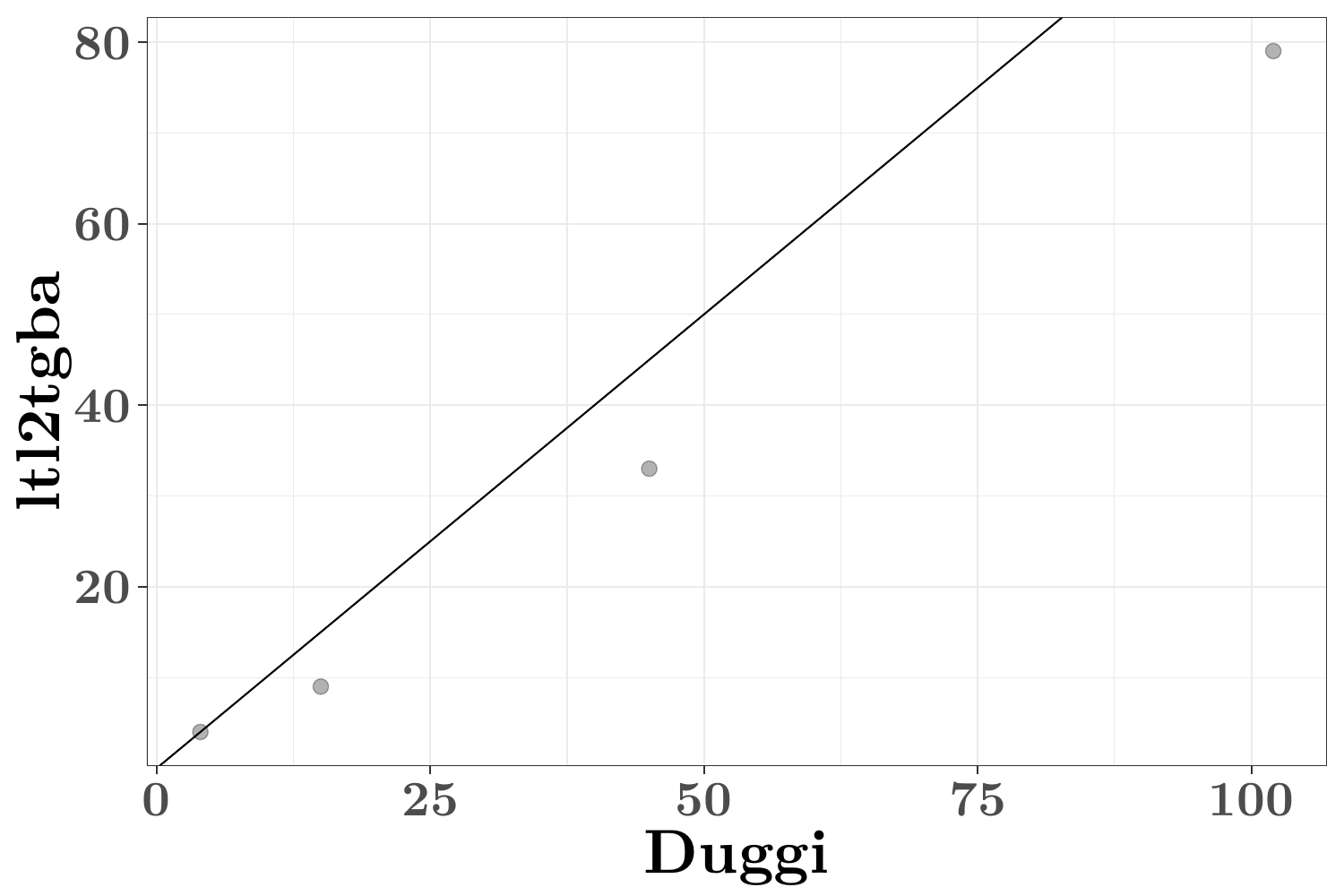}
    }
    \caption{round\_robin\_arbiter\_1-5, timeouts: \duggi{}: 6, \ltltotgba{}: 4}
  \end{subfigure}
  \hfill
  \begin{subfigure}[t]{0.47\textwidth}
    \centering
    \scalebox{0.37}{\includegraphics{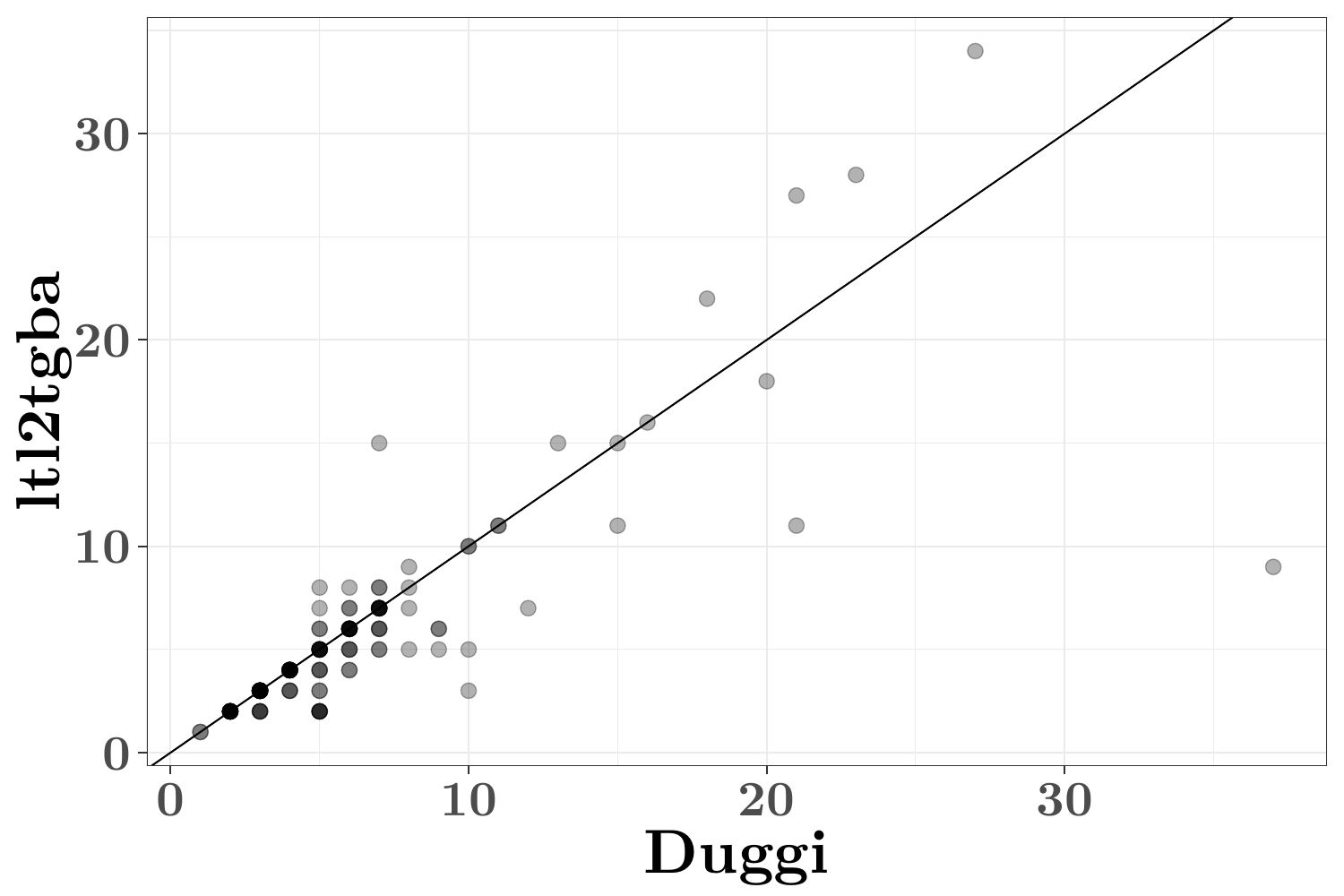}
    }
    \caption{real, timeouts: \duggi{}: 0, \ltltotgba{}: 0}
  \end{subfigure}
\end{figure}

\begin{figure}[h!]
  \begin{subfigure}[t]{0.47\textwidth}
    \centering
    \scalebox{0.37}{\includegraphics{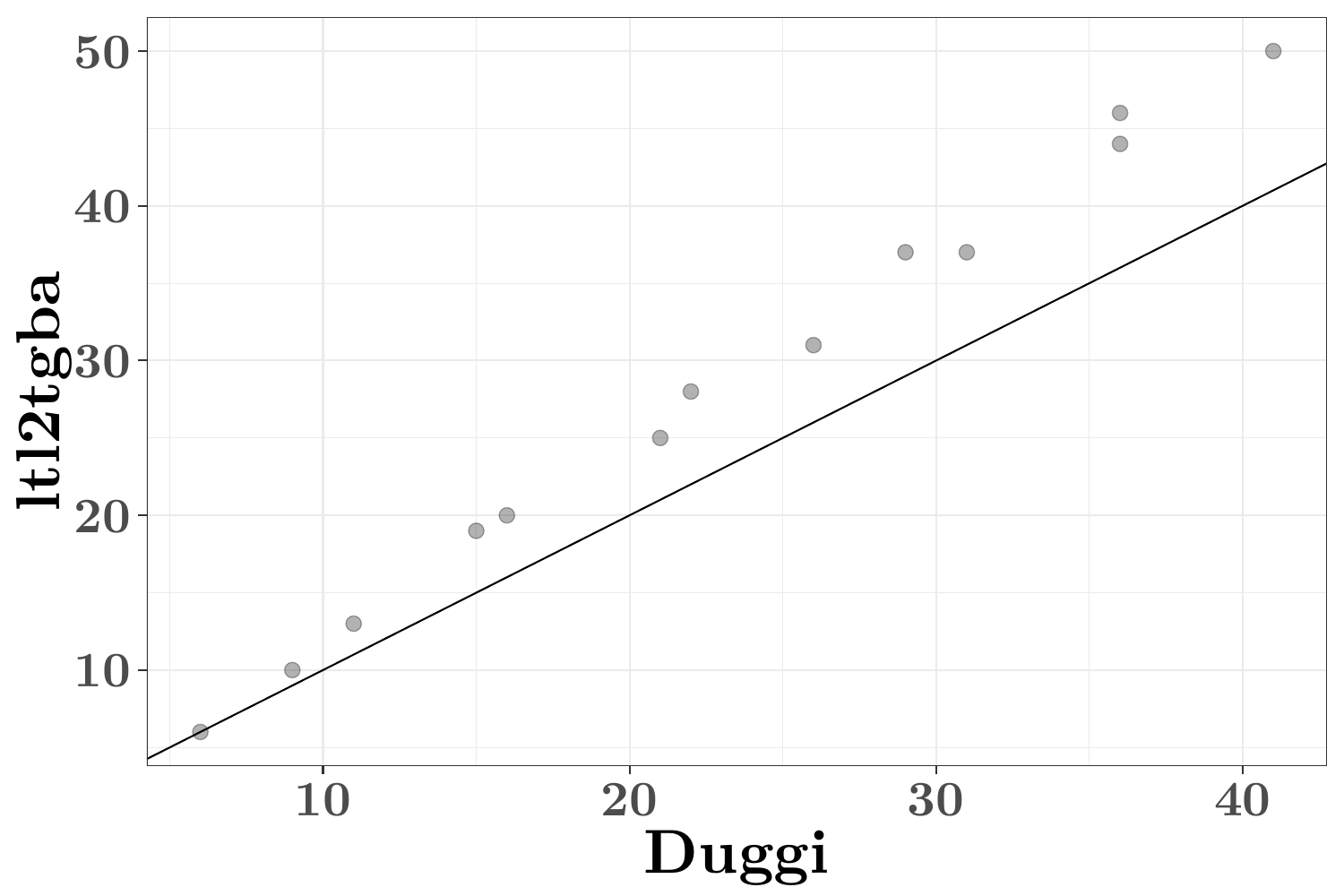}
    }
    \caption{scc, timeouts: \duggi{}: 2, \ltltotgba{}: 3}
  \end{subfigure}
  \hfill
  \begin{subfigure}[t]{0.47\textwidth}
    \centering
    \scalebox{0.37}{\includegraphics{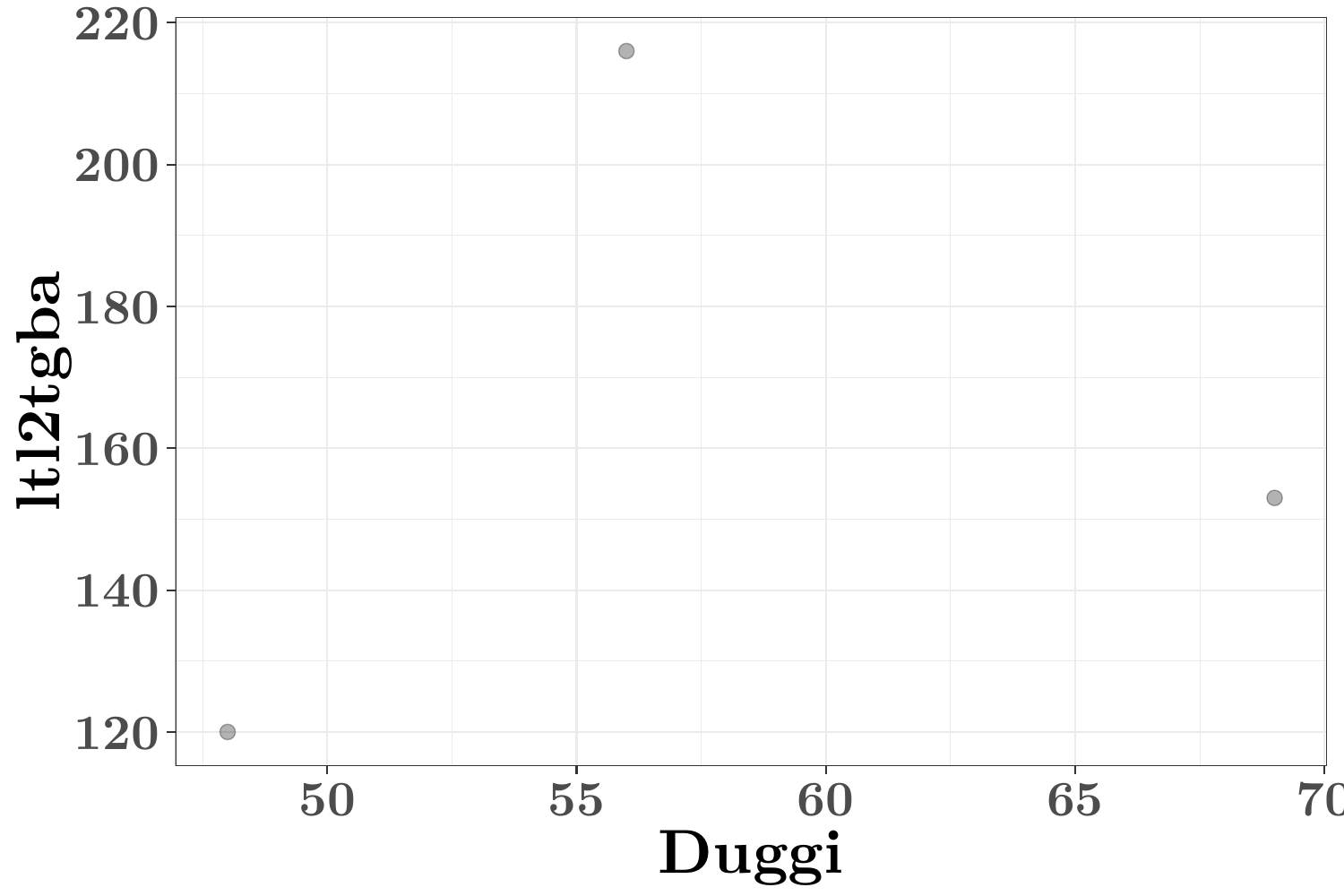}
    }
    \caption{suspension, timeouts: \duggi{}: 5, \ltltotgba{}: 4}
  \end{subfigure}
\end{figure}

\begin{figure}[h!]
  \begin{subfigure}[t]{0.47\textwidth}
    \centering
    \scalebox{.37}{\includegraphics{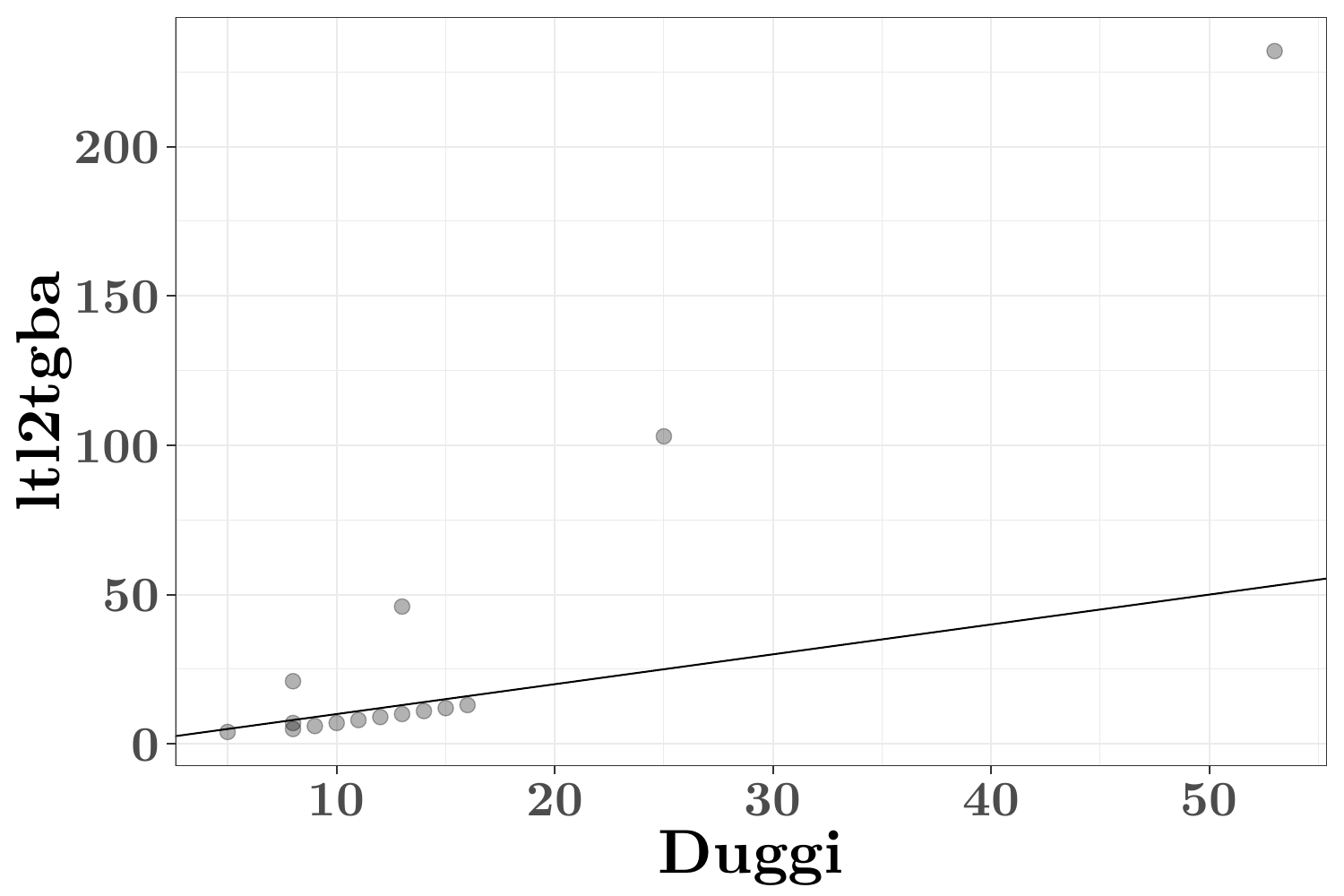}
    }
    \caption{theta, timeouts: \duggi{}: 5, \ltltotgba{}: 5}
  \end{subfigure}
  \hfill
  \begin{subfigure}[t]{0.47\textwidth}
    \centering
    \scalebox{.37}{\includegraphics{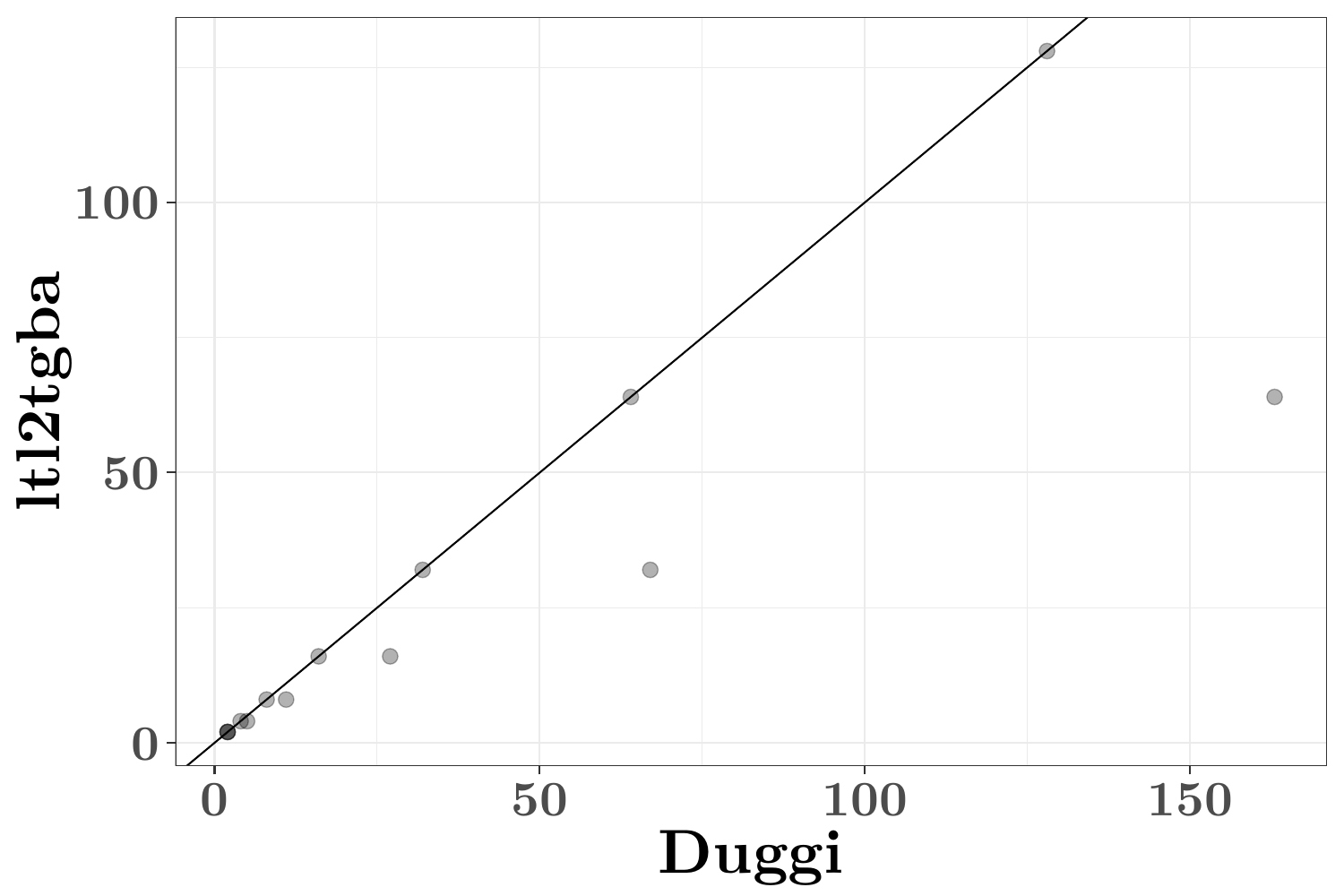}
    }
    \caption{U\_left, timeouts: \duggi{}: 5, \ltltotgba{}: 2}
  \end{subfigure}
\end{figure}

\begin{figure}[h!]
  \begin{subfigure}[t]{0.47\textwidth}
    \centering
    \scalebox{0.37}{\includegraphics{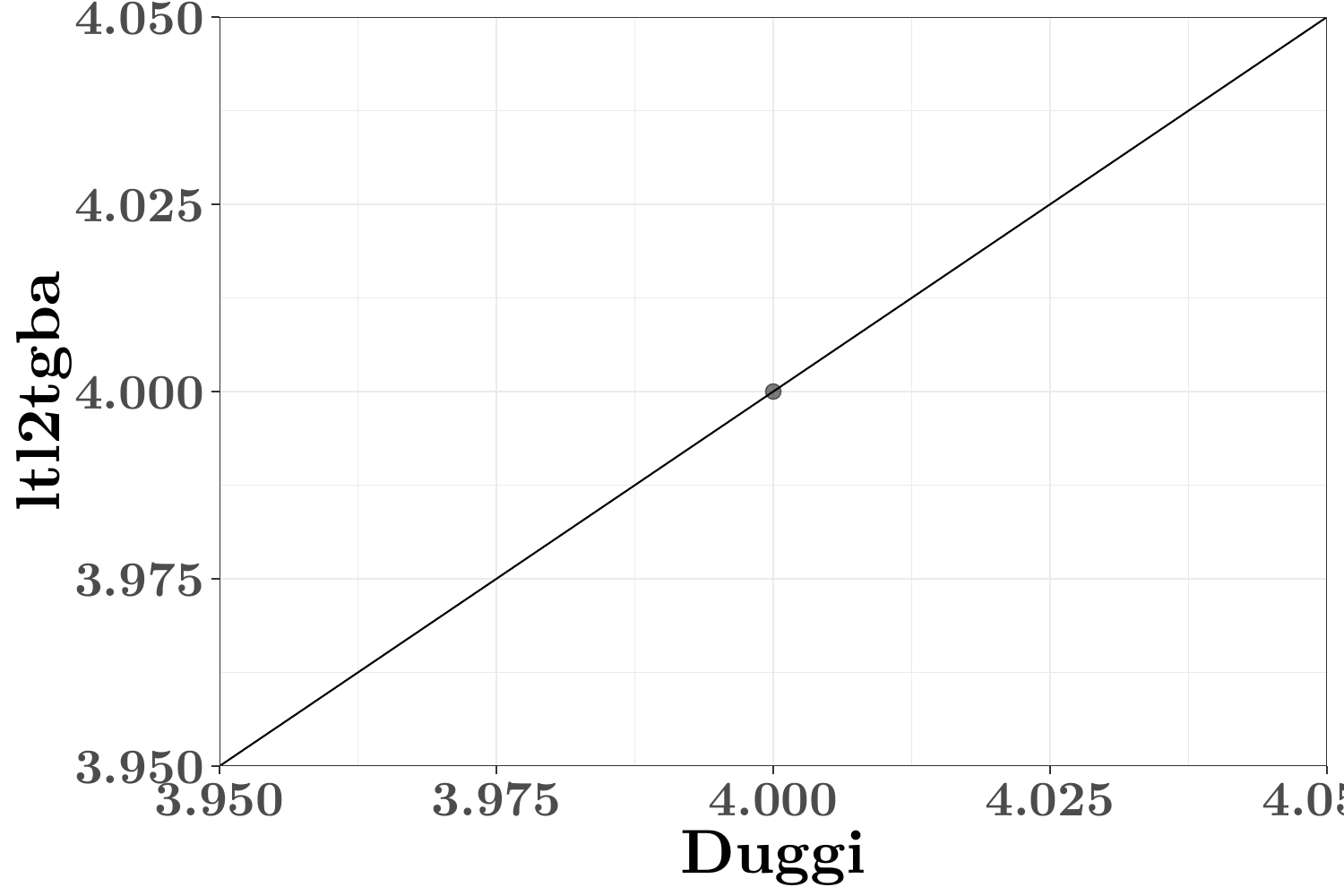}
    }
    \caption{lift, timeouts: \duggi{}: 30, \ltltotgba{}: 15}
  \end{subfigure}
  \hfill
   \begin{subfigure}[t]{0.47\textwidth}
     \centering
     \scalebox{0.37}{\includegraphics{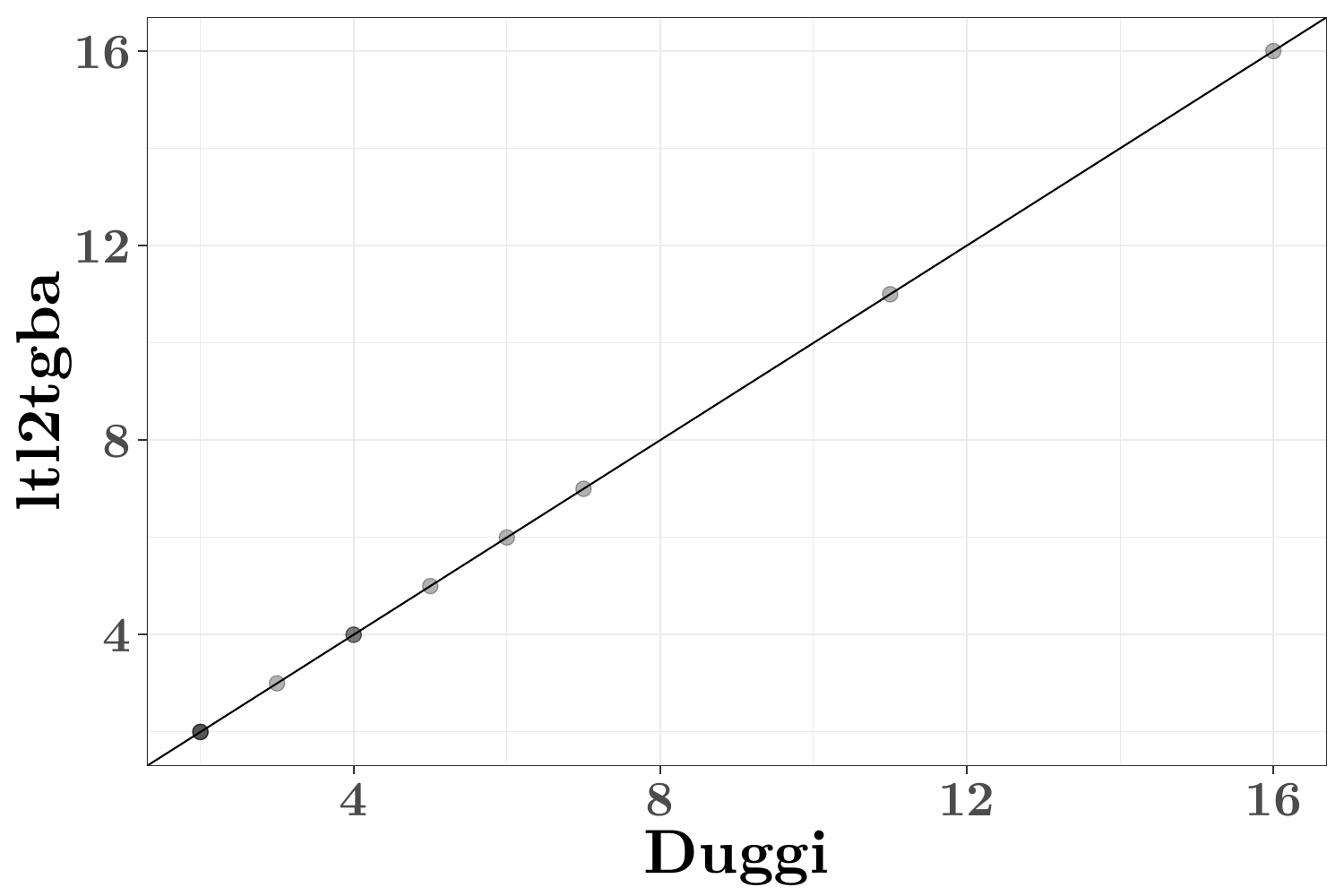}
     }
     \caption{U\_right, timeouts: \duggi{}: 8, \ltltotgba{}: 0}
   \end{subfigure}
\end{figure}

\FloatBarrier

\subsection{Markov chain analysis}
\label{app:pmc}

Here, we analyse shortly the missing formula for the cluster workstation
protocol described in \Cref{sec:pmc}. The formula is
\[\varphi^\until_k = \texttt{left} = n \, \until \, \Bigl(\texttt{left} = n-1 \, \until \, \bigl(\ldots \, \until \, (\texttt{left} = n-k \, \until \, \texttt{right} \neq n)\bigr)\Bigr)\]
and describes:
        ``The first \(k\) failures occur on the first/left cluster.''

\begin{figure}[htbp]
    \centering
  \resizebox{0.4 \textwidth}{!}{\includegraphics{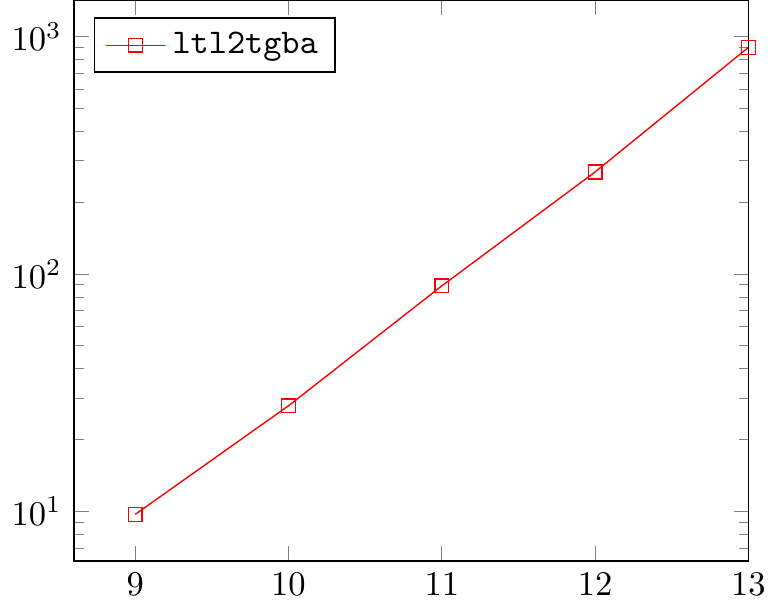}
   }
  \caption{Model checking times for the cluster workstation protocol for the
  formula \(\varphi^\until_k = \texttt{left} = n \, \until \, \Bigl(\texttt{left} = n-1 \, \until \, \bigl(\ldots \, \until \, (\texttt{left} = n-k \, \until \, \texttt{right} \neq n)\bigr)\Bigr)\)}
  \label{fig:cluster_until}
\end{figure}

For \(\varphi^\until_k\), the results (depicted in \Cref{fig:cluster_until}) shows
that in this case \ltltotgba{} performed significantly better, as \duggi{} was
not able to finish the model checking procedure for any \(k\). In particular the
automata generation took to long. However, the language described by
\(\varphi^\until_k\) is WDBA recognizable, which for this particular formula,
can be recognized already syntactically. Also, the WDBA generated by
\ltltotgba{} is smaller than the UBA generated (without WDBA minimization),
e.g., for \(k=9\) the (complete) WDBA has \(11\) states, whereas the (complete)
UBA has \(57\) states.

\end{document}